%% file: paper.tex
\pdfoutput=1

\documentclass[acmsmall,screen]{acmart}

\AtBeginDocument{%
  \providecommand\BibTeX{{%
    \normalfont B\kern-0.5em{\scshape i\kern-0.25em b}\kern-0.8em\TeX}}}

\setcopyright{rightsretained}
\acmJournal{PACMPL}
\acmYear{2020}
\acmVolume{4}
\acmNumber{OOPSLA}
\acmArticle{206}
\acmMonth{11}
\acmPrice{}
\acmDOI{10.1145/3428274}

\copyrightyear{2020}

\usepackage[utf8]{inputenc}
\usepackage[T1]{fontenc}
\usepackage{microtype}
\usepackage{mathpartir}
\usepackage{tikz}
\usepackage{graphicx}
\usetikzlibrary{arrows,automata, positioning, fadings, patterns}
\usepackage{wrapfig}
\usepackage{booktabs}
\usepackage{xspace}
\usepackage{stmaryrd}
\usepackage{mathtools}
\usepackage{adjustbox}
\usepackage{alltt}
\usepackage{natbib}
\usepackage{ottalt}
\input{macros}

\renewcommand\ottaltinferrule[4]{
  \inferrule*[narrower=0.6,right=\textsc{#1},#2]
    {#3}
    {#4}
}

\newtheorem{conjecture}{Conjecture}
\newtheorem{invariant}{Invariant}
\newtheorem{remark}{Remark}

\begin{document}

\title{Resolution as Intersection Subtyping via Modus Ponens}

\author{Koar Marntirosian}
\email{mardikoh15@gmail.com}
\author{Tom Schrijvers}
\affiliation{%
  \institution{KU Leuven}
  \city{Leuven}
  \country{Belgium}
}
\author{Bruno C. d. S. Oliveira}
\email{bruno@cs.hku.hk}
\affiliation{%
  \institution{The University of Hong Kong}
  \country{Hong Kong}
}
\author{Georgios Karachalias}
\email{georgios.karachalias@tweag.io}
\affiliation{%
  \institution{Tweag}
  \city{Paris}
  \country{France}
}
%
%
%
%
%

\begin{abstract}
Resolution and subtyping are two common mechanisms in
programming languages. Resolution is used by features such as
type classes or Scala-style implicits to synthesize values
automatically from contextual type information. Subtyping is
commonly used to automatically convert the type of a value
into another compatible type. So far the two mechanisms have been
considered independently of each other.
This paper shows that, with a small extension, subtyping with
intersection types can subsume resolution. This has 
three main consequences. Firstly, resolution does not need to
be implemented as a separate mechanism. Secondly, the
interaction between resolution and subtyping becomes apparent. Finally,
the integration of resolution into subtyping enables
\emph{first-class (implicit) environments}.
The extension that recovers the power of resolution
via subtyping is the \emph{modus
  ponens} rule of propositional logic.
While it is easily added to
declarative subtyping,
significant care needs to be taken to retain desirable properties,
such as \emph{transitivity} and \emph{decidability} of algorithmic subtyping, and \emph{coherence}.
To materialize these ideas we develop \calculus, a calculus that extends
a previous calculus with disjoint intersection types, and develop its
metatheory in the Coq theorem prover.
\end{abstract}



\maketitle

\input{sources/introduction.mng}

\input{sources/background.mng}


\input{sources/overview.mng}

\input{sources/calculus2.mng}

\input{sources/algorithm.mng}
\input{sources/coherence.mng}

\input{sources/related-work.mng}

\input{sources/conclusion.mng}


\begin{acks}                            
  We are grateful to anonymous reviewers that helped improving the presentation
of our work.
This work has been sponsored by Hong Kong Research Grant Council projects number
  17210617 and 17209519, Flemish Fund for Scientific Research project number 
G073816N and KU Leuven project number C14/20/079. 
\end{acks}

\newpage

\bibliographystyle{ACM-Reference-Format}
\bibliography{references}

\appendix

\input{sources/target-specification.mng}

\input{sources/definitions.mng}

\input{sources/context-typing.mng}

\input{sources/metafunctions.mng}

\input{sources/algorithm-example.mng}

\input{sources/coherence_appendix.mng}
\input{sources/termination-proof.mng}

\end{document}

%% file: macros.tex
\definecolor{BurntOrange}{rgb}{0.8, 0.33, 0.0}
\definecolor{ForestGreen}{rgb}{0.13, 0.55, 0.13}
\definecolor{RubineRed}{rgb}{0.88, 0.07, 0.37}
\definecolor{RoyalPurple}{rgb}{0.47, 0.32, 0.66}

\makeatletter
\DeclareRobustCommand{\leftloop}{%
  \mathrel{\mathpalette\left@loop\relax}%
}
\newcommand{\left@loop}[2]{%
  \vphantom{\looparrowright}
  \smash{\clipbox{0 {.43\height} {.64\width} 0}{$\m@th#1{\looparrowright}$}}%
}
\makeatother

\newcommand*{\substypingsymbol}{\prec:}
\newcommand{\lad}[1]{||#1||}
\newcommand*{\inAnd}{\in_{\scriptstyle \&}\hspace{-2pt}}
\newcommand*{\arrloopsymbol}{\hookrightarrow}
\newcommand*{\arrloop}[4]{(#1 \substypingsymbol #2)\arrloopsymbol (#3 \substypingsymbol #4)}
\newcommand*{\mploopsymbol}{\looparrowright}
\newcommand*{\mploop}[3]{#2\mathrel{\mploopsymbol^{\raisebox{2pt}{\hspace{-2mm}{$\scriptstyle #1 $}}}} #3}
\newcommand*{\mploopseqsymbol}{\mploopsymbol\hspace{-8pt}\mploopsymbol}
\newcommand*{\mploopseq}[3]{#2\,\mploopseqsymbol^{\raisebox{2pt}{\hspace{-1mm}{$\scriptstyle #1 $}}} #3}
\newcommand*{\SLookup}[3]{#2 \underset{\scriptstyle #1}{\bowtie} #3}
\newcommand*{\SBoundAux}[4]{#3 \underset{\scriptstyle #1}{\overset{#2}{\bowtie}} #4}
\newcommand*{\SBound}[3]{#2 \overset{#1}{\bowtie} #3}
\newcommand*{\tyrange}[1]{\llparenthesis #1 \rrparenthesis}

\newcommand{\necolus}{\textsf{NeColus}\xspace}
\newcommand{\calculus}{$\lambda_i^{\mathsf{MP}}$\xspace}

\newcommand{\target}{$\lambda_\mathit{c}^{\mathsf{MP}}$\xspace}

\newcommand{\cochis}{\textsc{Cochis}\xspace}
\newcommand{\ficalculus}{$F_i$\xspace}
\newcommand{\fiplus}{$F_i^+$\xspace}
\newcommand{\fsub}{System $F_\le$\xspace}

\newcommand{\bruno}[1]{\textcolor{blue}{\textbf{Bruno:} #1}}

\newsavebox{\circlehorlinebox}
\newcommand{\circlehorline}[0]{%
\begin{tikzpicture}[scale=0.20]%
\draw (0,0.5cm) -- (1.6cm,0.5cm);%
\draw (0.8cm,0.5cm) circle(0.4cm);
\end{tikzpicture}%
}
\savebox{\circlehorlinebox}{\circlehorline}
\newcommand*\modusponens{\,\vcenter{\hbox{\usebox{\circlehorlinebox}}}\,}

\newcommand{\ottnt}[1]{\mathit{#1}}
\newcommand{\ottsym}[1]{#1}
\newcommand{\ottmv}[1]{\mathit{#1}}

\newenvironment{fakedisplaymath}[0]
{\begin{center}$\displaystyle}
{$\end{center}}

\newsavebox{\fmbox}
\newenvironment{setofrules}[3][\textwidth]
{\begin{center}
\begin{minipage}{#1}\framebox{\mbox{#2}} \quad #3 \\[0pt]\begin{mathpar}}
{\end{mathpar}\end{minipage}
\end{center}
}

\newcommand{\facerule}[1]{{\tt\sc #1}}
\newcommand{\rlabel}[1]{{\small\mbox{\facerule{#1}}}}
\renewcommand{\drule}[3]{\inferrule*[vcenter,lab={\rlabel{#1}}]{#2}{#3}\label{rule:#1}}

\newcommand{\nrule}[3]{\inferrule*[vcenter,right={\rlabel{#1}}]{#2}{#3}}
\newcommand{\defeq}[0]{~\triangleq~}

\makeatletter
\newcommand{\midnextconst}[1]{ \mid #1\@ifnextchar\bgroup{\midnextconst}{}}
\newcommand{\checknextconst}{\@ifnextchar\bgroup{\midnextconst}{}}

\makeatother

\newcommand{\reducedstrut}{\textcolor{gray!20}{\vrule width 2pt height 1\ht\strutbox depth .8\dp\strutbox\relax}}
\newcommand{\elbox}[1]{%
  \begingroup
  \setlength{\fboxsep}{0pt}%
  \colorbox{gray!20}{\reducedstrut\(#1\)\/}%
  \endgroup
}

\newcommand{\wfd}[1]{\vdash_{d} #1}

\newcommand{\rellV}{\mathcal{V}}
\newcommand{\rellE}{\mathcal{E}}
\newcommand{\rellG}{\mathcal{G}}

\newcommand{\relV}{\(\rellV\)}
\newcommand{\relE}{\(\rellE\)}

\newcommand{\logeq}[2]{#1 \simeq_{\text{log}} #2}

\newcommand{\Then}{\Rightarrow}

\makeatletter
\newcommand\kleene{\@ifnextchar\bgroup{\@kleene}{}}
\newcommand\@kleene[1]{#1\@@kleene}
\newcommand\@@kleene{\@ifnextchar\bgroup{\@@@kleene}{}}
\newcommand\@@@kleene[1]{\simeq_{k}#1\@@kleene}
\makeatother
\makeatletter
\newcommand\Ckleene{\@ifnextchar\bgroup{\@Ckleene}{}}
\newcommand\@Ckleene[1]{#1\@@Ckleene}
\newcommand\@@Ckleene{\@ifnextchar\bgroup{\@@@Ckleene}{}}
\newcommand\@@@Ckleene[1]{\simeq_{\mathcal{C}}#1\@@Ckleene}
\makeatother

\newcounter{goal}
\newcounter{ih}
\newcounter{subcase}

\newcommand{\casebox}[1]{\textbf{Case:} \framebox[1.1\width]{\( #1 \)}}
\newcommand{\subcasebox}[1]{\textbf{Subcase:} \framebox[1.1\width]{\( #1 \)}}

\newcommand{\tagG}{\refstepcounter{goal}\tag{G\thegoal}}

\makeatletter
\newcommand{\gobblenextarg}[1]{, \eqref{#1}\@ifnextchar\bgroup{\gobblenextarg}{}}
\newcommand{\checknextarg}{\@ifnextchar\bgroup{\gobblenextarg}{}}

\makeatother

\newcommand{\subturnstyle}[1]{\mathrel{\vdash_{\text{#1}}}}
\newcommand*\turnstyleR{\subturnstyle{R}}
\newcommand*\turnstyleL{\subturnstyle{L}}

\newcommand{\algSubMain}[3]{{#1} \substypingsymbol {#2} \elbox{\rightsquigarrow {#3}}}
\newcommand{\algSubR}[4]{{#1} \turnstyleR {#2} \substypingsymbol {#3} \elbox{\rightsquigarrow {#4}}}
\newcommand{\algSubL}[7]{{#1}; {#2}; {#3}; \elbox{{#4}} \vdash_{\text{L}} {#5} \substypingsymbol {#6} \elbox{\rightsquigarrow {#7}}}

\newcommand{\noEvAlgSubMain}[2]{{#1} \substypingsymbol {#2}}
\newcommand{\noEvAlgSubR}[3]{{#1} \turnstyleR {#2} \substypingsymbol {#3}}
\newcommand{\noEvAlgSubL}[5]{{#1}; {#2}; {#3} \vdash_{\text{L}} {#4} \substypingsymbol {#5}}

\newcommand{\arrows}{\to\hspace{-8pt}\to}

\newcommand{\distArrowsCo}[4]{\llbracket{#1}\rrbracket_{\&}^{{#2},{#3},{#4}}}
\newcommand{\distArrowsTy}[3]{\llbracket{#1}\rrbracket_{\&}^{{#2},{#3}}}

\newcommand{\necolusAlgSub}[4]{{#1} \vdash {#2} \ll {#3} \elbox{\rightsquigarrow {#4}}}

%% file: sources/introduction.mng
\section{Introduction}
\label{sec:intro}

\emph{Subtyping} is a well-known programming language feature. It is most
notably employed in object-oriented languages, but there are many other uses of
subtyping in the literature~\citep{PierceS96,bcd-subtyping,Chen03,GaySJ}. 
Subtyping automatically converts the type of a value into
another compatible type. One particular form of subtyping arises from
\emph{intersection types}~\citep{bcd-subtyping,ReynoldsCoherence}. 
Intersection types support---among
others---a form of \emph{finitary overloading}~\citep{overloaded-functions-subtyping}. For
example, in various systems with intersection types, it is possible to
build overloaded functions such as:
\begin{verbatim}
  f : (Int -> Int) & (Char -> Char) = succ ,, toUpper
\end{verbatim}
The \emph{merge operator} (\verb|,,|) allows the function \verb|f| to behave as
both the \verb|succ| and the \verb|toUpper| functions.
Thus, \verb|f| can be applied both to an integer and to a
character. When applied to an integer, \verb|f| behaves like the
successor function, whereas when applied to a character \verb|f|
capitalizes its input. The automatic selection of 
the appropriate behavior (from the overloaded definition \verb|f|) 
is based on the type of the input, which triggers an upcast 
induced by the subtyping relation. For instance, in the 
case of the application \verb|f 3|, the input of the function 
must be \verb|Int| and therefore, the following upcast is
triggered in such an application:
\begin{verbatim}
  (Int -> Int) & (Char -> Char) <: Int -> Int
\end{verbatim}
\emph{Resolution} is another programming language mechanism. 
Haskell type classes~\citep{typeclasses} employ resolution to automatically
\emph{search} for and \emph{compose} instances of type classes. Resolution is
also used by Scala's
implicits~\citep{Oliveira:2010}. Many other programming
languages have mechanisms similar to type classes and
implicits~\citep{Devriese:2011gp,ModularImplicits,coqtypeclasses}, and
consequently employ a form of resolution. Type classes and 
implicits can also be used to define overloaded functions. For 
example, in Haskell, we may define:
\begin{verbatim}
  class Over a where  f :: a -> a
  instance Over Int where f = succ
  instance Over Char where f = toUpper
\end{verbatim}
This achieves an overloaded behavior similar to the
intersection types version of \verb|f|. Indeed, 
the Haskell version of \verb|f| can also 
be applied both to an integer and to a character with similar results.
However, in Haskell, the selection of the right function in the
application \verb|f 3| uses resolution. In Haskell 
(and languages with implicits) there is an implicit environment 
that stores information about type class instances. When class methods 
are used, resolution searches the implicit environment to find the
right instance or to combine existing instances to find a suitable
method implementation. 

The function \verb|f| illustrates the similarities and overlapping 
use cases for subtyping and resolution. However there are also various
use cases that are not in common. On the one hand, with subtyping we can, for instance, have 
\verb| Int -> Int <: Int -> Top|. Such a conversion is typically 
not allowed by resolution mechanisms documented in the
literature~\citep{HaHaJoWa1996,cochis,simplicity,typeclasses,implicitcalculus}. 
On the other hand, a key feature of resolution is 
its ability to compose instances (or implicits). For instance,
suppose that we want an overloaded version of \verb|f| for lists of integers. 
With type classes we can define the instance:
\begin{verbatim}
  instance Over Int => Over [Int] where
    f []       = []
    f (x : xs) = f x : f xs
\end{verbatim}
Note that here the context \verb|Over Int| is not necessary in Haskell, 
but for illustration purposes, we add the redundant context.
With this instance, if we call \verb|f [1,2,3]|, then the result is
\verb|[2,3,4]|. For this to work, 
resolution needs to find an instance for \verb|Over [Int]|. While 
no instance directly matches the required type, 
resolution composes the instance for lists with the instance for
integers to build an instance of the appropriate type. 
The ability to perform such compositions is lacking 
in traditional mechanisms with subtyping and intersection types.  

This paper shows that, with a small extension, subtyping with
intersection types can in fact subsume forms of resolution. 
The extension that is necessary to recover the power of resolution 
via subtyping is essentially the logical rule for \emph{modus
  ponens}: 
\begin{fakedisplaymath}\small
  \inferrule*[vcenter,right={\rlabel{Modus Ponens}}]
    {A \vdash B \to C \and A \vdash B}
    {A \vdash C}
\end{fakedisplaymath}
Modus ponens states that if from $A$ one can conclude both 
$B \to C$ and $B$ then one can also conclude $C$.
The key idea in this paper is that we can adapt modus ponens 
to subtyping, and thus enable its power (and the essential 
power of resolution). To accomplish this, we simply reinterpret 
entailment ($A \vdash B$) as subtyping ($A <: B$). 
Modus ponens captures the essence of composition that is necessary
to enable the following derivation:
\begin{verbatim}
  (Int -> Int) & ((Int -> Int) -> [Int] -> [Int]) <: [Int] -> [Int]
\end{verbatim}
which could be used in a language with intersection types 
and subtyping, to find an overloaded implementation of \verb|f| in a
call such as \verb|f [1,2,3]| (just like the version with type classes).

While modus ponens is easy to add in a declarative version
of subtyping, significant care needs to be taken to retain desirable
properties, such as transitivity and decidability of 
subtyping. Our work builds and extends upon the \necolus calculus~\citep{necolus}---or $\lambda_{i}^{+}$---which
has both transitivity and decidability of subtyping, as well as coherence of the elaboration semantics.
However, \necolus does 
not support modus ponens. Extending the proofs of transitivity, decidability
and coherence of \necolus 
to support modus ponens is highly non-trivial. The key difficulty 
is that modus ponens makes 
the context in which a type appears relevant for resolution
(which is not the case in \necolus).

Adding modus ponens to subtyping, and viewing 
resolution as a special case of subtyping, 
comes with important advantages. 
Firstly, resolution does not need to be implemented
as a completely separate mechanism. This can make implementations 
as well as metatheory more compact, because the two
mechanisms are generalized into a single relation.  
Secondly, the interaction between
resolution and subtyping becomes apparent, which contrasts 
with the current state-of-the-art where it is ignored.\footnote{For
example, GHC issue ticket
\href{https://gitlab.haskell.org/ghc/ghc/issues/17295}{17295} is a closely
related problem, caused by an unexpected interaction between resolution and
overlapping ``intersections''.} Indeed,
various foundational calculi related to Scala formalize either 
subtyping~\citep{dot1,dot2} or implicits~\citep{implicitcalculus,cochis,simplicity}, but usually not both.
The only exception is the recent work by~\citet{jeffery}, which studies a calculus 
with both features (but where resolution and subtyping are modeled independently).
Finally, the integration of
resolution into subtyping enables \emph{first-class (implicit)
  environments}. That is, we can create (implicit) environments
(encoded as intersection types), 
use them as arguments, or return them from functions. 
This is impossible in existing approaches for type classes and calculi with implicits, as
there is no explicit notion of a type for environments.

To materialize these ideas we develop \calculus: a calculus that extends 
the \necolus calculus (a calculus with disjoint intersection
types~\citep{DisjointIntersectionTypes}) 
and develop its metatheory in the Coq theorem prover.
In summary, the contributions of this work are:

\begin{itemize}

\item {\bf Resolution as subtyping:} We extend subtyping
  of intersection types with \emph{modus ponens}. 
  This enables composing components from intersections to
  solve subtyping problems, and (together with other standard
  intersection rules) subsumes forms of resolution.

\item {\bf The \calculus calculus:} We develop a concrete calculus 
  that illustrates the idea of subtyping as resolution. The calculus
  is an extension of \necolus~\citep{necolus}. We prove several basic properties, including 
  type-safety, which is proved via an elaboration into the target language \target, a simply
  typed lambda calculus extended with coercions.
  
\item {\bf Algorithmic subtyping for \calculus:} We incorporate
  modus ponens in the \necolus subtyping algorithm by (1) restructuring it
  in two mutually recursive judgments---one of which builds coercions
  by accumulating them in a continuation-passing form---and (2) adding
  a loop detector for termination.
  Our Coq proofs show that our algorithm
  subsumes that of \necolus, and is \emph{sound}, \emph{complete} and \emph{terminating}; this
  establishes that the subtyping of \calculus is decidable.

  The proofs of termination and transitivity, which is used to show
  completeness, are both highly non-trivial. The former uses different
  arguments for three different kinds of loops, including reasoning about
  the effectiveness of loop detection. The latter 
  consists of 7 mutually recursive lemmas with complex interdependencies. To 
  establish the well-foundness of their proof we employ the
  AProVE tool~\citep{aprove}.

\item {\bf Conditional Coherence for \calculus:} Establishing
  coherence of \calculus is also a challenging technical goal. The addition of
  modus ponens makes it a lot harder due to the ability to 
  compose information.
  We formulate a non-trivial extension of the
  canonicity relation of \necolus, based on which we build a mechanized partial
  coherence proof in the Coq theorem prover.
\end{itemize}

\noindent
The supplementary material contains a Coq mechanization of the definitions (4k lines)
and theorems (5.6k lines) in this paper, as well as a prototype implementation in Haskell (\textasciitilde 820 lines). It is available online at \url{https://zenodo.org/record/4043646#.X4gBAHVfhjs}.



%% file: sources/background.mng
\section{Background}
\label{sec:background}

This section introduces background on resolution,
coercive subtyping, intersection types and their use in programming languages.

\subsection{Coercive Subtyping, Resolution and their Computational Interpretation}

This paper brings together two different programming language features:
(implicit) \emph{resolution} as used in Haskell type classes and Scala implicits on the
one hand, and \emph{type-directed coercive subtyping} on the other hand.  While these
two features are very different on the surface, they are in fact both based on
similar computational interpretations of logical inference. The different rules
of logical inference they use are compatible. In this paper we show how to extend this
compatibility to the language design level and how to ensure properties that,
while typically of no concern for logical inference, are key for programming
languages, like coherence and decidable type checking.

In the sequent notation, the judgment $A \vdash B$ asserts that proposition $B$
can be inferred from proposition $A$. An inference rule provides a way to
construct a valid (or true) judgment. For instance, the inference rule
\vspace{-5pt}
\begin{fakedisplaymath}
  \inferrule*{ }{A \vdash A}
\end{fakedisplaymath}
states that the judgment $A \vdash A$ (from $A$ follows $A$) is valid for any
proposition $A$.

The $A \vdash B$ judgment can be given a computational interpretation following
the ``propositions as types'' approach \cite{proptypes}. This means that we take the types of a
programming language as the possible propositions. The judgment $A \vdash B$ then
means that there exists a computable function that turns values of type $A$
into values of type $B$. We call this function the (computational) evidence for the judgment. 
The evidence for a judgment follows from the inference rules that have been used
to establish that judgment. 
For instance, the evidence for the trivial $A \vdash A$ judgment
above is the identity function $\lambda x. x$.
A more advanced inference rule is that for transitivity
\begin{fakedisplaymath}
\inferrule{A \vdash B \\ B \vdash C}{A \vdash C}
\end{fakedisplaymath}
which allows to conclude $C$ from $A$ provided that $B$ can be concluded from
$A$ and $C$ from $B$. Given evidence $f$ for $B \vdash C$ and evidence $g$ for
$A \vdash B$, the evidence for $A \vdash C$ is $\lambda x. g\,(f\,x)$.

\paragraph{Coercive Subtyping}
A first programming language application of this mechanism is coercive
subtyping~\cite{luo:coercive:1999}.
In this case we write $A <: B$ instead of $A \vdash B$ and call $A$
and $B$ subtype and supertype respectively, rather than premise and conclusion.
Also we call the associated evidence of the judgment a coercion. Unlike other
forms of subtyping, coercive subtyping changes the runtime representation of a
value of type $A$ when upcasting to type $B$. This change is accomplished by
the coercion. Even though the coercion is executed at runtime, it is already
statically known at what point in the program it needs to be performed, namely
at the point in the static type system where an appeal is made to subtyping.
In order to introduce the coercion, coercive subtyping can therefore make use of
a static program transformation approach called (type-directed) elaboration.

Elaboration uses the information of static typing to transform a program with
implicit appeals to subtyping into a program with explicit applications of
coercions. For example, consider a function $f$ that expects a value of type
$B$ and a value $v$ of type $A$ where $A <: B$. Hence, in the function call
$f\,v$ there is a type mismatch that the type system bridges by appealing to
subtyping. Elaboration then transforms this function call into $f\,(c\,v)$
where $c$ is the coercion for $A <: B$. The explicit coercion application 
replaces the implicit appeal to subtyping.

%
%

\paragraph{Resolution}
A rather different application of logic inference in programming languages are
implicits in Scala. In their basic form, they 
involve an environment
of premises that the programmer can populate and extend with particular types
and values of those types as evidence. Functions can take implicit parameters
which the type system fills in by searching values of the appropriate type in the
implicit environment. Elaboration turns those implicit parameters into explicit
ones. This behavior is extended with the modus ponens inference rule,
to make it more powerful.

The computational evidence associated to this rule, shown in Section~\ref{sec:intro},
is $\lambda x. (f\,x)\,(g\,x)$ where $f$ is the evidence for $A \vdash B \to C$
and $g$ for $A \vdash B$.
The modus ponens rule can be applied recursively, e.g. to establish $B$ from $A$ by modus
ponens via $D \to B$ and $D$, and we call the general reasoning principle
\emph{recursive resolution}.


\paragraph{Resolution in Scala}
As a concrete example of resolution in programming languages,
consider the judgment
\vspace{-5pt}
\begin{fakedisplaymath}
  P \to Q, Q \to U, P \vdash U
\end{fakedisplaymath}
In this statement, $P$ implies $Q$, $Q$ implies $U$ and $P$ holds as a fact. Thus,
we can conclude that $U$ holds by straightforward logical reasoning.

\begin{figure}[t]
\begin{verbatim}
  object Test {
    class P                                        // proposition P
    class Q                                        // proposition Q
    class U                                        // proposition U

    implicit def f(implicit o : P) : Q = new Q()   // P -> Q
    implicit def g(implicit o : Q) : U = new U()   // Q -> U
    implicit val p : P = new P()                   // P

    val u : U = implicitly[U]                      // triggers resolution from 
                                                   // the implicit environment
  }
\end{verbatim}
\caption{Encoding of the judgment $P \to Q, Q \to U, P \vdash U$ as a Scala program.}
\label{fig:resolution}
\end{figure}

Figure \ref{fig:resolution} illustrates how to encode the judgment in Scala.
In this Scala program, we encode propositions as classes, and
we define a few implicit rules that are added to the "implicit environment"
of Scala. The definition of \verb|f| encodes $P\to Q$, the definition of \verb|g|
encodes $Q\to U$ and the definition of \verb|p| encodes that $P$ holds.
In the definition of the value \verb|u|, we trigger resolution (via the 
\verb|implicitly| keyword), which essentially searches the implicit environment 
and attempts to compose the rules and facts in the implicit environment, to find a 
value of type \verb|U|. This value will be given eventually by the synthesized expression 
\verb|g(f(p))|.

Essentially, in logic, resolution can be seen as a mechanism for
proof search. In programming languages, resolution is a mechanism that
searches for values of a given type by using the rules and facts of the implicit
environment. It can be viewed as a
type-directed form of program synthesis.

The connection to our work is that we interpret the
resolution relation as subtyping, and the conjunction of propositions
in the environment as intersections. Thus, the above judgment
corresponds to:
\vspace{-5pt}
\begin{fakedisplaymath}
  (P \to Q) \& (Q \to U) \& P <: U
\end{fakedisplaymath}

\noindent using subtyping and intersection types.

Haskell type classes also employ resolution,
but there are several superficial differences to Scala implicits. The name
spaces of implicit values---called type classes---is separate from that of
other types and has global scope. Moreover
type classes are used as a mechanism for overloading function names.
Yet, in essence, just like Scala implicits, type classes are
based on the modus ponens inference rule for constructing wanted evidence for 
types in terms of given evidence of other types. We will look at type classes
in more detail in Section~\ref{sec:overview}.

\subsection{Different Interpretations for Intersection Types in the Literature}

There are several different interpretations of intersection types
in the literature.
Intersection types date back to work from~\citet{Coppo1981FunctionalCO}
and~\citet{pottinger1980type} and were first used to characterize the strongly
normalizable terms of the $\lambda$-calculus.

Following Forsythe~\citep{forsythe}, intersection types were used in various other
programming languages, including CDuce~\citep{cduce} and Object Oriented languages like
TypeScript\footnote{\url{https://www.typescriptlang.org/}}, Flow~\citep{raskin1974flow},
Ceylon\footnote{\url{https://ceylon-lang.org/}} and Scala~\citep{scala-implicits}, enabling multiple
inheritance~\citep{PierceThesis}.
They were also studied as part of the syntax of several calculi, to enable
various forms of polymorphism such as finitary function
overloading~\citep{overloaded-functions-subtyping}, bounded
polymorphism~\citep{pierce_1997} and evaluation-order
polymorphism~\citep{EvaluationOrderPoly}.
Refinement types is yet another
feature where intersection types has been
studied~\citep{freeman1991refinement, DaviesPfenning2000,stardust}.
However, in those systems the
type \textsf{Int \& Bool} is invalid, since \textsf{Int} and \textsf{Bool} are
not refinements of each other.

Note that intersection types alone only increase the expressive power of types.
While languages with intersection types do not necessarily have a merge operator,
the latter adds expressive power to the terms.
This term construct has been used in various, but usually limited, forms like in
Forsythe or in the work of~\citet{overloaded-functions-subtyping}, discussed in
Section~\ref{sec:related}.

The use of intersection types in this paper follows the interpretation of
~\citet{DunfieldIntersectionElaboration}
who has proposed a particular flavor of intersection
types with an unrestricted merge operator, allowing the construction of values that
inhabit 
intersection types such as \textsf{Int \& Bool}. In her work, Dunfield uses coercive
subtyping, based on the inference rules for the logical conjunction operator.
In particular, conjunction elimination is expressed in two rules:
\begin{fakedisplaymath}
  \inferrule*{ }{A\land B\vdash A}\quad\quad\quad\quad\quad\quad\quad
  \inferrule*{ }{A\land B\vdash B}
\end{fakedisplaymath}
Here, we write $A \& B$, instead of $A \wedge B$, to denote an intersection type.
The runtime representation of values of type $A \& B$ is a
tuple $(v,w)$ where $v$ has type $A$ and $w$ has type $B$. The coercions
associated with the two inference rules are respectively $\pi_1$ and $\pi_2$,
the projection on respectively the first and second component of the tuple.
Dunfield observes that her formulation of the calculus does not eliminate ambiguous
expressions. However, recent work by~\citet{DisjointIntersectionTypes} remedies this issue
by introducing a notion of disjoint intersection types.

Our proposed design fits well with this interpretation, i.e. intersection types as a
mechanism to implicitly provide coercions on pairs. 
Our source language provides a layer of convenience on top of
an ordinary target language with pairs,
by generating various types of coercions involving pairs.
In particular, the modus ponens
rule is just one of the possible coercions involving pairs.

%% file: sources/overview.mng
\section{Overview}\label{sec:overview}

This section gives an overview of our work.
We are mainly interested in the \emph{type-based overloading} and \emph{resolution} capability at
the core of type classes or implicits. In what follows, we use Haskell
type classes for illustration, but we could have used Scala implicits
(or other similar mechanisms) just as well.
We start by establishing
the relationship between type-based overloading based on type classes
and/or implicits and disjoint intersection types. 
Then we explain the key idea of our work to
add modus ponens to the subtyping relation for a calculus with
(disjoint) intersection types to recover resolution. Finally, we
discuss several technical challenges that need to be addressed for
this addition to work.

\subsection{Type-based Overloading with Type Classes}

Type classes in Haskell were initially meant to provide
a less ad-hoc mechanism to overloading. In Section~\ref{sec:intro} we
have already seen a simple example of an overloaded function
\texttt{f} defined via a type class. Overloading with type classes is based on
\emph{parametric polymorphism}~\citep{Re1974}. All type classes require
at least one type argument.
This type argument is used to specify the places in the signatures
of overloaded methods where different types can occur.
For the type class \texttt{Over} the method \texttt{f\,::\,a -> a} can
be overloaded both on the input type and the output type and
those two types need to be the same for a particular overloading.

Conventionally, type class instances cannot overlap. That is, two instances for a type class must 
use distinct types. This property is satisfied by the
three instances in Section~\ref{sec:intro}:
\begin{verbatim}
  instance Over Int where ...
  instance Over Char where ...
  instance Over Int => Over [Int] where ...
\end{verbatim}
Adding a fourth instance for a type that is distinct
from those in the previous instances (i.e. \texttt{Int}, or \texttt{Char}
or \texttt{[Int]}) is allowed. For example, the following
\verb|Bool| instance is allowed:
\begin{verbatim}
  instance Over Bool where f = not
\end{verbatim}
However adding another instance for integers is not allowed:
\begin{verbatim}
  instance Over Int where f x = x + 2  -- rejected!
\end{verbatim}
The compiler complains that there is already
another instance in scope for \texttt{Over\,Int}.

\paragraph{Overlapping instances and ambiguity}
The restriction of non-overlapping instances is put into place to ensure
that there is no ambiguity in a program. If the second instance for integers
were accepted then the definition:
\begin{verbatim}
  p = f 1        -- ambiguous: could be 2 or 3
\end{verbatim}
would be ambiguous as it could choose either of the two integer instances.

\subsection{The Merge Operator and Disjoint Intersection Types}

The merge operator, originally proposed by~\citet{forsythe}, adds expressive power to 
intersection types. In most calculi with intersection types, a type
such as \texttt{Int\&Char} is not inhabited, since these two types are
\emph{disjoint} (i.e. have no common values), so their
intersection is empty. With the merge operator, however, it is
possible to build values that inhabit such an intersection type.
For example:
\begin{verbatim}
  v : Int & Char = 1 ,, 'c'
\end{verbatim}
In the definition of \texttt{v}, the merge operator (\verb|,,|),
is used to build a value of type \texttt{Int\&Char} from an integer and a
character. In calculi with a merge operator, intersection types are 
similar to conventional pair types. The merge operator is used to
build values of the intersection types, similarly to how we build
values for pairs. The main difference to pairs and pair types
is that while for pairs the eliminations (i.e. projections such as
\texttt{fst} and \texttt{snd}) are explicit, for intersection types
they 
are implicit and driven by the type system.

\paragraph{Disjointness and ambiguity} The merge operator is powerful,
but without any restrictions it is easy to have ambiguous programs.
For instance, consider the following value:
\begin{verbatim}
  n : Int & Int = 1 ,, 2
\end{verbatim}
If we use \texttt{n} in an expression such as \texttt{n + 1}
there could be two possible results (\texttt{2} and \texttt{3}) depending
on what integer gets selected from the merge.

Disjoint intersection types~\citep{DisjointIntersectionTypes} were proposed to solve ambiguity
problems in calculi with intersection types and a merge
  operator. The key idea is to introduce a restriction on
merges: a valid merge can only be built from two values with disjoint
types. Two types are disjoint if the only supertypes that they have
in common are top-like types~\citep{DisjointIntersectionTypes}. This restriction prevents the
merged value \texttt{n} from type-checking, while accepting \texttt{v}.
Furthermore, for previous calculi with disjoint intersection types
it has been proven that the disjointness restriction is sufficient
to guarantee \emph{coherence}~\citep{ReynoldsCoherence}, a
correctness property that rules out any ambiguity.
This property is important for calculi with non-deterministic semantics,
where well-formed expressions can take on multiple meanings, and ensures that
these meanings are equivalent, in the sense that the program which the initial
expression models will always behave the same.

\subsection{Disjoint Merges, Non-Overlapping Instances and Resolution}

Disjointness checks for merges and non-overlap checks for type
class instances are closely related. The non-overlap check for
type classes can be viewed as a simple form of disjointness checking.
If we ignore polymorphic types, then the non-overlap check
is simply testing whether two types are distinct\footnote{
With polymorphism we also need to account for instantiation.
For example the type \texttt{[Int]} is not equal to \texttt{[a]}, but it
does instantiate \texttt{[a]}. So 
two instances \texttt{C\,[a]} and \texttt{C\,[Int]} (for some
type class \texttt{C}) overlap.}, which is subsumed by a disjointness
check for merges.

\paragraph{Type class instances as merges}
The fact that disjointness checking relates/subsumes to
non-overlapping checking is crucial for expressing code analogous to type class instances
as merges. Consider a variant of \verb|f| in Section~\ref{sec:intro}:
\begin{verbatim}
  f : (Int -> Int) & (Char -> Char) & ((Int -> Int) -> [Int] -> [Int])
    = succ ,, toUpper ,, listInstance
\end{verbatim}
where \verb|listInstance| is analogous to the
type class instance \verb|Over [Int]| from Section~\ref{sec:intro}, with the type
\texttt{((Int -> Int) -> [Int] -> [Int])}. Here, the merge creates an overloaded function
and checks that the three implementations have disjoint types.
This is similar to checking whether \verb|Over Int|,
\verb|Over Char| and \verb|Over [Int]| overlap.
Therefore, we can essentially express
type class instances through merges, whose components
define the methods of the instances for overloading.

\paragraph{Resolution via Modus Ponens}
While we can express some type class instances with disjoint
intersection types, an important ``class'' of instances cannot be expressed
because previous work on disjoint intersection types lacks resolution. That is,
in previous work a program such as:
\begin{verbatim}
  p : [Int] = f [1,2,3]   
\end{verbatim}
would not type check because \verb|[Int] -> [Int]|, which is the
type that \verb|f| needs to have, \emph{is not} a supertype of the
type of \verb|f|. However the addition of modus ponens changes this
and \verb|[Int] -> [Int]| becomes a supertype of the type of \verb|f|,
thus enabling the above use of \verb|f|.

\subsection{First-Class Environments}
\label{subsec:FCE}
Viewing instances/implicits as merges opens up the possibility of new
and interesting features for programming with implicit values and resolution. In
particular, unlike type class instances (which are second class),
merges are \emph{first-class}. In other words, we can pass merged
values as arguments or return them as results. This is not possible
with type class instances. An instance declaration is top-level, but
we cannot create instances on the fly.\footnote{Though recent work
by~\citet{WinantDevriese2018} aims to rectify this.}

To illustrate the use of first-class environments, consider 
a program that processes data from a sensor measuring temperatures. 
The temperatures are collected at fixed time intervals, and the 
sensor may fail to read temperatures (for example due to 
temperatures being too high, low or to some unknown error). 
The sensor data is collected in a text file using a sequence 
of numbers or strings (for errors) separated by commas. 
For instance:
\begin{verbatim}
  32,-9,"LOW",10,"ERROR"
\end{verbatim}
With Haskell type classes one may be tempted to use 
the \texttt{Read} type class:
\begin{verbatim}
  class Read a where
     read :: String -> a
\end{verbatim} 
to process the String using the existing instances of the \texttt{Read} class:
\begin{verbatim}
  instance Read Int where ...
  instance Read String where ...
  instance Read a => Read [a] where ...
  instance (Read a, Read b) => Read (Either a b) where ...
\end{verbatim}
The programmer's hope here would be that we could 
interpret the type of sensor values as \texttt{[Either String Int]}, 
and then simply employ resolution to compose the above instances: 
\begin{verbatim}
  read "32,-9,\"LOW\",10,\"ERROR\"" :: [Either String Int]
\end{verbatim}
However this does not work because the input string 
is not in the correct format to be parsed by the \texttt{Read} 
instances. In Haskell the \texttt{Read} instances for types
such as \texttt{Either a b} and \texttt{[a]} use the Haskell 
syntax%
\footnote{The Haskell format requires that strings to be read as \texttt{Either a b} values start with a \texttt{Left}
  or \texttt{Right} substring. For list values, the string should start and end with opening and closing brackets.}%
.
The \texttt{read} method provided by \texttt{Read [Either String Int]} would
require the string:
\begin{verbatim}
  "[Right 32, Right -9, Left \"LOW\", Right 10, Left \"ERROR\"]"
\end{verbatim}
Because Haskell does not allow more than one
instance of a type class for the same type, it is not 
possible to use alternative instances for \texttt{Either a b} and
\texttt{[a]}. There are various workarounds for this issue, but none 
of these is ideal. 

\paragraph{A solution using first-class environments} An
alternative approach to solve this problem is to use merges 
to encode first-class environments for instances of \texttt{Read}.
We present such a solution using the \calculus calculus (assuming
a few convenience source language features).
For the purposes of illustration and readability we assume the 
following definition of \texttt{Read}: 
\begin{verbatim}
  type Read a = String -> a
\end{verbatim}
and we also assume built-in types for lists and sum types (i.e. modeling,
respectively, Haskell's \texttt{List a} and \texttt{Either a b}). Given those,
we can model four Haskell instances for \texttt{Read}: 
\begin{verbatim}
  readInt : Read Int = ...
  readString : Read String = ...
  readEitherSI : Read String -> Read Int -> Read (Either String Int) = ...
  readSensor : Read (Either String Int) -> Read [Either String Int] = ...
\end{verbatim}
Here, \texttt{readInt} and \texttt{readString} behave just
like the corresponding Haskell instances, while \texttt{readEitherSI} 
and \texttt{readSensor} are custom instances that read 
values in the sensor format. 
With intersection types and the merge operator we can define
both the type of environments for instances that read sensor data, and 
the environment that collects the above instances: 
\begin{verbatim}
  type SensorEnv = Read Int & Read String
                 & (Read String -> Read Int -> Read (Either String Int))
                 & (Read (Either String Int) -> Read [(Either String Int)])

  env : SensorEnv = readInt ,, readString ,, readEitherSI ,, readSensor
\end{verbatim}
Finally, the \texttt{parseSensor} function 
parametrizes over a suitable sensor environment:
\begin{verbatim}
  parseSensor (read : SensorEnv) (s: String) : [Either String Int] = read s
\end{verbatim}
Calling \texttt{parseSensor} with \texttt{env} can parse
strings in the expected format for sensor data. Moreover, an
additional advantage of the \texttt{parseSensor} function is that 
it is \emph{decoupled} from concrete implementations of
\texttt{Read}. For instance, if later we want to express sensor 
data using Haskell's own format for \texttt{Read} instances, we just have to provide an
alternative environment:
\begin{verbatim}
  haskEnv : SensorEnv = readInt ,, readString ,, readEither ,, readList
\end{verbatim}
where \texttt{readEither} and \texttt{readList} behave 
just like the Haskell instances for \texttt{Read (Either a b)} and 
\texttt{Read [a]}.
By calling \verb|parseSensor| with \verb|haskEnv| as its first argument, the system
will infer an instance of the same type, but taken from the newly provided
environment encoding, \verb|haskEnv|.

Without the modus ponens rule in subtyping, the same functionality would require
a more verbose program, where both the provided environment and the parser function
should be parametrized
over the alternative instances for \texttt{Either String Int} and
\texttt{[Either String Int]}, and where application of instances should be
explicit:
\begin{verbatim}
  type PSensor =  (Read Int -> Read String -> Read (Either String Int)))
               -> (Read (Either String Int) -> Read [Either String Int])
               -> Read Int & Read String 
                  & Read (Either String Int) & Read [Either String Int]

  env' (rEitherSI : Read Int -> Read String -> Read (Either String Int))
       (rList     : Read (Either String Int) -> Read [Either String Int])
   : Read Int & Read String & Read (Either String Int) & Read [Either String Int]
   = readInt ,, readString ,, (rEitherSI readInt readString)
     ,, (rList (rEitherSI readInt readString))

  parseSensor' (rEither : Read Int -> Read String -> Read (Either String Int )))
               (rList   : Read (Either String Int) -> Read [Either String Int])
               (readArg : PSensor) (s : String)
    : [Either String Int] 
    = readArg rEither rList s
\end{verbatim}

In summary, modus ponens captures the essence of resolution and
brings convenience to programmers by automatically
composing instances. This is also the reason
why resolution is heavily used by Haskell and Scala programmers.
In addition, the integration of modus ponens into a system
with intersection subtyping adds expressive power that is not
available in Haskell or Scala. In contrast to Haskell, implicit
environments are first-class. With respect to Scala, viewing
resolution as subtyping makes obvious how the two features
interact, which we discuss in more detail next.

\subsection{Interaction Between Subtyping and Resolution}

In languages like Scala, which supports both subtyping and resolution, 
we expect that the two mechanisms interact. For example, here is a simple Scala program that 
requires both mechanisms: 
\begin{verbatim}
  class A {}
  class B extends A {}

  implicit val v : B = new B() {}; // adds v to the implicit environment
  val u : A = implicitly  // resolves to v
\end{verbatim}
This program defines two classes \texttt{A} and \texttt{B}. The value
\texttt{v} is marked with an \texttt{implicit} keyword, which has the effect of
adding \texttt{v} to the implicit 
environment, so that it can be used by resolution later. 
In the definition of \texttt{u}, \texttt{implicitly} triggers 
resolution to search for a value of type \texttt{A}. The search
succeeds and resolves to \texttt{v} which has type
\texttt{B}, a subtype of \texttt{A}. Clearly this only works if resolution 
accounts for subtyping. 

The Scala compiler accepts this program and its implementation 
of resolution does indeed account for subtyping. However there is 
very little work formalizing the (non-trivial) interaction between resolution and
subtyping. Most foundational calculi for Scala formalize subtyping,
but not resolution~\citep{dot1,dot2}. Similarly, works that formalize 
resolution typically do not account for subtyping~\citep{implicitcalculus,cochis,simplicity}. 

One advantage of making resolution a part of subtyping is that the
interaction between the two mechanisms then becomes trivial. 
For instance, the following program:
\begin{verbatim}
  a : Int                  = 1
  b : Int & Char           = a,,'c'
  u (env : Int&Char) : Int = env  
\end{verbatim}
is roughly analogous to the previous Scala program, except 
that we use a first-class environment (the argument \verb|env|
of function \verb|u|)
instead of an implicit
environment. In this case resolution triggers a search for an integer 
in the body of \texttt{u}, which is successful because \texttt{Int\&Char <:
Int}. Other more complex interactions between subtyping and
resolution, such as in:
\begin{verbatim}
  (Int -> Top -> Bool) & Int  <: String -> Bool
\end{verbatim}
work as well. In this case the two types in the intersection 
need to be composed with modus ponens to obtain \texttt{Top -> Bool} and 
\texttt{Top -> Bool <: String -> Bool} via subtyping.

\subsection{Technical challenges}

To formalize our integration of resolution and subtyping, we create the
\calculus calculus by extending the \necolus calculus with the modus
ponens rule.  While the extension is seemingly a small change, it 
has far-reaching consequences with respect to the metatheory. Conceptually,
this is due to the fact that modus ponens invalidates much of the 
reasoning about types in the \necolus metatheory. In \necolus it is possible to
consider the syntactic components of types in isolation.
This is no longer possible in \calculus because one component may interact with
another through modus ponens.

This affects coherence: the \necolus disjointness constraint is insufficient
for \calculus and needs to be tightened. Also the coherence proof requires a
considerably more complex canonicity relation that features a form of context
dependency to capture the possible interaction between type components. 

Algorithmic subtyping is also affected by this. The algorithm no longer
terminates naturally and requires a loop detector. This makes the termination
proof a lot harder. Moreover, the size-based induction used to prove transitivity
for the algorithmic subtyping of \necolus breaks. Instead, our transitivity
proof involves 7 mutually recursive lemmas and relies on the AProVE 
automatic termination checker to establish their well-foundedness.

The following sections set up \calculus and explain these aspects in detail.

%% file: sources/calculus2.mng
\section{The \calculus Calculus}
\label{sec:calculus}

\paragraph*{Syntax}

\input{figures/source_grammar.mng}

Figure~\ref{fig:source-syntax:mp} shows the syntax of \calculus. It is the same as
that of \necolus, but---for the sake of conciseness---without records.
Types $\ottnt{A}, \ottnt{B}, C$ include naturals $ \mathsf{Nat} $, the top type $ \top $,
function types $\ottnt{A}  \rightarrow  \ottnt{B}$ and intersection types $\ottnt{A}  \, \& \,  \ottnt{B}$.
Terms $E$ include variables $\ottmv{x}$, natural numbers $\mathit{i}$, the top value $ \top $,
function abstractions $ \lambda \ottmv{x} .\, E $, function applications $E_{{\mathrm{1}}} \, E_{{\mathrm{2}}}$, merges $E_{{\mathrm{1}}}  \,,,\,  E_{{\mathrm{2}}}$ and
(type-)annotated terms $E  \ottsym{:}  \ottnt{A}$.

\input{figures/source_typing.mng}

\paragraph{Bidirectional Typing} 
The type system of \calculus, shown in Figure~\ref{fig:typing}, is bidirectional and thus has two modes:
the \emph{synthesis} mode ($\Rightarrow$) and the \emph{checking} mode ($\Leftarrow$).
The synthesis judgment $ \Gamma   \vdash   E    \Rightarrow    \ottnt{A}  \elbox{ \rightsquigarrow   \ottnt{e} } $ means that we can synthesize type $\ottnt{A}$
from expression $E$ in context $\Gamma$, while the checking judgment $ \Gamma   \vdash   E    \Leftarrow    \ottnt{A}  \elbox{ \rightsquigarrow   \ottnt{e} } $
checks whether $E$ has given type $\ottnt{A}$. As usual, type annotations switch from
synthesis to checking mode (rule \rlabel{T-anno}).
We reserve further discussion over the parts highlighted in gray for
Section~\ref{sec:calculus:elaboration-semantics}.


The rules are identical to those of \necolus with one notable difference. Rule
\rlabel{T-merge} not only requires the types $\ottnt{A_{{\mathrm{1}}}}$ and $\ottnt{A_{{\mathrm{2}}}}$ of the two
subterms to be disjoint, $\ottnt{A_{{\mathrm{1}}}}  *  \ottnt{A_{{\mathrm{2}}}}$, but also requires them to be
\emph{internally} disjoint, $ \wfd{  \ottnt{A_{{\mathrm{1}}}}  } $ and $ \wfd{  \ottnt{A_{{\mathrm{2}}}}  } $, all summarised in the
internal disjointness premise \( \wfd{  \ottnt{A_{{\mathrm{1}}}}  \, \& \,  \ottnt{A_{{\mathrm{2}}}}  } \) of the rule.
The two disjointness conditions are discussed in Section~\ref{sec:calculus:disjointness}.

The judgment \(\vdash  \Gamma\) used in the premise of rules \rlabel{T-top},
\rlabel{T-nat} and \rlabel{T-var} expresses that the typing context \(\Gamma\)
is well-formed, i.e. it holds no duplicate variables. Its definition can be
found in Appendix~\ref{appendix:definitions}. 

\subsection{Declarative Subtyping}
\label{sec:calculus:subtyping}

\input{figures/source_subtyping.mng}

Figure~\ref{fig:subtyping} declaratively specifies the subtyping relation of
\calculus.  The judgment $ \ottnt{A}   \ottsym{<:}   \ottnt{B}  \elbox{ \rightsquigarrow   \ottnt{c} } $ states that $\ottnt{A}$ is a
subtype of $\ottnt{B}$ and produces coercion $\ottnt{c}$.
Intuitively, $\ottnt{c}$ captures the way to convert values of type \(\ottnt{A}\) into
values of type \(\ottnt{B}\).

\paragraph{BCD-Style Subtyping}
All but the last subtyping rule are the same as those of \necolus and originate from
the BCD type system~\citep{bcd-subtyping}.
Rules \rlabel{S-refl}, \rlabel{S-trans} and \rlabel{S-top} ensure that subtyping is a partial order with $ \top $ as
greatest element. Rule \rlabel{S-arr} is the standard subtyping rule for function types, while rules \rlabel{S-andl},
\rlabel{S-andr} and \rlabel{S-and} are standard for intersection types. Rules \rlabel{S-distArr} and \rlabel{S-topArr}
are standard BCD subtyping rules; the former enables \emph{nested composition}~\citep{necolus}, while the latter
provides a more unified treatment of top-like types in the meta-theory of the language: without 
that rule, we cannot have \(\ottnt{A}  \rightarrow  \ottnt{B}  \ottsym{<:}  \top  \rightarrow  \top\), if \(\ottnt{A}\) is not top-like itself.


\paragraph{Modus Ponens} Our calculus extends the subtyping relation of
\necolus with one new rule, \rlabel{S-mp}, which corresponds to the Modus
Ponens inference rule of propositional logic. It states that if $\ottnt{A}$ is
a subtype of both $\ottnt{B_{{\mathrm{1}}}}  \rightarrow  \ottnt{B_{{\mathrm{2}}}}$ and of $\ottnt{B_{{\mathrm{1}}}}$, then it is also a subtype of
$\ottnt{B_{{\mathrm{2}}}}$. While this extension of subtyping is conceptually small,
it has far-reaching consequences for the meta-theory of the language.

\subsection{Elaboration Semantics}
\label{sec:calculus:elaboration-semantics}

Following \necolus we assign a semantics to \calculus by means of an elaboration.
The target of the elaboration is \target, which is the simply typed
lambda-calculus extended with products and coercions.
Pairs are the
elaboration target of merges (rule \rlabel{T-merge}),
while coercions explicitly witness the
implicit use of subtyping in \calculus (rule \rlabel{T-sub}).

\input{figures/target_grammar.mng}

\paragraph{The \target Calculus}
Figure~\ref{fig:target-syntax:mp} shows the syntax of \target. This syntax is nearly
identical to that of \necolus' target language. There are only two differences,
both situated in the coercions $\ottnt{c}$. These coercions express the conversion
of a term from one type to another. The reason why we use a
separate syntactic sort of coercions for this---rather than encoding the
conversions as regular function terms---is that it facilitates the coherence
proof. Each coercion form corresponds to a
particular subtyping rule (Figure~\ref{fig:subtyping}).
The first and main difference to \necolus' target calculus is that, because
we have a new modus ponens subtyping rule in \calculus, there is also a new
corresponding coercion form $\ottnt{c_{{\mathrm{1}}}}  \modusponens  \ottnt{c_{{\mathrm{2}}}}$ in \target to witness it.
The second and minor difference is that we have annotated several existing
coercion forms with type information in order to facilitate type checking.

Figure~\ref{fig:target:new} lists the new inference rules in the type system
and small-step operational semantics of \target to support the new coercion
form. The full definitions of these relations can be found in Appendix~\ref{appendix:target-specification}. 
The coercion typing rule \rlabel{CT-mp} essentially replicates the
subtyping rule \rlabel{S-mp} of Figure~\ref{fig:subtyping}, but now with
\target types. The new small-step rule, \rlabel{step-mp}, splits up an
application of a modus ponens coercion $\ottnt{c_{{\mathrm{2}}}}  \modusponens  \ottnt{c_{{\mathrm{1}}}}$ to value $\ottnt{v}$ into
separate coercion applications $\ottnt{c_{{\mathrm{2}}}} \, \ottnt{v}$ and $\ottnt{c_{{\mathrm{1}}}} \, \ottnt{v}$ where the former should
yield a function that is applied to the latter.

\input{figures/target_new.mng}

How the features of \calculus are mapped to \target is given in the grey parts
of Figures \ref{fig:subtyping} and \ref{fig:typing}.  In those figures the meta-function
$|\cdot|$ transforms \calculus types to \target types, and extends naturally to
typing contexts. Its definition is in Appendix~\ref{appendix:definitions}. 

\subsection{Target and Elaboration Metatheory}

We have formally investigated the metatheory of
our calculus, \calculus, and of its target language, \target, and mechanised it in Coq.

\paragraph{Target Language} Firstly, we have proven that the key metatheoretical
properties of \target are not invalidated by the new coercion form. Indeed, \target
is type safe:

\begin{theorem}[Preservation]
\label{thm:preservation}
If $ \bullet   \vdash  \ottnt{e}  \ottsym{:}  \tau$ and $\ottnt{e}  \longrightarrow  \ottnt{e'}$, then $ \bullet   \vdash  \ottnt{e'}  \ottsym{:}  \tau$.
\end{theorem}
\begin{theorem}[Progress]
\label{thm:progress}
If $ \bullet   \vdash  \ottnt{e}  \ottsym{:}  \tau$, then either $\ottnt{e}$ is a value or there is an $\ottnt{e'}$ such that $\ottnt{e}  \longrightarrow  \ottnt{e'}$.
\end{theorem}

\noindent
Moreover, every well-typed \target term is normalising.

\begin{theorem}[Normalisation]
\label{thm:normalisation}
If $ \bullet   \vdash  \ottnt{e}  \ottsym{:}  \tau$ then there exists a value $\ottnt{v}$ such that $\ottnt{e}  \longrightarrow^{*}  \ottnt{v}$.
\end{theorem}
Here \( \longrightarrow^{*} \), is the reflexive transitive closure of \( \longrightarrow \).


\paragraph{Elaboration}
Also, the elaboration of \calculus to \target preserves well-typing.

\begin{theorem}[Coercions Preserve Types]
\label{thm:coercions-preserve-types}
If $ \ottnt{A}   \ottsym{<:}   \ottnt{B}  \elbox{ \rightsquigarrow   \ottnt{c} } $, then $\ottnt{c}  \vdash  \ottsym{\mbox{$\mid$}}  \ottnt{A}  \ottsym{\mbox{$\mid$}} \, \triangleright \, \ottsym{\mbox{$\mid$}}  \ottnt{B}  \ottsym{\mbox{$\mid$}}$.
\end{theorem}

\begin{theorem}[Elaboration Preserves Types]
 \label{thm:type-preserving-elaboration}
~\\[-4mm]
\begin{itemize}
\item
If $ \Gamma   \vdash   E    \Rightarrow    \ottnt{A}  \elbox{ \rightsquigarrow   \ottnt{e} } $, then $\ottsym{\mbox{$\mid$}}  \Gamma  \ottsym{\mbox{$\mid$}}  \vdash  \ottnt{e}  \ottsym{:}  \ottsym{\mbox{$\mid$}}  \ottnt{A}  \ottsym{\mbox{$\mid$}}$.
\item
If $ \Gamma   \vdash   E    \Leftarrow    \ottnt{A}  \elbox{ \rightsquigarrow   \ottnt{e} } $, then $\ottsym{\mbox{$\mid$}}  \Gamma  \ottsym{\mbox{$\mid$}}  \vdash  \ottnt{e}  \ottsym{:}  \ottsym{\mbox{$\mid$}}  \ottnt{A}  \ottsym{\mbox{$\mid$}}$.
\end{itemize}
\end{theorem}

The meta-function \(\ottsym{\mbox{$\mid$}}  \Gamma  \ottsym{\mbox{$\mid$}}\) translates a source typing context \(\Gamma\) to a target one.
The full definition can be found in Appendix~\ref{appendix:definitions}. 

\subsection{Disjointness Conditions}
\label{sec:calculus:disjointness}

\input{figures/source_disjointness.mng}

Figure~\ref{fig:disjointness} shows the rules for disjointness (top) and internal disjointness (bottom).

\paragraph{Disjointness} The purpose of the disjointness condition is to avoid
ambiguous expressions like $1 ,, 2$. Judgment $\ottnt{A}  *  \ottnt{B}$ conservatively captures the notion that
the common supertypes of $\ottnt{A}$ and $\ottnt{B}$ are
all top-like. The first 4 rules of Figure~\ref{fig:disjointness}, for the top type and
intersection types, are the same as those of \necolus. We turn our focus to the new rules
\rlabel{D-mpL} and \rlabel{D-mpR}, added to accommodate the
new modus ponens subtyping rule.
While they can be combined
to recover \necolus' \rlabel{D-arr} rule%
\footnote{This rule states that two arrow types are disjoint if their codomains are disjoint.},
\begin{fakedisplaymath}\small
\nrule{D-arr}{\ottnt{B_{{\mathrm{1}}}}  *  \ottnt{B_{{\mathrm{2}}}}}{\ottsym{(}  \ottnt{A_{{\mathrm{1}}}}  \rightarrow  \ottnt{B_{{\mathrm{1}}}}  \ottsym{)}  *  \ottsym{(}  \ottnt{A_{{\mathrm{2}}}}  \rightarrow  \ottnt{B_{{\mathrm{2}}}}  \ottsym{)}}
\end{fakedisplaymath}
they cannot be used to recover the \necolus axioms $\mathsf{Nat}  *  \ottsym{(}  \ottnt{A}  \rightarrow  \ottnt{B}  \ottsym{)}$ and
$\ottsym{(}  \ottnt{A}  \rightarrow  \ottnt{B}  \ottsym{)}  *  \mathsf{Nat}$. On the contrary, \calculus does not consider the types $ \mathsf{Nat} $
and $ \mathsf{Bool}   \rightarrow  \mathsf{Nat}$ to be disjoint.  For the sake of conciseness and
compositionality, the disjointness definition is conservative. Indeed, merging
any expressions of those two types does not lead to ambiguity. Yet, if we merge
two expressions whose types contain the above two types, ambiguity can in fact
arise. Consider the example $\ottsym{(}  \ottsym{(}   \mathsf{true}   \,,,\,  \ottsym{(}   \lambda \ottmv{x} .\,  1    \ottsym{)}  \ottsym{:}   \mathsf{Bool}   \rightarrow  \mathsf{Nat}  \ottsym{)}  \,,,\,   2   \ottsym{)}  \ottsym{:}  \mathsf{Nat}$.
The elaborated form
of the merge expression on the left of the outer type annotation
is the tuple \(\langle  \langle   \mathsf{true}   \ottsym{,}   \lambda \ottmv{x} :  \mathsf{Bool}  .\,  1    \rangle  \ottsym{,}   2   \rangle\).
The outer type annotation will trigger the subtyping mechanism, first via rule \rlabel{T-anno}
and then through rule \rlabel{T-sub}, in order to produce a value of type \( \mathsf{Nat} \).
This example
is ambiguous, because it can produce---through subtyping---two different values of this type: $ 2 $
by projecting on the right component of the outer tuple, and $ 1 $ by projecting on the left component of the outer
tuple and using modus ponens to apply $\ottsym{(}   \lambda \ottmv{x} :  \mathsf{Bool}  .\,  1    \ottsym{)}$ to $ \mathsf{true} $.
The respective subtyping derivations and the coercions they generate are shown in
Figure~\ref{fig:subtypingExample}.
\input{figures/example_subtyping.mng}
For similar reasons, expressions $\ottsym{(}   \lambda \ottmv{x} .\,  1    \ottsym{)}  \ottsym{:}   \mathsf{Bool}   \rightarrow  \mathsf{Nat}  \,,,\,  \ottsym{(}   2   \,,,\,   \mathsf{true}   \ottsym{)}$ and
$\ottsym{(}   2   \,,,\,  \ottsym{(}   \lambda \ottmv{x} .\,  1    \ottsym{)}  \ottsym{:}   \mathsf{Bool}   \rightarrow  \mathsf{Nat}  \ottsym{)}  \,,,\,   \mathsf{true} $ are also ambiguous. In all three cases, the component types of a merge
will not be disjoint.

The two new rules \rlabel{D-mpL} and \rlabel{D-mpR} bring the notion of disjointness between two types
closer to the notion of no-overlaps between two (or more) instances. Types $\ottnt{B}$ and
$\ottnt{A}  \rightarrow  \ottnt{B}$ are not disjoint, similarly to how two Haskell instances \verb|C Int| and \verb|D Int => C Int|
overlap.

\paragraph{Harmless overlaps and (non-)ambiguity}
Note that not all forms of overlap introduce semantic ambiguity.
In particular, we identify two such \emph{harmless} cases.
Firstly, in the presence of an overlap, it can still be the case that there is only one way to
resolve a value of a given (overlapping) type. For example,
in the term \( 2   \,,,\,  \ottsym{(}   \lambda \ottmv{x} .\,  1    \ottsym{)}  \ottsym{:}   \mathsf{Bool}   \rightarrow  \mathsf{Nat}\) of type \(\mathsf{Nat}  \, \& \,  \ottsym{(}   \mathsf{Bool}   \rightarrow  \mathsf{Nat}  \ottsym{)}\),
the types \( \mathsf{Nat} \) and \( \mathsf{Bool}   \rightarrow  \mathsf{Nat}\) overlap, but there is only one applicable way to
resolve a value of type \( \mathsf{Nat} \), because there
is no \( \mathsf{Bool} \) value to feed to the function \(\ottsym{(}   \lambda \ottmv{x} .\,  1    \ottsym{)}  \ottsym{:}   \mathsf{Bool}   \rightarrow  \mathsf{Nat}\).
Secondly, expressions of an overlapping type are harmless when all possible resolution paths for
a wanted type result in the same inferred value.
For example, the term \(\ottsym{(}   1   \,,,\,   1   \ottsym{)}\), while containing an overlap, is unambiguous.

Because of the disjointness condition imposed over merges, the source calculus conservatively rejects
expressions that fall under either of the two harmless forms of overlap, when they are explicitly
written by the programmer.
However, they are still possible implicitly, through a type annotation, which invokes subtyping.
For example, we can encode the rejected term \(\ottsym{(}   1   \,,,\,   1   \ottsym{)}\) as \( 1   \ottsym{:}  \mathsf{Nat}  \, \& \,  \mathsf{Nat}\),
which is accepted and semantically same.
Its full typing derivation is in Appendix~\ref{appendix:algorithm-example}. 

\paragraph{Internal Disjointness}
The internal disjointness predicate, \( \wfd{  \ottnt{A}  } \), requires that the two
components of any intersection type that appear in \(\ottnt{A}\)
are disjoint. Intuitively, $ \wfd{  \ottnt{A}  } $ ensures that if $\ottnt{A}  \ottsym{<:}  \ottnt{B}$, then
$\ottnt{B}$ can only contain harmless overlaps.


The internal disointness condition
is imposed on the subterms of a merge in rule \rlabel{T-merge}.
Again this is a conservative measure to avoid ambiguity, this time of a more
convoluted nature that involves nested merges interleaved with subtyping.
Here is a minimal ambiguous example to illustrate the problem:
\begin{fakedisplaymath}
\ottsym{(}  \ottsym{(}   \mathsf{true}   \,,,\,   \mathsf{b2n}   \ottsym{)}  \ottsym{:}  \mathsf{Nat}  \, \& \,  \ottsym{(}   \mathsf{Bool}   \rightarrow  \mathsf{Nat}  \ottsym{)}  \,,,\,   \mathsf{false}   \ottsym{)}  \ottsym{:}  \mathsf{Nat}
\end{fakedisplaymath}
This example features $ \mathsf{b2n} $, which we assume to be a built-in function
of type $ \mathsf{Bool}   \rightarrow  \mathsf{Nat}$ that maps $ \mathsf{true} $ to $ 1 $ and $ \mathsf{false} $ to
$ 0 $.  The inner merge has type $ \mathsf{Bool}   \, \& \,  \ottsym{(}   \mathsf{Bool}   \rightarrow  \mathsf{Nat}  \ottsym{)}$, but is converted
to type $\mathsf{Nat}  \, \& \,  \ottsym{(}   \mathsf{Bool}   \rightarrow  \mathsf{Nat}  \ottsym{)}$ by means of the type annotation, which triggers the
subtyping mechanism. The
latter produces a coercion that is then applied on (the elaborated form of) the subterm
\( \mathsf{true}   \,,,\,   \mathsf{b2n} \).
The
corresponding value is $ 1   \,,,\,   \mathsf{b2n} $, which could not be written
explicitly because $ \mathsf{Nat} $ and $ \mathsf{Bool}   \rightarrow  \mathsf{Nat}$ are not considered disjoint.
Yet, because this value is not explicitly written by the programmer, but implicitly
produced through subtyping, we know that the overlap is harmless. Indeed, while the expression
$ 1   \,,,\,   \mathsf{b2n} $ potentially produces two different values of type
$ \mathsf{Nat} $, one of those requires a $ \mathsf{Bool} $, which is not available at this point.

However, the outer merge with $ \mathsf{false} $ provides the missing $ \mathsf{Bool} $ value, 
making the outer $ \mathsf{Nat} $ type annotation ambiguously yield both $ 0 $ and $ 1 $ as
possible results.
The disjointness check rejects none
of the two merges in the example.
Here, the internal disjointness constraint comes in.
It catches the ambiguity problem at the outer merge where it identifies $ 1   \,,,\,   \mathsf{b2n} $
as unfit to merge with.

Internal disjointness inherits the conservative behavior of the (binary) disjointness
condition, as witnessed in rule \rlabel{ID-arr}. This rule
regards the case of a function type \(\ottnt{A}  \rightarrow  \ottnt{B}\) that potentially is a
supertype of some internally disjoint type, say \(\ottnt{A'}\) (formally, \(\ottnt{A'}  \ottsym{<:}  \ottnt{A}  \rightarrow  \ottnt{B}\)).
If $\ottnt{A'}$ is not a subtype of $\ottnt{A}$ (thus, \(\ottnt{A'}  \ottsym{<:}  \ottnt{A}\) does not hold), then rule
\rlabel{S-mp} does not apply. In fact, it becomes impossible to implicitly produce
\(\ottnt{B}\) from
\(\ottnt{A'}\) and, therefore, unnecessary to demand internal disjointness of \(\ottnt{B}\).

The indifference of rule \rlabel{ID-arr} around domains of arrow types
is justified by the intuition
that if a supertype of an internally disjoint type contains overlaps, then those are
harmless.
A function of type $\ottnt{A}  \rightarrow  \ottnt{B}$ is unambiguous if it outputs unambiguous values
of type $\ottnt{B}$, given unambiguous input values of type $\ottnt{A}$. This implies the
requirement that $\ottnt{A}$ is also internally disjoint. However, at this point, only
values produced implicitly from $\ottnt{A'}$ are considered, and these can only contain
harmless ambiguity. Indeed, explicit values can be added
only through the merge operator, thus falling under rule \rlabel{ID-and}.
Thus, unambiguity is ensured for $\ottnt{A}$.

%% file: figures/source_grammar.mng
\begin{figure}[t]
  \centering
\begin{tabular}{llll} 
  Types & $\ottnt{A}, \ottnt{B}, C$ & $\Coloneqq$ & $ \mathsf{Nat}  \mid  \top  \mid \ottnt{A}  \rightarrow  \ottnt{B} \mid \ottnt{A}  \, \& \,  \ottnt{B}$ \\
  Terms     & $E$           & $\Coloneqq$ & $\ottmv{x} \mid  \top  \mid \mathit{i} \mid  \lambda \ottmv{x} .\, E  \mid E_{{\mathrm{1}}} \, E_{{\mathrm{2}}} \mid E_{{\mathrm{1}}}  \,,,\,  E_{{\mathrm{2}}} \mid E  \ottsym{:}  \ottnt{A}$  \\
  Typing Contexts & $\Gamma$     & $\Coloneqq$ & $ \bullet  \mid \Gamma  \ottsym{,}  \ottmv{x}  \ottsym{:}  \ottnt{A}$ \\
\end{tabular}
  \caption{\calculus Syntax}
  \label{fig:source-syntax:mp}
\end{figure}

%% file: figures/target_grammar.mng
\begin{figure}[t]
  \centering
\begin{tabular}{@{\hspace{0mm}}l@{\hspace{2mm}}l@{\hspace{2mm}}l@{\hspace{2mm}}l@{\hspace{0mm}}} 
  Types           & $\tau$ & $\Coloneqq$ & $ \mathsf{Nat}  \mid  \langle \rangle  \mid \tau_{{\mathrm{1}}}  \rightarrow  \tau_{{\mathrm{2}}} \mid \tau_{{\mathrm{1}}}  \times  \tau_{{\mathrm{2}}}$ \\
  Terms           & $\ottnt{e}$ & $\Coloneqq$ & $\ottmv{x} \mid  \langle \rangle  \mid \mathit{i} \mid  \lambda \ottmv{x} : \tau .\, \ottnt{e}  \mid \ottnt{e_{{\mathrm{1}}}} \, \ottnt{e_{{\mathrm{2}}}} \mid \langle  \ottnt{e_{{\mathrm{1}}}}  \ottsym{,}  \ottnt{e_{{\mathrm{2}}}}  \rangle \mid \ottnt{c} \, \ottnt{e}$ \\
  Coercions       & $\ottnt{c}$ & $\Coloneqq$ & $ \mathsf{id} _{  \tau  }  \mid  \ottnt{c_{{\mathrm{1}}}}  \circ  \ottnt{c_{{\mathrm{2}}}}  \mid  \mathsf{top} _{  \tau  }  \mid \ottnt{c_{{\mathrm{1}}}}  \rightarrow  \ottnt{c_{{\mathrm{2}}}} \mid  \pi_1 ^{  \tau_{{\mathrm{1}}} , \tau_{{\mathrm{2}}}  }  \mid  \pi_2 ^{  \tau_{{\mathrm{1}}} , \tau_{{\mathrm{2}}}  }  \mid \langle  \ottnt{c_{{\mathrm{1}}}}  \ottsym{,}  \ottnt{c_{{\mathrm{2}}}}  \rangle \mid  \mathsf{top}_{\rightarrow}  \mid  \mathsf{dist}_{\rightarrow} ^{  \tau_{{\mathrm{1}}} , \tau_{{\mathrm{2}}} , \tau_{{\mathrm{3}}}  }  \mid$ \\
                  &         &             & $\ottnt{c_{{\mathrm{1}}}}  \modusponens  \ottnt{c_{{\mathrm{2}}}} $   \\
  Typing contexts & $\Delta$ & $\Coloneqq$ & $ \bullet  \mid \Delta  \ottsym{,}  \ottmv{x}  \ottsym{:}  \tau$ \\
  Values          & $\ottnt{v}$ & $\Coloneqq$ & $ \langle \rangle  \mid \mathit{i} \mid  \lambda \ottmv{x} : \tau .\, \ottnt{e}  \mid \langle  \ottnt{v_{{\mathrm{1}}}}  \ottsym{,}  \ottnt{v_{{\mathrm{2}}}}  \rangle \mid \ottsym{(}  \ottnt{c_{{\mathrm{1}}}}  \rightarrow  \ottnt{c_{{\mathrm{2}}}}  \ottsym{)} \, \ottnt{v} \mid  \mathsf{top}_{\rightarrow}  \, \ottnt{v} \mid \ottsym{(}   \mathsf{dist}_{\rightarrow} ^{  \tau_{{\mathrm{1}}} , \tau_{{\mathrm{2}}} , \tau_{{\mathrm{3}}}  }   \ottsym{)} \, \ottnt{v}$ \\
\end{tabular}
  \caption{\target Syntax}
  \label{fig:target-syntax:mp}
\end{figure}

%% file: sources/algorithm.mng
\section{Algorithmic Subtyping}
\label{sec:algorithm}

In this section, we present an algorithm that decides the subtyping relation
of Figure~\ref{fig:subtyping}. 

In the declarative specification of the subtyping relation (Figure~\ref{fig:subtyping}),
rules~\rlabel{S-trans} and~\rlabel{S-mp} hinder the decidability of subtyping:
they both require guessing an appropriate intermediate
type ($\ottnt{A_{{\mathrm{2}}}}$ in the former, and $\ottnt{B_{{\mathrm{1}}}}$ in the latter).
We address this by employing a technique originating from proof
search known as \emph{focusing}~\citep{Focusing}. The general strategy for
checking a constraint of the form $\ottnt{A}  \ottsym{<:}  \ottnt{B}$ is to structurally analyze
(focus on) $\ottnt{B}$, and when it is stripped down to a base type do the same for
$\ottnt{A}$. This task is performed by two mutually-recursive judgments, the first
focusing on the \emph{right} type $\ottnt{B}$, and the second focusing on the
\emph{left} type $\ottnt{A}$.
This technique has already been employed by~\citet{necolus}---and,
before that, by~\citet{PierceDecidability}---though with a single
all-in-one judgment. Our algorithm separates the two phases for clarity (and
ease of reasoning) and also deals with the additional complexity introduced by
rule~\rlabel{S-mp}.
Unfortunately, the support for modus ponens greatly complicates the algorithm's
proofs of transitivity and termination. For the former, we had to come up with
an approach external to Coq, involving the AProVE termination checker~\citep{aprove}, to establish
well-foundedness.
In the remainder of the section we describe the subtyping algorithm in detail
(Section~\ref{sec:algorithm:definition}), as well as its most important
meta-theoretical properties (Section~\ref{subsec:algorithm:metatheory}). We
follow the earlier convention to
\elbox{\text{highlight}} the parts of the rules that concern coercion
generation.



\subsection{The Subtyping Algorithm}
\label{sec:algorithm:definition}

\input{figures/algorithm.mng}
In Figure~\ref{fig:algorithmic-subtyping}, we present the algorithmic subtyping
judgment $\algSubMain{\ottnt{A}}{\ottnt{B}}{\ottnt{c}}$; the algorithmic counterpart of the
declarative judgment $ \ottnt{A}   \ottsym{<:}   \ottnt{B}  \elbox{ \rightsquigarrow   \ottnt{c} } $ of Section~\ref{sec:calculus:subtyping}.
The main judgment is defined in terms of two auxiliary, mutually-recursive
judgments $\algSubR{\mathcal{L}}{\ottnt{A}}{\ottnt{B}}{\ottnt{c}}$ and
$\algSubL{\mathcal{L}}{\mathcal{M}}{\ottnt{A_{{\mathrm{0}}}}}{\mathcal{X}}{A}{B}{\mathcal{X}'}$, each focusing on the
right or the left component of a type inequality, respectively. We explain the
former in Section~\ref{sec:right-focusing-complete} and the latter in
Section~\ref{sec:left-focusing-complete}.


\paragraph{Preliminaries}

Meta-variables $\mathcal{L}$ and $\mathcal{M}$ stand for queues of types. The
algorithm uses them to capture the left-hand sides of arrow types. Operation
$ \mathcal{L}  \to\hspace{-8pt}\to  \ottnt{B} $ converts a queue
$\mathcal{L}$ back into a type, given the right-hand side type $\ottnt{B}$. It is
inductively defined as follows:
\begin{fakedisplaymath}
\small
   []  \arrows \ottnt{B} = \ottnt{B} \qquad\qquad ( \ottnt{A} ,  \mathcal{L} ) \arrows \ottnt{B} = \ottnt{A}  \rightarrow  \ottsym{(}   \mathcal{L}  \to\hspace{-8pt}\to  \ottnt{B}   \ottsym{)}
\end{fakedisplaymath}

\noindent
Meta-variable $\mathcal{X}$ denotes \emph{algorithmic contexts}. An algorithmic
context is merely the Cayley representation of a coercion: a
function from coercions to coercions. Given a coercion $\ottnt{c}$, we can convert
an algorithmic context $\mathcal{X}$ into a coercion by function application,
denoted as $ \mathcal{X} [  \ottnt{c}  ] $.
The role of algorithmic contexts is explained in
Section~\ref{sec:left-focusing-complete} where we discuss the left-focusing
rules.\footnote{In both our mechanization and
implementation of the algorithm we use a
defunctionalized~\citep{defunctionalization} representation of algorithmic
contexts, which we then interpret as actual functions between coercions.}


\subsubsection{Right Focusing}
\label{sec:right-focusing-complete}

As mentioned earlier, the first judgment
($\algSubR{\mathcal{L}}{\ottnt{A}}{\ottnt{B}}{\ottnt{c}}$) structurally analyzes
$\ottnt{B}$, until it is stripped down to a base type ($ \mathsf{Nat} $ or $ \top $). The
accumulating parameter $\mathcal{L}$ captures the 
domains of arrows encountered in $\ottnt{B}$, such that the following holds:

\begin{lemma}[Soundness of Right Focusing]
\label{lemma:right-rules-soundness}
If $\algSubR{\mathcal{L}}{\ottnt{A}}{\ottnt{B}}{\ottnt{c}}$
then $ \ottnt{A}   \ottsym{<:}   \ottsym{(}   \mathcal{L}  \to\hspace{-8pt}\to  \ottnt{B}   \ottsym{)}  \elbox{ \rightsquigarrow   \ottnt{c} } $.
\end{lemma}

\noindent
The first judgment is mostly borrowed from
\necolus;
we explain it only briefly here.

\paragraph{Subtyping} Ignoring coercion generation for now, the judgment works as follows:
Rule~\rlabel{AR-and} decomposes an intersection type, similar to its declarative
counterpart (rule \rlabel{S-and}).
Rule~\rlabel{AR-arr} deals with arrow types: the left sides of arrow types are
inserted in the rear of type queue $\mathcal{L}$ and removed again later,
left-to-right,
as arrows are encountered in $\ottnt{A}$ (see rule~\rlabel{AL-arr} below).
There are two base cases: either $\ottnt{B}= \top $ or
$\ottnt{B}= \mathsf{Nat} $. Rule~\rlabel{AR-top} handles the first case, where subtyping
directly produces a coercion with the use of the meta-function \( \llbracket{ \mathcal{L} }\rrbracket_{\top} \), discussed in the next paragraph.
If $\ottnt{B}= \mathsf{Nat} $, then we perform
structural analysis on $\ottnt{A}$, using judgment
$\algSubL{\mathcal{L}}{\mathcal{M}}{\ottnt{A_{{\mathrm{0}}}}}{\mathcal{X}}{\ottnt{A}}{\ottnt{B}}{\mathcal{X}'}$,
which we elaborate on in Section~\ref{sec:left-focusing-complete}.

\paragraph{Coercion Generation} The elaboration side is a bit more involved.
Coercion generation can be understood via
Lemma~\ref{lemma:right-rules-soundness} and works as follows.
Rule~\rlabel{AR-and} deals with intersection types. By induction we have that
\[
\left.
\begin{array}{l}
   \ottnt{A}   \ottsym{<:}   \ottsym{(}   \mathcal{L}  \to\hspace{-8pt}\to  \ottnt{B_{{\mathrm{1}}}}   \ottsym{)}  \elbox{ \rightsquigarrow   \ottnt{c_{{\mathrm{1}}}} }  \\
   \ottnt{A}   \ottsym{<:}   \ottsym{(}   \mathcal{L}  \to\hspace{-8pt}\to  \ottnt{B_{{\mathrm{2}}}}   \ottsym{)}  \elbox{ \rightsquigarrow   \ottnt{c_{{\mathrm{2}}}} }  \\
\end{array}
\right\} \implies
 \ottnt{A}   \ottsym{<:}   \ottsym{(}   \mathcal{L}  \to\hspace{-8pt}\to  \ottnt{B_{{\mathrm{1}}}}   \ottsym{)}  \, \& \,  \ottsym{(}   \mathcal{L}  \to\hspace{-8pt}\to  \ottnt{B_{{\mathrm{2}}}}   \ottsym{)}  \elbox{ \rightsquigarrow   \langle  \ottnt{c_{{\mathrm{1}}}}  \ottsym{,}  \ottnt{c_{{\mathrm{2}}}}  \rangle } 
\]
We finish the transition to $ \mathcal{L}  \to\hspace{-8pt}\to  \ottsym{(}  \ottnt{B_{{\mathrm{1}}}}  \, \& \,  \ottnt{B_{{\mathrm{2}}}}  \ottsym{)} $ using meta-function
$ \distArrowsTy{ \mathcal{L} }{ \ottnt{B_{{\mathrm{1}}}} }{ \ottnt{B_{{\mathrm{2}}}} } $, whose meaning is given by the following
lemma:\footnote{\label{foot:metafunctions}Its definition is straightforward and
thus omitted; it can be found in Appendix~\ref{appendix:metafunctions}. 
}
\begin{lemma}
$\forall \mathcal{L}, \ottnt{B_{{\mathrm{1}}}}, \ottnt{B_{{\mathrm{2}}}}$,
$ \ottsym{(}   \mathcal{L}  \to\hspace{-8pt}\to  \ottnt{B_{{\mathrm{1}}}}   \ottsym{)}  \, \& \,  \ottsym{(}   \mathcal{L}  \to\hspace{-8pt}\to  \ottnt{B_{{\mathrm{2}}}}   \ottsym{)}   \ottsym{<:}    \mathcal{L}  \to\hspace{-8pt}\to  \ottsym{(}  \ottnt{B_{{\mathrm{1}}}}  \, \& \,  \ottnt{B_{{\mathrm{2}}}}  \ottsym{)}   \elbox{ \rightsquigarrow    \distArrowsTy{ \mathcal{L} }{ \ottnt{B_{{\mathrm{1}}}} }{ \ottnt{B_{{\mathrm{2}}}} }  } $.
\end{lemma}

\noindent
Rule~\rlabel{AR-top} deals with the same problem that rule~\rlabel{AR-and}
does: according to Lemma~\ref{lemma:right-rules-soundness}, the coercion needs
to witness $\ottnt{A}  \ottsym{<:}  \ottsym{(}   \mathcal{L}  \to\hspace{-8pt}\to  \top   \ottsym{)}$, and not simply $\ottnt{A}  \ottsym{<:}  \top$. As 
$\ottsym{(}   \mathcal{L}  \to\hspace{-8pt}\to  \top   \ottsym{)}$ is {\em top-like}~\citep{DisjointPolymorphism} we can
always cast $ \top $ to it, which is what $ \llbracket{ \mathcal{L} }\rrbracket_{\top} $ witnesses:$^{\ref{foot:metafunctions}}$
\begin{lemma}
$\forall \mathcal{L},
   \top   \ottsym{<:}   \ottsym{(}   \mathcal{L}  \to\hspace{-8pt}\to  \top   \ottsym{)}  \elbox{ \rightsquigarrow    \llbracket{ \mathcal{L} }\rrbracket_{\top}  } $.
\end{lemma}

\noindent
Rule~\rlabel{AR-arr} is straightforward.
Finally, rule~\rlabel{AR-nat} handles the base case where $\ottnt{B}$ is $ \mathsf{Nat} $.
In this case we resort to the auxiliary judgment
$\algSubL{\mathcal{L}}{\mathcal{M}}{\ottnt{A_{{\mathrm{0}}}}}{\mathcal{X}}{A}{B}{\mathcal{X}'}$,
which we discuss next.

\subsubsection{Left Focusing}
\label{sec:left-focusing-complete}


\begin{figure}[t]
  \small
  \centering
\begin{fakedisplaymath}
\begin{array}{r@{\qquad}ccc@{\qquad}l}
  \rlabel{AR-nat} \Longrightarrow & \mathcal{L}_0; \mathcal{M}_0; \ottnt{A_{{\mathrm{0}}}}; \mathcal{X}_0 \vdash_L \ottnt{A_{{\mathrm{0}}}} \prec:  \mathsf{Nat}  & \rightsquigarrow & \mathcal{X}_n & (\mathcal{M}_0 =  [] , \mathcal{X}_0 =  \Box ) \\
  & \downarrow & & \uparrow & \\
  & \mathcal{L}_1; \mathcal{M}_1; \ottnt{A_{{\mathrm{0}}}}; \mathcal{X}_1 \vdash_L \ottnt{A}_1 \prec:  \mathsf{Nat}  & \rightsquigarrow & \mathcal{X}_n & \\
  & \downarrow & & \uparrow \\
  & \cdots & & \cdots & \\
  & \downarrow & & \uparrow & \\
  \rlabel{AL-nat} \Longrightarrow & \mathcal{L}_n; \mathcal{M}_n; \ottnt{A_{{\mathrm{0}}}}; \mathcal{X}_n \vdash_L \ottnt{A}_n \prec:  \mathsf{Nat}  & \rightsquigarrow & \mathcal{X}_n & (\ottnt{A}_n =  \mathsf{Nat} , \mathcal{L}_n =  [] ) \\
\end{array}
\end{fakedisplaymath}
  \caption{Visual Representation of the Execution of ($\algSubL{\mathcal{L}}{\mathcal{M}}{\ottnt{A_{{\mathrm{0}}}}}{\mathcal{X}}{A}{B}{\mathcal{X}'}$)}
  \label{fig:algorithm-left-visualization}
\end{figure}

The intuition behind
$\algSubL{\mathcal{L}}{\mathcal{M}}{\ottnt{A_{{\mathrm{0}}}}}{\mathcal{X}}{A}{B}{\mathcal{X}'}$
is that the (input) algorithmic context $\mathcal{X}$ {\em witnesses} the
transition from $\ottnt{A_{{\mathrm{0}}}}$ to $\ottsym{(}   \mathcal{M}  \to\hspace{-8pt}\to  \ottnt{A}   \ottsym{)}$, the component of $\ottnt{A_{{\mathrm{0}}}}$ that we are
currently focusing  on. The (output) algorithmic context $\mathcal{X}'$ is meant
to capture the complete transition from $\ottnt{A_{{\mathrm{0}}}}$ to $( \mathcal{M}  \to\hspace{-8pt}\to  \ottsym{(}   \mathcal{L}  \to\hspace{-8pt}\to  \ottnt{B}   \ottsym{)} )$.
Essentially, starting with rule~\rlabel{AR-nat} as the entry point, the
judgment continuously decomposes $\ottnt{A}$ (via rules \rlabel{AL-and1},
\rlabel{AL-and2}, \rlabel{AL-arr}, and \rlabel{AL-mp}) until it equals
$\ottnt{B}$ (rule~\rlabel{AL-nat}); then the accumulated context $\mathcal{X}$
witnesses the complete transition from $\ottnt{A_{{\mathrm{0}}}}$ to $( \mathcal{M}  \to\hspace{-8pt}\to  \ottsym{(}   \mathcal{L}  \to\hspace{-8pt}\to  \ottnt{B}   \ottsym{)} )$ and the job is done. 
Figure~\ref{fig:algorithm-left-visualization} visualizes this process.

%
%
%
%

\paragraph{Subtyping}

Ignoring coercion generation, the judgment works as follows:
Rule~\rlabel{AL-nat} constitutes the base case where subtyping holds trivially.
Rules \rlabel{AL-and1} and~\rlabel{AL-and2} non-deterministically focus on the
left or the right component of an intersection, similarly to their declarative
counterparts (rules~\rlabel{S-andl} and~\rlabel{S-andr}, respectively).
Rule~\rlabel{AL-arr} is the algorithmic version of rule~\rlabel{S-arr}: we dequeue
$\ottnt{B_{{\mathrm{1}}}}$ from $\mathcal{L}$ in a first-in-first-out manner, in order to match the order
in which arrows are encountered in $\ottnt{A}$. After we prove subtyping for the
domains of the functions, we record that we are going under a binder of type
$\ottnt{B_{{\mathrm{1}}}}$ in the queue $\mathcal{M}$ and proceed recursively with the function
images. Queue $\mathcal{M}$ thus captures all the arrow domains in $\ottnt{B}$
that have been removed in lockstep with arrow domains in $\ottnt{A}$ using
rule~\rlabel{S-arr}.\footnote{In Figure~\ref{fig:algorithm-left-visualization},
it is easy to see that $\mathcal{M}_i \,\mathtt{++}\,
\mathcal{L}_i = \mathcal{L}_0$, where \(\mathtt{++}\) denotes list
concatenation. At the end, we have
$\mathcal{M}_n = \mathcal{L}_0$.} This context becomes useful when we need to
use Modus Ponens under a binder, e.g. to prove
\[
\mathsf{Nat}  \rightarrow  \ottsym{(}   \mathsf{Bool}   \, \& \,  \ottsym{(}   \mathsf{Bool}   \rightarrow   \mathsf{String}   \ottsym{)}  \ottsym{)}  \ottsym{<:}  \mathsf{Nat}  \rightarrow   \mathsf{String} 
\]
(or even $\top  \rightarrow  \ottsym{(}   \mathsf{Bool}   \, \& \,  \ottsym{(}   \mathsf{Bool}   \rightarrow   \mathsf{String}   \ottsym{)}  \ottsym{)}  \ottsym{<:}  \mathsf{Nat}  \rightarrow   \mathsf{String} $). Lastly,
rule~\rlabel{AL-mp} is the algorithmic version of the~\rlabel{S-mp} rule. The
first premise reveals the synergistic relationship between this rule and
rule~\rlabel{AL-arr}: when we need to synthesize an argument for Modus Ponens,
we restart from the original type $\ottnt{A_{{\mathrm{0}}}}$---thus capturing the complete
implicit context of $\ottnt{A_{{\mathrm{1}}}}  \rightarrow  \ottnt{A_{{\mathrm{2}}}}$---and use the contextualized version of
$\ottnt{A_{{\mathrm{1}}}}$, ($ \mathcal{M}  \to\hspace{-8pt}\to  \ottnt{A_{{\mathrm{1}}}} $). For the above example, this amounts to
proving
\[
\mathsf{Nat}  \rightarrow  \ottsym{(}   \mathsf{Bool}   \, \& \,  \ottsym{(}   \mathsf{Bool}   \rightarrow   \mathsf{String}   \ottsym{)}  \ottsym{)}  \ottsym{<:}  \mathsf{Nat}  \rightarrow   \mathsf{Bool} 
\]
The second premise is straightforward.

Note that rules \rlabel{AL-arr} and \rlabel{AL-mp} overlap and, to ensure completeness
with respect to the declarative specification of subtyping, backtracking is required.
However, it is never the case that both rules can apply for resolving a subtyping
judgment. That is, backtracking will always fail in at least one of the two choices.


\paragraph{Coercion Generation}

The elaboration side, again, requires a bit more explanation.
The complication arises in rule~\rlabel{AL-mp}, where the first premise
restarts from the original type $\ottnt{A_{{\mathrm{0}}}}$. This is essential for recovering the
unfocused context needed for Modus Ponens, but it goes against the inductive,
top-down approach to coercion generation that we follow in the right-focusing
rules. Coercions produced by the recursive calls can no longer be directly
combined. This is why the left focusing rules use algorithmic
contexts instead of coercions.
The essence of the algorithmic contexts is captured in the following invariant,
which is preserved by each step of the left focusing algorithm:
\begin{invariant}
\label{invariant:algorithmic-contexts}
$\forall \ottnt{c}~\ottnt{A'},
\ottnt{A}_i <: \ottnt{A}' \elbox{\rightsquigarrow \ottnt{c}} \implies \ottnt{A_{{\mathrm{0}}}} <: (\mathcal{M}_i \arrows \ottnt{A}') \elbox{\rightsquigarrow \mathcal{X}_i [c]}$
\end{invariant}

Metavariables $\ottnt{A}_i$, $\ottnt{A_{{\mathrm{0}}}}$, $\mathcal{M}_i$, and $\mathcal{X}_i$
refer to the corresponding parameters at the $i$-th step in
Figure~\ref{fig:algorithm-left-visualization}.
Informally, $\mathcal{X}$ captures the transition from $\ottnt{A_{{\mathrm{0}}}}$ to
$\ottnt{A}_i$: $\mathcal{X}_i [c]$ witnesses the complete transition from $\ottnt{A_{{\mathrm{0}}}}$
to {\em any} type $\ottnt{A}'$, if we have a way to perform the remaining
transition from $\ottnt{A}_i$ to $\ottnt{A}'$.
Hence, every rule of the judgment strips a layer off $\ottnt{A}$ and extends the
``input'' algorithmic context accordingly, so that
Invariant~\ref{invariant:algorithmic-contexts} is preserved.

Rule~\rlabel{AL-mp} is by far the most interesting rule.
It commits to using the~\rlabel{S-mp} rule: the left premise
produces a coercion $\ottnt{c_{{\mathrm{2}}}}$ that provides the argument (of type
$ \mathcal{M}  \to\hspace{-8pt}\to  \ottnt{A_{{\mathrm{1}}}} $), while the appropriate instantiation of the current
context $\mathcal{X}$ gives us the function (of type $ \mathcal{M}  \to\hspace{-8pt}\to  \ottsym{(}  \ottnt{A_{{\mathrm{1}}}}  \rightarrow  \ottnt{A_{{\mathrm{2}}}}  \ottsym{)} $):
\[
\begin{array}{l}
 \ottnt{A_{{\mathrm{0}}}}   \ottsym{<:}    \mathcal{M}  \to\hspace{-8pt}\to  \ottnt{A_{{\mathrm{1}}}}   \elbox{ \rightsquigarrow   \ottnt{c_{{\mathrm{2}}}} }  \\
 \ottnt{A_{{\mathrm{0}}}}   \ottsym{<:}    \mathcal{M}  \to\hspace{-8pt}\to  \ottsym{(}  \ottnt{A_{{\mathrm{1}}}}  \rightarrow  \ottnt{A_{{\mathrm{2}}}}  \ottsym{)}   \elbox{ \rightsquigarrow    \mathcal{X} [   \mathsf{id} _{  \ottsym{\mbox{$\mid$}}  \ottnt{A_{{\mathrm{1}}}}  \rightarrow  \ottnt{A_{{\mathrm{2}}}}  \ottsym{\mbox{$\mid$}}  }   ]  }  \\
\end{array}
\]
Yet, the above coercions cannot be directly combined, due to the type queue
$\mathcal{M}$. We remedy this using meta-function
$ \distArrowsCo{ \mathcal{M} }{ \ottnt{c} }{ \ottnt{B_{{\mathrm{1}}}} }{ \ottnt{B_{{\mathrm{2}}}} } $, whose meaning is given by the following
lemma:\footnote{Its definition is straightforward and
thus omitted; it can be found in Appendix~\ref{appendix:metafunctions}. 
}

\begin{lemma}
\small
$\forall \mathcal{L}\,\ottnt{B_{{\mathrm{1}}}}\,\ottnt{B_{{\mathrm{2}}}}\,\ottnt{B}\,\ottnt{c},
 \ottnt{B_{{\mathrm{1}}}}  \, \& \,  \ottnt{B_{{\mathrm{2}}}}   \ottsym{<:}   \ottnt{B}  \elbox{ \rightsquigarrow   \ottnt{c} }  \implies
    \ottsym{(}   \mathcal{L}  \to\hspace{-8pt}\to  \ottnt{B_{{\mathrm{1}}}}   \ottsym{)}  \, \& \,  \ottsym{(}   \mathcal{L}  \to\hspace{-8pt}\to  \ottnt{B_{{\mathrm{2}}}}   \ottsym{)}   \ottsym{<:}    \mathcal{L}  \to\hspace{-8pt}\to  \ottnt{B}   \elbox{ \rightsquigarrow    \distArrowsCo{ \mathcal{L} }{ \ottnt{c} }{ \ottnt{B_{{\mathrm{1}}}} }{ \ottnt{B_{{\mathrm{2}}}} }  } $.
\end{lemma}

\noindent
The general idea is that if $( \ottnt{A_{{\mathrm{0}}}}   \ottsym{<:}    \mathcal{M}  \to\hspace{-8pt}\to  \ottnt{A_{{\mathrm{1}}}}   \elbox{ \rightsquigarrow   \ottnt{c_{{\mathrm{2}}}} } )$ and $( \ottnt{A_{{\mathrm{0}}}}   \ottsym{<:}    \mathcal{M}  \to\hspace{-8pt}\to  \ottsym{(}  \ottnt{A_{{\mathrm{1}}}}  \rightarrow  \ottnt{A_{{\mathrm{2}}}}  \ottsym{)}   \elbox{ \rightsquigarrow   \ottnt{c_{{\mathrm{1}}}} } )$
then
\[
  \ottnt{A_{{\mathrm{0}}}}  \ottsym{<:}   \mathcal{M}  \to\hspace{-8pt}\to  \ottnt{A_{{\mathrm{2}}}}  \elbox{\rightsquigarrow (\lambda c.   \distArrowsCo{ \mathcal{M} }{ \ottsym{(}   \ottnt{c}  \circ  \ottsym{(}   \pi_1 ^{  \ottsym{\mbox{$\mid$}}  \ottsym{(}  \ottnt{A}  \rightarrow  \ottnt{B}  \ottsym{)}  \ottsym{\mbox{$\mid$}} , \ottsym{\mbox{$\mid$}}  \ottnt{A}  \ottsym{\mbox{$\mid$}}  }   \modusponens   \pi_2 ^{  \ottsym{\mbox{$\mid$}}  \ottsym{(}  \ottnt{A}  \rightarrow  \ottnt{B}  \ottsym{)}  \ottsym{\mbox{$\mid$}} , \ottsym{\mbox{$\mid$}}  \ottnt{A}  \ottsym{\mbox{$\mid$}}  }   \ottsym{)}   \ottsym{)} }{ \ottsym{(}  \ottnt{A}  \rightarrow  \ottnt{B}  \ottsym{)} }{ \ottnt{A} }   \circ  \langle  \ottnt{c_{{\mathrm{1}}}}  \ottsym{,}  \ottnt{c_{{\mathrm{2}}}}  \rangle )~[\null{} \mathsf{id} _{  \ottsym{\mbox{$\mid$}}  \ottnt{A_{{\mathrm{2}}}}  \ottsym{\mbox{$\mid$}}  } \null]}
\]
Returning to rule~\rlabel{AL-mp}, the second premise uses the above
algorithmic context (named $\mathcal{X}'$ for convenience) to further focus on
$\ottnt{A_{{\mathrm{2}}}}$.

\subsection{Algorithm Metatheory}
\label{subsec:algorithm:metatheory}

\paragraph{Subsumption of \necolus' Algorithmic Subtyping}
First, we have proven that the subtyping algorithm of
Figure~\ref{fig:algorithmic-subtyping} subsumes the subtyping algorithm of
\necolus~\cite[Fig.~12]{necolus}:
%
\begin{theorem}[Subsumption of \necolus' Algorithm]
If $\necolusAlgSub{ [] }{\ottnt{A}}{\ottnt{B}}{\ottnt{c}}$ then $\algSubMain{\ottnt{A}}{\ottnt{B}}{\ottnt{c}}$.
\end{theorem}
where $\necolusAlgSub{\mathcal{L}}{\ottnt{A}}{\ottnt{B}}{\ottnt{c}}$ denotes the
subtyping algorithm of \necolus.

\paragraph{Transivity and Modus Ponens}

A major difference between algorithmic and declarative subtyping is that the
latter has an explicit transitivity rule while the former does not. Hence, 
transitivity of algorithmic subtyping requires a separate proof. Unfortunately,
this proof is considerably more involved than in \necolus. Instead of 
one inductive core lemma, it consists of 7 mutually recursive lemmas. 
The well-foundedness of the induction used in the proofs of those 7 lemmas is no longer
a straightforward size-based induction that can be easily understood
by Coq. To overcome this problem we have used a new approach, external to Coq, to establish
well-foundedness. We have encoded the proof---which is essentially an algorithm that
transforms derivation trees---as a term rewriting system. Then we have used the AProVE
tool~\citep{aprove} to verify its termination automatically. AProVE reports success after 40 steps of transformation,
decomposition and basic termination arguments.

\begin{theorem}[Transitivity of Algorithmic Subtyping]\label{theorem:transivity}
If $\algSubMain{\ottnt{A_{{\mathrm{1}}}}}{\ottnt{A_{{\mathrm{2}}}}}{\ottnt{c_{{\mathrm{1}}}}}$ and $\algSubMain{\ottnt{A_{{\mathrm{2}}}}}{\ottnt{A_{{\mathrm{3}}}}}{\ottnt{c_{{\mathrm{2}}}}}$,
then $\algSubMain{\ottnt{A_{{\mathrm{1}}}}}{\ottnt{A_{{\mathrm{3}}}}}{\ottnt{c_{{\mathrm{3}}}}}$ for some $\ottnt{c_{{\mathrm{3}}}}$.
\end{theorem}

The modus ponens property for algorithmic subtyping is closely tied to 
transitivity proof and is another consequence of the 7 mutually recursive lemmas.
\begin{corollary}[Modus Ponens of Algorithmic Subtyping]
If $\algSubMain{\ottnt{A}}{\ottnt{B_{{\mathrm{1}}}}  \rightarrow  \ottnt{B_{{\mathrm{2}}}}}{\ottnt{c_{{\mathrm{1}}}}}$ and $\algSubMain{\ottnt{A}}{\ottnt{B_{{\mathrm{1}}}}}{\ottnt{c_{{\mathrm{2}}}}}$,
then $\algSubMain{\ottnt{A}}{\ottnt{B_{{\mathrm{2}}}}}{\ottnt{c_{{\mathrm{3}}}}}$ for some $\ottnt{c_{{\mathrm{3}}}}$.
\end{corollary}

\paragraph{Soundness and Completeness}

We have proven that the subtyping algorithm
(Figure~\ref{fig:algorithmic-subtyping}) is sound and complete with respect to its
declarative specification (Figure~\ref{fig:subtyping}). The soundness of the
overall algorithm is a direct corollary of the soundness of right focusing
(Lemma~\ref{lemma:right-rules-soundness}):
\begin{theorem}[Soundness]
If $\algSubMain{\ottnt{A}}{\ottnt{B}}{\ottnt{c}}$
then $ \ottnt{A}   \ottsym{<:}   \ottnt{B}  \elbox{ \rightsquigarrow   \ottnt{c} } $.
\end{theorem}

\noindent
The completeness proof largely follows the same strategy as for \necolus. The main 
challenge was the new transitivity proof discussed above. 

\begin{theorem}[Completeness]
If $ \ottnt{A}   \ottsym{<:}   \ottnt{B}  \elbox{ \rightsquigarrow   \ottnt{c} } $ then there exists $\ottnt{c'}$ such that $\algSubMain{\ottnt{A}}{\ottnt{B}}{\ottnt{c'}}$.
\end{theorem}

\paragraph{Termination}
The final key property of the algorithm is termination. Actually, the algorithm
presented in Figure~\ref{fig:algorithmic-subtyping} is not terminating. 
Consider for example the trace of $\noEvAlgSubMain{\mathsf{Nat}  \rightarrow  \mathsf{Nat}}{ \mathsf{Nat} }$:
\[
\begin{array}{l}
  (\noEvAlgSubR{ [] }{\mathsf{Nat}  \rightarrow  \mathsf{Nat}}{ \mathsf{Nat} }) \xrightarrow{\rlabel{AR-nat}} (\noEvAlgSubL{ [] }{ [] }{\mathsf{Nat}  \rightarrow  \mathsf{Nat}}{\mathsf{Nat}  \rightarrow  \mathsf{Nat}}{ \mathsf{Nat} }) \xrightarrow{\rlabel{AL-mp}} \\
  (\noEvAlgSubR{ [] }{\mathsf{Nat}  \rightarrow  \mathsf{Nat}}{ \mathsf{Nat} }) \xrightarrow{\rlabel{AR-nat}} (\noEvAlgSubL{ [] }{ [] }{\mathsf{Nat}  \rightarrow  \mathsf{Nat}}{\mathsf{Nat}  \rightarrow  \mathsf{Nat}}{ \mathsf{Nat} }) \xrightarrow{\rlabel{AL-mp}} \dots \\
\end{array}
\]
%
Such a subtyping judgment can result from an erroneous
type annotation by the user. Nonetheless, the language does not statically
detect annotations that may cause non-termination
in the subtyping algorithm.
Another example, perhaps more plausible as a programming error, is an annotation
that leads to the judgment
$\noEvAlgSubMain{\ottsym{(}  \mathsf{Nat}  \rightarrow   \mathsf{Bool}   \ottsym{)}  \, \& \,  \ottsym{(}   \mathsf{Bool}   \rightarrow  \mathsf{Nat}  \ottsym{)}}{ \mathsf{Nat} }$.
One simple ingredient needs to be added to the algorithm to make it
terminating: a loop detection on the $\noEvAlgSubR{ [] }{A}{B}$ call in
rule \rlabel{AL-mp}. This loop detection should reject any calls
that repeat one of their ancestor calls.

We can show that (1) any infinite derivation involves an infinite number of these
nested calls and that (2) detecting repeated calls is effective at
stopping the non-termination. 

A priori there are multiple possible ways in which the algorithm can
diverge. It might involve an infinite sequence of recursive calls within one of the two judgments.
However, following the focusing approach, the two judgments structurally
decompose their focal argument. Hence, on their own they are clearly
terminating. Any divergence must thus involve an infinite number of mutually recursive
calls between the two judgments. There are two ``gatekeepers'' of the mutual recursion,
calling back from the left-focusing into the right-focusing judgment. These two are the 
problematic one in rule~\rlabel{AL-mp} and the one in rule~\rlabel{AL-arr}.
Yet, we can show that an infinite number of recursion steps through rule
\rlabel{AL-Arr} is impossible and thus we can eliminate it from consideration.
Indeed, the \emph{left-arrow depth} $\lad{\cdot}$ defined below
decreases strictly in every \rlabel{AL-arr} loop step, and non-strictly in
every other step. Moreover, a value of 0 inhibits rule \rlabel{AL-arr}. Thus,
after a finite number of \rlabel{AL-arr} steps, the rule can no longer be used.
\begin{equation*}
\begin{array}{rcl@{\hspace{2cm}}rcl}
\lad{ \mathsf{Nat} } & = & 0 & \lad{\ottnt{A}  \rightarrow  \ottnt{B}} & = & \max{(\lad{A} + 1,\lad{B})} \\
\lad{ \top } & = & 0 & \lad{\ottnt{A}  \, \& \,  \ottnt{B}} & = & \max{(\lad{A},\lad{B})} \\
\multicolumn{6}{c}{\lad{\noEvAlgSubR{ [] }{A}{B}} = \max{(\lad{A},\lad{B})}}
\end{array}
\end{equation*}

That leaves only the problematic mutually recursive call in rule~\rlabel{AL-mp}, which we
have already demonstrated to be a source of non-termination. Fortunately, we
can show that the number of distinct nested calls is finite. Hence, on any
infinite derivation path, a repeated call has to occur after a finite number of
steps. This makes the loop detection effective at curtailing divergence.

A more extensive account can be found in Appendix~\ref{appendix:termination-proof}. 

\begin{theorem}[Termination]
The algorithm extended with loop detection terminates.
\end{theorem}

\paragraph{Decidability}
In summary, if we take the soundness, completeness and termination of the subtyping
together, we can conclude that the declarative subtyping is decidable.
\begin{corollary}[Decidable Subtyping]
In \calculus, \(\ottnt{A}  \ottsym{<:}  \ottnt{B}\) is decidable for any types \(\ottnt{A}\) and \(\ottnt{B}\).
\end{corollary}

%% file: sources/coherence.mng
\section{Coherence}

A desired property for calculi with a merge operator is \emph{coherence}.
This ensures that no semantic ambiguity arises in programs, i.e. programs and
their subparts always have a single, unambiguous meaning. Recall that the semantics of
\calculus is given indirectly, through an elaboration to 
\target  and its operational semantics.
Hence, coherence for \calculus means that all possible elaborations of a
well-typed source expression for a given type are contextually equivalent in the target language. 

Two target expressions, that may be syntactically different, are
\emph{contextually equivalent} iff, when used in the context of any program,
there is no difference in that program's behavior. In our purely functional and
strongly normalising setting, we can reduce a program's behavior to the value
it returns. Moreover, as usual, we can restrict ourselves without loss of
generality to programs of type \( \mathsf{Nat} \) to facilitate the comparison of the
results.

\emph{Program contexts} express the above notion of using an expression in the context of
a program and are defined as follows:
\begin{center}
\begin{tabular}{@{\hspace{0mm}}l@{\hspace{2mm}}l@{\hspace{2mm}}l@{\hspace{2mm}}l@{\hspace{0mm}}}
  Program contexts & \(\mathcal{C}\) & \(\Coloneqq\) & \( [\cdot]  \mid  \lambda \ottmv{x} .\, \mathcal{C}  \mid \mathcal{C} \, E \mid E \, \mathcal{C} \mid \mathcal{C}  \,,,\,  E \mid E  \,,,\,  \mathcal{C} \mid \mathcal{C}  \ottsym{:}  \ottnt{A}\)\\
\end{tabular}
\end{center}
Intuitively,
a program context is an expression with a missing subexpression. This hole can be
filled in with an appropriate term to form an actual expression. Which terms are eligible
to be used in a program context is captured by the context typing judgment that
consists of judgments of the form \( \mathcal{C}  : (  \Gamma    \Rightarrow    \ottnt{A} )  \mapsto  ( \Gamma'    \Leftarrow    \ottnt{B} ) \elbox{ \rightsquigarrow   \mathcal{D} } \).\footnote{
There are also judgments for the three other combinations of the two typing directions
that appear in the judgment. The full specification can be found in
Appendix~\ref{appendix:context-typing}. 
}
Essentially, the judgment expresses that program context \(\mathcal{C}\)
expects a term of type \(\ottnt{A}\) under typing context \(\Gamma\) to form an expression
of type \(\ottnt{B}\) under context \(\Gamma'\). Moreover, the source context \(\mathcal{C}\) elaborates
straightforwardly to a target context \(\mathcal{D}\). Indeed, for the coherence of \calculus we do
not account for all possible target contexts, but only for those that are the
image of a source context \(\mathcal{C}\).
This restriction is important,
as~\citet{necolus} have
demonstrated that there are other target contexts that can distinguish between
alternative elaborations.



We can now formally state
coherence%
\footnote{There
is a second 
statement for the type checking direction of the two hypotheses.}
.
\begin{theorem}[Coherence of \calculus]
  \label{prop:coherence}
  For any~~\( \mathcal{C}  : (  \Gamma    \Rightarrow    \ottnt{A} )  \mapsto  (  \bullet     \Rightarrow    \mathsf{Nat} ) \elbox{ \rightsquigarrow   \mathcal{D} } \),\\
  if\, \( \Gamma   \vdash   E    \Rightarrow    \ottnt{A}  \elbox{ \rightsquigarrow   \ottnt{e_{{\mathrm{1}}}} } \) and \( \Gamma   \vdash   E    \Rightarrow    \ottnt{A}  \elbox{ \rightsquigarrow   \ottnt{e_{{\mathrm{2}}}} } \),
  then there is a value \( \bullet   \vdash  \ottnt{v}  \ottsym{:}  \mathsf{Nat}\) such that
  \(\mathcal{D}  \ottsym{\{}  \ottnt{e_{{\mathrm{1}}}}  \ottsym{\}}  \longrightarrow^{*}  \ottnt{v}\) and \(\mathcal{D}  \ottsym{\{}  \ottnt{e_{{\mathrm{2}}}}  \ottsym{\}}  \longrightarrow^{*}  \ottnt{v}\).
\end{theorem}

Here, \(\mathcal{D}  \ottsym{\{}  \ottnt{e}  \ottsym{\}}\) denotes the term that results from filling in
the hole of \(\mathcal{D}\) with expression \(\ottnt{e}\). 


The remainder of this section summarizes our proof approach for coherence
and discusses the issues for our proof's partiality.
A more extensive account can be found in Appendix~\ref{appendix:coherence}. 

\subsection{A Context Dependent Logical Relation}

We follow the general approach for showing coherence as \necolus. This approach
is based on a rather atypical heterogeneous binary logical relation, called
\emph{canonicity}, by~\citet{bi2019distributive}.

\paragraph{Canonicity}
The canonicity relation is in fact two families of relations, one for values
$\mathcal{V}$ and one for expressions $\mathcal{E}$, indexed by types.
Two values $\ottnt{v_{{\mathrm{1}}}}$ and $\ottnt{v_{{\mathrm{2}}}}$ of types $\tau_{{\mathrm{1}}}$ and $\tau_{{\mathrm{2}}}$ are related by
canonicity, written $(\ottnt{v_{{\mathrm{1}}}}, \ottnt{v_{{\mathrm{2}}}}) \in \mathcal{V}(\tau_{{\mathrm{1}}},\tau_{{\mathrm{2}}})$, iff any
parts of the same type have the same meaning. By parts we mean values
$\ottnt{v'_{{\mathrm{1}}}}$ and $\ottnt{v'_{{\mathrm{2}}}}$ derivable from respectively $\ottnt{v_{{\mathrm{1}}}}$ and
$\ottnt{v_{{\mathrm{2}}}}$ by means of subtyping coercions. If two values have non-trivial
(i.e., non-top-like) parts with the same type, we say that they overlap.  For
instance, values $\langle   1   \ottsym{,}   \mathsf{true}   \rangle$ and $ 1 $ overlap since both have
a part of type $ \mathsf{Nat} $; because their overlapping part is the
same, namely $ 1 $, these values are related by canonicity. In contrast,
$\langle   1   \ottsym{,}   \mathsf{true}   \rangle$ and $ 2 $ are not related since their overlapping
$ \mathsf{Nat} $ parts are different.

While the \calculus canonicity relation follows the above principles just like
that of \necolus, there are considerable differences in the actual definition
because of the addition of modus ponens.

\paragraph{Context Dependency}

A major complicating factor in the design of the new logical relation is
that the modus ponens subtyping rule introduces a form of context dependency.
Indeed, when reasoning about overlapping parts, the \necolus approach of
decomposing terms and their types, and looking at their syntactic components
in isolation no longer works. 
Instead, in \calculus the components have to be
considered in the context of the whole value they appear in. For
example, the target term $ \lambda \ottmv{x} : \mathsf{Nat} .\,  \mathsf{true}  $ does not overlap with
$ \mathsf{false} $, since one is an arrow type and the other a boolean. Yet,
when $ \lambda \ottmv{x} : \mathsf{Nat} .\,  \mathsf{true}  $ is part of the larger
term $\langle   \lambda \ottmv{x} : \mathsf{Nat} .\,  \mathsf{true}    \ottsym{,}   1   \rangle$, a problematic
overlap arises: via modus ponens, the function applied
to $ 1 $ to yields the boolean value $ \mathsf{true} $ that
differs from $ \mathsf{false} $.

To account for that dependency on the whole value, or contextual dependency for
short, we equip the canonicity relation of \calculus with two additional
indices: the whole values and their types.
Thus, we write 
\begin{equation*}
 (  \ottnt{v_{{\mathrm{1}}}}  ,  \ottnt{v_{{\mathrm{2}}}}  ) \in \rellV ^{(  \ottnt{v'_{{\mathrm{1}}}}  :  \tau'_{{\mathrm{1}}}  ;  \ottnt{v'_{{\mathrm{2}}}}  :  \tau'_{{\mathrm{2}}}  )} \llbracket  \tau_{{\mathrm{1}}}  ;  \tau_{{\mathrm{2}}} \rrbracket 
\end{equation*}
to express that $\ottnt{v_{{\mathrm{1}}}}$ and $\ottnt{v_{{\mathrm{2}}}}$ are related values of types $\tau_{{\mathrm{1}}}$ and
$\tau_{{\mathrm{2}}}$ that are components of respectively the values $\ottnt{v'_{{\mathrm{1}}}}$ of type
$\tau'_{{\mathrm{1}}}$ and $\ottnt{v'_{{\mathrm{2}}}}$ of type $\tau'_{{\mathrm{2}}}$.

We refer to Appendix~\ref{appendix:coherence} 
for the full definition of this
context-dependent canonicity relation $\mathcal{V}$ for values and its 
companion $\mathcal{E}$ for expressions.

\paragraph{Properties}

A key property of the canonicity relation is \emph{coercion compatibility}.
This states that any coercions of related values are related as well.

\begin{lemma}[Coercion Compatibility] \ \\
\label{lemma:coercion-compatibility}
If $ (  \ottnt{v_{{\mathrm{1}}}}  ,  \ottnt{v_{{\mathrm{2}}}}  ) \in \rellV ^{(  \ottnt{v_{{\mathrm{1}}}}  :  \tau_{{\mathrm{1}}}  ;  \ottnt{v_{{\mathrm{2}}}}  :  \tau_{{\mathrm{2}}}  )} \llbracket  \tau_{{\mathrm{1}}}  ;  \tau_{{\mathrm{2}}} \rrbracket $, then
for all coercions $\ottnt{c_{{\mathrm{1}}}}$, $\ottnt{c_{{\mathrm{2}}}}$ and types $\tau'_{{\mathrm{1}}}$, $\tau'_{{\mathrm{2}}}$ such that
$\ottnt{c_{{\mathrm{1}}}}  \vdash  \tau_{{\mathrm{1}}} \, \triangleright \, \tau'_{{\mathrm{1}}}$ and $\ottnt{c_{{\mathrm{2}}}}  \vdash  \tau_{{\mathrm{2}}} \, \triangleright \, \tau'_{{\mathrm{2}}}$ it holds that
$ (  \ottnt{c_{{\mathrm{1}}}} \, \ottnt{v_{{\mathrm{1}}}}  ,  \ottnt{c_{{\mathrm{2}}}} \, \ottnt{v_{{\mathrm{2}}}}  ) \in \rellE^{(  \ottnt{c_{{\mathrm{1}}}} \, \ottnt{v_{{\mathrm{1}}}}  :  \tau'_{{\mathrm{1}}}  ;  \ottnt{c_{{\mathrm{2}}}} \, \ottnt{v_{{\mathrm{2}}}}  :  \tau'_{{\mathrm{2}}}  )} \llbracket \tau'_{{\mathrm{1}}}  ;  \tau'_{{\mathrm{2}}} \rrbracket $.
\end{lemma}

The following lemma asserts another key property of the value relation, which
establishes its connection to disjointness. This follows from the design
principle that any overlap in types should concern indistinguishable values.
If there is no overlap, then this is trivially satisfied.

\begin{lemma}[Disjointly-Typed Values are Related]
\label{lemma:disjoint-values-related}
If $ \bullet   \vdash  \ottnt{v_{{\mathrm{1}}}}  \ottsym{:}  \ottsym{\mbox{$\mid$}}  \ottnt{A_{{\mathrm{1}}}}  \ottsym{\mbox{$\mid$}}$ and $ \bullet   \vdash  \ottnt{v_{{\mathrm{1}}}}  \ottsym{:}  \ottsym{\mbox{$\mid$}}  \ottnt{A_{{\mathrm{2}}}}  \ottsym{\mbox{$\mid$}}$ where $\ottnt{A_{{\mathrm{1}}}}  *  \ottnt{A_{{\mathrm{2}}}}$, then $ (  \ottnt{v_{{\mathrm{1}}}}  ,  \ottnt{v_{{\mathrm{2}}}}  ) \in \rellV ^{(  \ottnt{v_{{\mathrm{1}}}}  :  \ottsym{\mbox{$\mid$}}  \ottnt{A_{{\mathrm{1}}}}  \ottsym{\mbox{$\mid$}}  ;  \ottnt{v_{{\mathrm{2}}}}  :  \ottsym{\mbox{$\mid$}}  \ottnt{A_{{\mathrm{2}}}}  \ottsym{\mbox{$\mid$}}  )} \llbracket  \ottsym{\mbox{$\mid$}}  \ottnt{A_{{\mathrm{1}}}}  \ottsym{\mbox{$\mid$}}  ;  \ottsym{\mbox{$\mid$}}  \ottnt{A_{{\mathrm{2}}}}  \ottsym{\mbox{$\mid$}} \rrbracket $.
\end{lemma}

\subsection{Coherence Proof}

With the canonicity relation in place, we can turn to the actual coherence
proof. This proof is split in two parts by the canonicity relation.
The first and hardest part is to show that any two elaborations of the same
source term are related by canonicity; this is the so-called \emph{fundamental
property}, . The second part is to show that related programs of type \( \mathsf{Nat} \)
return the same value; it follows almost trivially from the definition of the
canonicity relation.

\paragraph{Logical Equivalence}
To express the fundamental property, we need to introduce a notion of logical
equivalence for open expressions on top of the canonicity relations for closed values
(\relV) and closed terms (\relE). Open expressions are considered in some
typing context \(\Delta\).
A \emph{closing substitution} \(\gamma\) of a typing context \(\Delta\) maps variables
to values of the type specified in \(\Delta\).
Two open expressions are related iff applying any two related closing
substitutions to \(\ottnt{e_{{\mathrm{1}}}}\) and \(\ottnt{e_{{\mathrm{2}}}}\), respectively, results in
\relE-related closed terms:
\begin{equation*}
   \Delta_{{\mathrm{1}}}  ;  \Delta_{{\mathrm{2}}}   \vdash  \logeq{ \ottnt{e_{{\mathrm{1}}}} }{ \ottnt{e_{{\mathrm{2}}}} } :  \tau_{{\mathrm{1}}}  ;  \tau_{{\mathrm{2}}}  \defeq \forall~ (  \gamma_{{\mathrm{1}}}  ;  \gamma_{{\mathrm{2}}}  ) \in \rellG\llbracket \Delta_{{\mathrm{1}}}  ;  \Delta_{{\mathrm{2}}} \rrbracket ,~~ (  \ottnt{e_{{\mathrm{1}}}}  \ottsym{[}  \gamma_{{\mathrm{1}}}  \ottsym{]}  ,  \ottnt{e_{{\mathrm{2}}}}  \ottsym{[}  \gamma_{{\mathrm{2}}}  \ottsym{]}  ) \in \rellV ^{(  \ottnt{e_{{\mathrm{1}}}}  \ottsym{[}  \gamma_{{\mathrm{1}}}  \ottsym{]}  :  \tau_{{\mathrm{1}}}  ;  \ottnt{e_{{\mathrm{2}}}}  \ottsym{[}  \gamma_{{\mathrm{2}}}  \ottsym{]}  :  \tau_{{\mathrm{2}}}  )} \llbracket  \tau_{{\mathrm{1}}}  ;  \tau_{{\mathrm{2}}} \rrbracket 
\end{equation*}

Two closing substitutions \(\gamma_{{\mathrm{1}}}\) and \(\gamma_{{\mathrm{2}}}\) of \(\Delta_{{\mathrm{1}}}\) and \(\Delta_{{\mathrm{2}}}\)
are related iff the
two typing contexts differ only in the types of the variables (otherwise,
they contain the same variables and in the same order) and
the values that \(\gamma_{{\mathrm{1}}}\) and \(\gamma_{{\mathrm{2}}}\) contain for each variable are \relV-related.

\paragraph{Fundamental Property}
Now we can state the fundamental property:\footnote{There
is a second 
statement for the type checking direction of the two hypotheses.}
\begin{theorem}[Fundamental property]
  \label{thm:fundamental-property}
  If \( \Gamma   \vdash   E    \Rightarrow    \ottnt{A}  \elbox{ \rightsquigarrow   \ottnt{e_{{\mathrm{1}}}} } \) and \( \Gamma   \vdash   E    \Rightarrow    \ottnt{A}  \elbox{ \rightsquigarrow   \ottnt{e_{{\mathrm{2}}}} } \),\\
  then \( \ottsym{\mbox{$\mid$}}  \Gamma  \ottsym{\mbox{$\mid$}}  ;  \ottsym{\mbox{$\mid$}}  \Gamma  \ottsym{\mbox{$\mid$}}   \vdash  \logeq{ \ottnt{e_{{\mathrm{1}}}} }{ \ottnt{e_{{\mathrm{2}}}} } :  \ottsym{\mbox{$\mid$}}  \ottnt{A}  \ottsym{\mbox{$\mid$}}  ;  \ottsym{\mbox{$\mid$}}  \ottnt{A}  \ottsym{\mbox{$\mid$}} \).
\end{theorem}
Due to the complexity of the canonicity relation, we have proven
this theorem under Conjectures~\ref{conj:abstraction-compatibility}
and~\ref{conj:merge-compatibility}.
The proof proceeds by induction on the first hypothesis and
calls coercion compatibility for the \rlabel{T-sub} case. The conjectures
are used in the inductive cases of \rlabel{T-abs} and \rlabel{T-merge}.

\begin{conjecture}[Compatibility of abstractions]
  \label{conj:abstraction-compatibility}
  If~~\( \Delta  \ottsym{,}  \ottmv{x}  \ottsym{:}  \tau'_{{\mathrm{1}}}  ;  \Delta  \ottsym{,}  \ottmv{x}  \ottsym{:}  \tau'_{{\mathrm{2}}}   \vdash  \logeq{ \ottnt{e_{{\mathrm{1}}}} }{ \ottnt{e_{{\mathrm{2}}}} } :  \tau_{{\mathrm{1}}}  ;  \tau_{{\mathrm{2}}} \),\\
  then~~\( \Delta  ;  \Delta   \vdash  \logeq{  \lambda \ottmv{x} : \tau'_{{\mathrm{1}}} .\, \ottnt{e_{{\mathrm{1}}}}  }{  \lambda \ottmv{x} : \tau'_{{\mathrm{2}}} .\, \ottnt{e_{{\mathrm{2}}}}  } :  \tau'_{{\mathrm{1}}}  \rightarrow  \tau_{{\mathrm{1}}}  ;  \tau'_{{\mathrm{2}}}  \rightarrow  \tau_{{\mathrm{2}}} \).
\end{conjecture}
\begin{conjecture}[Compatibility of pairs]
  \label{conj:merge-compatibility}
  If~~\( \Delta  ;  \Delta   \vdash  \logeq{ \ottnt{e_{{\mathrm{1}}}} }{ \ottnt{e_{{\mathrm{2}}}} } :  \tau_{{\mathrm{1}}}  ;  \tau_{{\mathrm{2}}} \)\\and~~\( \Delta  ;  \Delta   \vdash  \logeq{ \ottnt{e'_{{\mathrm{1}}}} }{ \ottnt{e'_{{\mathrm{2}}}} } :  \tau'_{{\mathrm{1}}}  ;  \tau'_{{\mathrm{2}}} \)~~and~~\(\tau_{{\mathrm{1}}}  *  \tau'_{{\mathrm{2}}}\)~~and~~\(\tau_{{\mathrm{2}}}  *  \tau'_{{\mathrm{1}}}\),
  then~~\( \Delta  ;  \Delta   \vdash  \logeq{ \langle  \ottnt{e_{{\mathrm{1}}}}  \ottsym{,}  \ottnt{e'_{{\mathrm{1}}}}  \rangle }{ \langle  \ottnt{e_{{\mathrm{2}}}}  \ottsym{,}  \ottnt{e'_{{\mathrm{2}}}}  \rangle } :  \tau_{{\mathrm{1}}}  \times  \tau'_{{\mathrm{1}}}  ;  \tau_{{\mathrm{2}}}  \times  \tau'_{{\mathrm{2}}} \).
\end{conjecture}

Regarding compatibility of pairs, a tuple value can result either from an explicit merge or
from a coercion applied on some value. In the first case, the subcomponents of the tuple
are disjoint with each other and thus related (by Lemma~\ref{lemma:disjoint-values-related}).
In the second case, given that the original expression is unambiguous and
since coercions do not introduce ambiguity, the coercion application is also unambiguous.
However, 
because of the context dependency
it is not easy to recover the original expression and inspect it.
For a similar reason, compatibility of abstraction is also hard to prove.

\paragraph{Logical Equivalence Entails Contextual Equivalence}

Lastly, we need to show that any logically related target expressions
are contextually equivalent. The proof is fairly straightforward.
\begin{theorem}[Logical Equivalence Entails Contextual Equivalence]
  \label{prop:contextual-equivalence}
  \mbox{}\\
  If \( \ottsym{\mbox{$\mid$}}  \Gamma  \ottsym{\mbox{$\mid$}}  ;  \ottsym{\mbox{$\mid$}}  \Gamma  \ottsym{\mbox{$\mid$}}   \vdash  \logeq{ \ottnt{e_{{\mathrm{1}}}} }{ \ottnt{e_{{\mathrm{2}}}} } :  \ottsym{\mbox{$\mid$}}  \ottnt{A}  \ottsym{\mbox{$\mid$}}  ;  \ottsym{\mbox{$\mid$}}  \ottnt{A}  \ottsym{\mbox{$\mid$}} \), then
  for any~~\( \mathcal{C}  : (  \Gamma    \Rightarrow    \ottnt{A} )  \mapsto  (  \bullet     \Rightarrow    \mathsf{Nat} ) \elbox{ \rightsquigarrow   \mathcal{D} } \),
  there is a value \( \bullet   \vdash  \ottnt{v}  \ottsym{:}  \mathsf{Nat}\) such that
  \(\mathcal{D}  \ottsym{\{}  \ottnt{e_{{\mathrm{1}}}}  \ottsym{\}}  \longrightarrow^{*}  \ottnt{v}\) and \(\mathcal{D}  \ottsym{\{}  \ottnt{e_{{\mathrm{2}}}}  \ottsym{\}}  \longrightarrow^{*}  \ottnt{v}\).
\end{theorem}

Taken together, the (conditionally proven) fundamental property and the entailment of contextual
equivalence establish Theorem~\ref{prop:coherence}, the coherence of \calculus.

%% file: sources/related-work.mng
\section{Related Work}
\label{sec:related}



\paragraph{Resolution and Subtyping}

The main contribution of our work is to unify resolution and subtyping into 
a single mechanism. Resolution is a common technique for automatically constructing implicit
values, such as Haskell's {\em type class dictionaries}~\citep{typeclasses} or
Scala's {\em implicits}~\citep{scala-implicits}. Subtyping is widely
used in Object-Oriented programming and various other type 
systems~\citep{PierceS96,bcd-subtyping,Chen03,GaySJ}. In our work 
we interpret resolution as a special case of coercive
subtyping~\citep{Chen03}, where coercions are generated automatically in a
type-directed fashion and then ``inserted'' into the program. 
This fits well with the common type-directed elaboration style usually employed by resolution mechanisms, which work in a similar way.   

As far as we are aware, the unification of subtyping and resolution is
novel. Despite the fact that languages like Scala integrate both mechanisms,
only~\citet{jeffery} combines both mechanisms within his DIF calculus.
Unfortunately---in contrast to \calculus---DIF's subtyping is known to be
undecidable. Furthermore, Jeffery does not study the coherence of DIF.
Scala implicits~\citep{scala-implicits} have motivated other researchers 
to look into the foundations of the mechanism. However, all previous
work on implicit calculi, including the \emph{implicit calculus}~\citep{implicitcalculus}, 
\emph{Cochis}~\citep{cochis} and the \emph{SI calculus}~\citep{simplicity}, do 
not account for subtyping. Instead, they all focus on a System F~\citep{Re1974} based
language (without subtyping), and model resolution for such a
language. One aspect that these implicit calculi preserve
from Scala is a biased form of resolution where innermost implicits
are preferred. In contrast, our disjointness
approach is more reminiscent of Haskell's non-overlapping 
type class instances, which are unbiased.
 
The most common technique to
algorithmically construct values by resolution, attributed
to~\citet{Kowalski74}, is commonly known as {\em SLD resolution}, or
{\em backwards chaining}.
Another technique, mostly used as a proof search method, is {\em
focusing}~\citep{Focusing,FocusingLatest,UniformProofs} and has been used to
great effect in similar calculi (e.g. \cochis~\citep{cochis} and  Haskell with quantified
class constraints~\citep{quantcs}) to create coherent resolution mechanisms. Our
subtyping algorithm of Section~\ref{sec:algorithm} 
closely resembles
the focusing technique.

\paragraph{Coherence}



Since~\citet{ReynoldsCoherence} proved coherence for a
calculus with intersection types, based on denotational semantics, 
many researchers have studied it in a
variety of typed calculi. Below we summarize two commonly-found approaches in
the literature: normal forms, and logical relations.


\citet{InheritanceImplicitCoercion} proved the coherence of
a coercion translation from Fun~\citep{OnUnderstandingTypes} extended with
recursive types to System F, by showing that any two typing derivations of the
same judgment are normalizable to the same form. \citet{CurienAndGhelli}
presented a translation of \fsub into a calculus
with explicit coercions and showed that any derivations of the same judgment
are translated to terms that are normalizable to a unique normal form.
Following the same approach,~\citet{MonadicSubsumptionCoherence}
proved the coherence of coercion translation from Moggi's computational lambda
calculus~\citep{Moggi} with subtyping.


Central to the first approach is to find a unique normal form for 
derivations. However, this approach cannot be directly applied to Curry-style
calculi, i.e., where the lambda abstractions are not type-annotated. Also, it does not work
when the calculus has general recursion.
\citet{BiernackiAndPolesiuk} considered the coherence
problem of coercion semantics. Their criterion for coherence of the translation
is {\em contextual equivalence} in the target calculus. They used a 
logical relation for establishing coherence for coercion
semantics, applicable in a variety of calculi, including delimited
continuations and control-effect subtyping.

As far as we know, the first work to use logical relations to show the
coherence for intersection types and the merge operator was that of~\citet{necolus}
for \necolus, and later for
\fiplus~\citep{bi2019distributive}. The BCD subtyping in both calculi (as well
as our own) poses a non-trivial complication over Bernacki and Polesiuk's
simple structural subtyping. In their system any two
coercions between given types are behaviorally equivalent in the target
language, so their coherence reasoning can all take place in the target
language. In \necolus, \fiplus, and \calculus this is not the case; coercions
can be distinguished by arbitrary target programs but not those that are
elaborations of source programs. Hence, reasoning is restricted to the latter
class, which is reflected in a more complicated notion of contextual
equivalence and the logical relation's non-trivial treatment of pairs. In
addition to these complications, our logical relation (and consequently the
coherence proof) is significantly more convoluted than that of \necolus and
\fiplus, due to the non-compositionality of Modus Ponens.

\paragraph{Intersection Types and the Merge Operator}

Forsythe~\citep{forsythe} is one of the first languages with intersection types
and a merge-like operator. However, merges in Forsythe cannot contain more than
one function; this restriction ensures that the language is coherent.
Similarly, \citeauthor{overloaded-functions-subtyping}'s $\lambda\&$
calculus~\citep{overloaded-functions-subtyping} includes a special merge
operator that works on functions only.
More recently,~\citet{DunfieldIntersectionElaboration} showed
significant expressiveness of type systems with intersection types and a merge
operator, at the cost of coherence. The limitation was addressed
by~\citet{DisjointIntersectionTypes}, who introduced disjointness to ensure
coherence.
The combination of intersection types, a
merge operator and parametric polymorphism---while achieving coherence---was
first studied in the \ficalculus calculus~\citep{DisjointPolymorphism}. Its
successor, \fiplus~\citep{bi2019distributive} additionally incorporates BCD
subtyping~\citep{bcd-subtyping}, while remaining coherent.
\necolus~\citep{necolus} has been the starting point for our design of \calculus
and incorporates intersection types, a merge operator, and BCD-style
distributivity. Though neither \necolus nor \calculus accounts for parametric
polymorphism, this is a direction we would like to explore in the future.

\paragraph{BCD Type System and Decidability}

The BCD type system was first introduced by~\citet{bcd-subtyping}.
It is derived from a filter lambda model in order to
characterize exactly the strongly normalizing terms. The BCD type system
features a powerful subtyping relation, which serves as a base for ours.
\citet{classes-mixins-intersections} showed
how to type classes and mixins in a BCD-style record calculus with
the merge operator of~\citet{merge-operator}. Their merge only
operates on records, and they only study a type assignment system. The
decidability of BCD subtyping has been shown in several
works~\citep{PierceDecidability,KurataTakahashi,Rehof:2011:FCL:2021953.2021970,Statman}.
\citet{Laurent} has formalized the relation in Coq in order to eliminate
transitivity cuts from it, but he does not deliver an algorithm.
Based on the work of~\citet{Statman},~\citet{types2016} show a
formally verified subtyping algorithm in Coq.
Our algorithm follows the decision procedure of~\citet{necolus}
and~\citet{PierceDecidability}. However the addition of modus ponens
requires significant changes to the algorithm, and complicates
the metatheory.

\paragraph{Coq Type Classes}
Coq employs a flexible type class system~\citep{coqclasses} that permits overlapping instances and provides
the user with the means to control resolution: Coq instances are named and can be
explicitly passed as arguments. The system poses no restrictions over the implicit
environment, like non-overlapping instances in Haskell or disjointness in \calculus.
Therefore, coherence becomes a choice and responsibility of the user.

In contrast, \calculus employs a stricter system that provides less control over
subtyping. First, implicit environments are not possible if not internally disjoint.
Second, going against the intuition which relates classes to types and instances to
appropriately typed terms, expressions in \calculus can be rather viewed as
typed implicit environments, which can be composed with the use of the merge operator.
Passing a value to a function in \calculus can be seen as bringing an implicit
environment in scope within the body of the function. The function, then, can use that
environment to infer values from it, but it does not have access
to the names of the instances contained in it. This is
conceptually different from the named instances of Coq that can be optionally passed
explicitly by the user. 
Last, Coq uses a global environment to which instances are inserted at their point
of declaration.

%% file: sources/conclusion.mng
\section{Conclusion and Future Work}

This paper has shown that, with a small extension, subtyping with
intersection types can be used to model resolution. We have materialized our
ideas in the design of \calculus, a type-safe extension of \necolus with the
logical rule for Modus Ponens. This small extension significantly increases the
expressive power of \calculus's predecessor, allowing us to model features such
as type class resolution and first-class environments.

\paragraph{Future Work}
A natural avenue for future work is the extension of \calculus with parametric
polymorphism. When done naively, it can make the algorithm diverge and,
like in Haskell, additional conditions will have to be imposed to recover
termination. Moreover, we believe that there are various applications for
our work that we wish to explore in the future. 

In a Haskell-like setting, our design would enable multiple
implicit environments, instead of a single
global one. Such environments
could be first-class, could be explicitly passed as arguments
or returned as a result, or could be composed to form larger implicit
environments. The programmer would then be able to choose which of
the implicit environments to use for resolution. 

In a Scala-like setting, where subtyping is present, our design
helps clarify the interaction between resolution and subtyping.
This is important, from the practical point of view, since the interaction
between the two mechanisms has been a continuous source of compiler bugs\footnote{
See for example \url{https://github.com/scala/bug/issues/2509} or \url{https://github.com/scala/bug/issues/9764}.}.
Furthermore, our work could also be interesting to offer a
unified view of subtyping and resolution, which could significantly
reduce the implementation effort for the two mechanisms.

Finally, languages like TypeScript can support a merge operator for intersection
types via an encoding~\citep{DisjointPolymorphism}. Since Typescript
has subtyping, intersection types and the merge operator (via an encoding), 
%
it would be quite interesting to explore the potential to extend
the current formulation of intersection types in TypeScript with the
ideas of our work.

%
%

%% file: sources/definitions.mng
\section{Additional Definitions}
\label{appendix:definitions}

\begin{definition}[Type Translation]
\[
\begin{array}{@{\hspace{0mm}}l@{\hspace{2mm}}c@{\hspace{2mm}}l@{\hspace{0mm}}}
  \ottsym{\mbox{$\mid$}}  \ottnt{A}  \ottsym{\mbox{$\mid$}}      & = &  \mathsf{Nat}  \\
  \ottsym{\mbox{$\mid$}}  \top  \ottsym{\mbox{$\mid$}}    & = &  \langle \rangle  \\
  \ottsym{\mbox{$\mid$}}  \ottnt{A}  \rightarrow  \ottnt{B}  \ottsym{\mbox{$\mid$}} & = & \ottsym{\mbox{$\mid$}}  \ottnt{A}  \ottsym{\mbox{$\mid$}}  \rightarrow  \ottsym{\mbox{$\mid$}}  \ottnt{B}  \ottsym{\mbox{$\mid$}} \\
  \ottsym{\mbox{$\mid$}}  \ottnt{A}  \, \& \,  \ottnt{B}  \ottsym{\mbox{$\mid$}}  & = & \ottsym{\mbox{$\mid$}}  \ottnt{A}  \ottsym{\mbox{$\mid$}}  \times  \ottsym{\mbox{$\mid$}}  \ottnt{B}  \ottsym{\mbox{$\mid$}} \\
\end{array}
\]
\end{definition}

\begin{definition}[Typing Context Translation]
\[
\begin{array}{@{\hspace{0mm}}l@{\hspace{2mm}}c@{\hspace{2mm}}l@{\hspace{0mm}}}
  \ottsym{\mbox{$\mid$}}   \bullet   \ottsym{\mbox{$\mid$}}      & = &  \bullet  \\
  \ottsym{\mbox{$\mid$}}  \Gamma  \ottsym{,}  \ottmv{x}  \ottsym{:}  \ottnt{A}  \ottsym{\mbox{$\mid$}}    & = & \ottsym{\mbox{$\mid$}}  \Gamma  \ottsym{\mbox{$\mid$}}  \ottsym{,}  \ottmv{x}  \ottsym{:}  \ottsym{\mbox{$\mid$}}  \ottnt{A}  \ottsym{\mbox{$\mid$}}
\end{array}
\]
\end{definition}

\begin{definition}\hfill
  \begin{setofrules}{\(\vdash  \Gamma\)}{Well formed typing context}
      \nrule{WFC-empty}{ }{ \vdash   \bullet  }\and
      \nrule{WFC-cons}{\vdash  \Gamma\and  \ottmv{x}  \notin  \Gamma }{\vdash  \Gamma  \ottsym{,}  \ottmv{x}  \ottsym{:}  \ottnt{A}}
  \end{setofrules}
\end{definition}

%% file: sources/algorithm-example.mng
\section{Complete Derivation Examples of Typing and Algorithmic Subtyping}
\label{appendix:algorithm-example}
\paragraph{Typing Derivation Example}
Below we show the typing derivation for the example expression \( 1   \ottsym{:}  \mathsf{Nat}  \, \& \,  \mathsf{Nat}\),
taken from Section~\ref{sec:calculus:disjointness}.
\begin{mathpar}
  \inferrule*[right=\rlabel{T-anno}]
             {
               \inferrule*[Right=\rlabel{T-sub}]
                          {
                            \inferrule*[right=\rlabel{T-nat}]
                                       {\vdash   \bullet }
                                       {  \bullet    \vdash    1     \Rightarrow    \mathsf{Nat}  \elbox{ \rightsquigarrow    1  } }
                            \and
                            \inferrule*[Right=\rlabel{S-and}]
                                       {
                                         \inferrule*[right=\rlabel{S-refl}]{ }{ \mathsf{Nat}   \ottsym{<:}   \mathsf{Nat}  \elbox{ \rightsquigarrow    \mathsf{id} _{  \mathsf{Nat}  }  } } \and
                                         \inferrule*[Right=\rlabel{S-refl}]{ }{ \mathsf{Nat}   \ottsym{<:}   \mathsf{Nat}  \elbox{ \rightsquigarrow    \mathsf{id} _{  \mathsf{Nat}  }  } }
                                       }
                                       { \mathsf{Nat}   \ottsym{<:}   \mathsf{Nat}  \, \& \,  \mathsf{Nat}  \elbox{ \rightsquigarrow   \langle   \mathsf{id} _{  \mathsf{Nat}  }   \ottsym{,}   \mathsf{id} _{  \mathsf{Nat}  }   \rangle } }
                          }
                          {  \bullet    \vdash    1     \Leftarrow    \mathsf{Nat}  \, \& \,  \mathsf{Nat}  \elbox{ \rightsquigarrow   \langle   \mathsf{id} _{  \mathsf{Nat}  }   \ottsym{,}   \mathsf{id} _{  \mathsf{Nat}  }   \rangle \,  1  } }
             }
             {  \bullet    \vdash    1   \ottsym{:}  \mathsf{Nat}  \, \& \,  \mathsf{Nat}    \Rightarrow    \mathsf{Nat}  \, \& \,  \mathsf{Nat}  \elbox{ \rightsquigarrow   \langle   \mathsf{id} _{  \mathsf{Nat}  }   \ottsym{,}   \mathsf{id} _{  \mathsf{Nat}  }   \rangle \,  1  } }
\end{mathpar}
The generated target term is \(\langle   \mathsf{id} _{  \mathsf{Nat}  }   \ottsym{,}   \mathsf{id} _{  \mathsf{Nat}  }   \rangle \,  1 \), which reduces to \(\langle   1   \ottsym{,}   1   \rangle\),
following the operational semantics of \target.

\paragraph{Algorithmic Subtyping Derivation Example}
Here we illustrate how the algorithm of Section~\ref{sec:algorithm:definition}
works with an example derivation. Due to space restrictions, we shorten
$ \mathsf{Nat} $, $ \mathsf{String} $, and $ \mathsf{Bool} $ to $ \mathsf{N} $, $ \mathsf{S} $, and $\ottnt{B}$,
respectively, and omit all mention of coercions.

\begin{mathpar}
  \inferrule*[vcenter,right=\rlabel{A-main}]
    { \inferrule*[vcenter,right=\rlabel{AR-arr}]
        { \inferrule*[vcenter,right=\rlabel{AR-nat}]
            { \inferrule*[vcenter,right=\rlabel{AL-arr}]
                { 
                  \inferrule*[vcenter,right=\rlabel{AR-top}]
                    { }
                    { \noEvAlgSubR{ [] }{ \mathsf{S} }{ \top } }
                  \\ D_2
                }
                { \noEvAlgSubL{( [] ,  \mathsf{S} )}{ [] }{\top  \rightarrow  \ottsym{(}   \mathsf{B}   \, \& \,  \ottsym{(}   \mathsf{B}   \rightarrow   \mathsf{N}   \ottsym{)}  \ottsym{)}}{\top  \rightarrow  \ottsym{(}   \mathsf{B}   \, \& \,  \ottsym{(}   \mathsf{B}   \rightarrow   \mathsf{N}   \ottsym{)}  \ottsym{)}}{ \mathsf{N} } }
            }
            { \noEvAlgSubR{( [] ,  \mathsf{S} )}{\top  \rightarrow  \ottsym{(}   \mathsf{B}   \, \& \,  \ottsym{(}   \mathsf{B}   \rightarrow   \mathsf{N}   \ottsym{)}  \ottsym{)}}{ \mathsf{N} } }
        }
        { \noEvAlgSubR{ [] }{\top  \rightarrow  \ottsym{(}   \mathsf{B}   \, \& \,  \ottsym{(}   \mathsf{B}   \rightarrow   \mathsf{N}   \ottsym{)}  \ottsym{)}}{ \mathsf{S}   \rightarrow   \mathsf{N} } }
    }
    { \noEvAlgSubMain{\top  \rightarrow  \ottsym{(}   \mathsf{B}   \, \& \,  \ottsym{(}   \mathsf{B}   \rightarrow   \mathsf{N}   \ottsym{)}  \ottsym{)}}{ \mathsf{S}   \rightarrow   \mathsf{N} } }
\end{mathpar}


\noindent where $D_2$ is the following derivation:
\begin{mathpar}
  \inferrule*[vcenter,right=\rlabel{AL-and2}]
    { \inferrule*[vcenter,right=\rlabel{AL-mp}]
        { D_3 \\
          \inferrule*[vcenter,right=\rlabel{AL-nat}]
            { }
            { \noEvAlgSubL{ [] }{( \mathsf{S} ,  [] )}{\top  \rightarrow  \ottsym{(}   \mathsf{B}   \, \& \,  \ottsym{(}   \mathsf{B}   \rightarrow   \mathsf{N}   \ottsym{)}  \ottsym{)}}{ \mathsf{N} }{ \mathsf{N} } }
        }
        { \noEvAlgSubL{ [] }{( \mathsf{S} ,  [] )}{\top  \rightarrow  \ottsym{(}   \mathsf{B}   \, \& \,  \ottsym{(}   \mathsf{B}   \rightarrow   \mathsf{N}   \ottsym{)}  \ottsym{)}}{ \mathsf{B}   \rightarrow   \mathsf{N} }{ \mathsf{N} } }
    }
    { \noEvAlgSubL{ [] }{( \mathsf{S} ,  [] )}{\top  \rightarrow  \ottsym{(}   \mathsf{B}   \, \& \,  \ottsym{(}   \mathsf{B}   \rightarrow   \mathsf{N}   \ottsym{)}  \ottsym{)}}{ \mathsf{B}   \, \& \,  \ottsym{(}   \mathsf{B}   \rightarrow   \mathsf{N}   \ottsym{)}}{ \mathsf{N} } }
\end{mathpar}

\noindent where $D_3$ is the following derivation:
\begin{mathpar}
  \inferrule*[vcenter,right=\rlabel{AR-arr}]
    { \inferrule*[vcenter,right=\rlabel{AR-bool}] 
        { \inferrule*[vcenter,right=\rlabel{AL-arr}]
            { 
              \inferrule*[vcenter,right=\rlabel{AR-top}]
                { }
                { \noEvAlgSubR{ [] }{ \mathsf{S} }{ \top } }
              \\ D_4
            }
            { \noEvAlgSubL{( [] ,  \mathsf{S} )}{ [] }{\top  \rightarrow  \ottsym{(}   \mathsf{B}   \, \& \,  \ottsym{(}   \mathsf{B}   \rightarrow   \mathsf{N}   \ottsym{)}  \ottsym{)}}{\top  \rightarrow  \ottsym{(}   \mathsf{B}   \, \& \,  \ottsym{(}   \mathsf{B}   \rightarrow   \mathsf{N}   \ottsym{)}  \ottsym{)}}{ \mathsf{B} } }
        }
        { \noEvAlgSubR{( [] ,  \mathsf{S} )}{\top  \rightarrow  \ottsym{(}   \mathsf{B}   \, \& \,  \ottsym{(}   \mathsf{B}   \rightarrow   \mathsf{N}   \ottsym{)}  \ottsym{)}}{ \mathsf{B} } }
    }
    { \noEvAlgSubR{ [] }{\top  \rightarrow  \ottsym{(}   \mathsf{B}   \, \& \,  \ottsym{(}   \mathsf{B}   \rightarrow   \mathsf{N}   \ottsym{)}  \ottsym{)}}{ \mathsf{S}   \rightarrow   \mathsf{B} } }
\end{mathpar}



\noindent where $D_4$ is the following derivation:
\begin{mathpar}
  \inferrule*[vcenter,right=\rlabel{AL-and1}]
    { \inferrule*[vcenter,right=\rlabel{AL-bool}]
        { }
        { \noEvAlgSubL{( [] ,  \mathsf{S} )}{ [] }{\top  \rightarrow  \ottsym{(}   \mathsf{B}   \, \& \,  \ottsym{(}   \mathsf{B}   \rightarrow   \mathsf{N}   \ottsym{)}  \ottsym{)}}{ \mathsf{B} }{ \mathsf{B} } }
    }
    { \noEvAlgSubL{( [] ,  \mathsf{S} )}{ [] }{\top  \rightarrow  \ottsym{(}   \mathsf{B}   \, \& \,  \ottsym{(}   \mathsf{B}   \rightarrow   \mathsf{N}   \ottsym{)}  \ottsym{)}}{ \mathsf{B}   \, \& \,  \ottsym{(}   \mathsf{B}   \rightarrow   \mathsf{N}   \ottsym{)}}{ \mathsf{B} } }
\end{mathpar}

%% file: sources/coherence_appendix.mng
\section{The Logical Relation for \calculus}\label{appendix:coherence}

\begin{figure}[t]
\fbox{
\begin{minipage}{.95\textwidth}
\begin{equation*}
\begin{array}{rcl}
\multicolumn{3}{l}{\boxed{ (  \ottnt{v_{{\mathrm{1}}}}  ,  \ottnt{v_{{\mathrm{2}}}}  ) \in \rellV ^{(  \ottnt{v'_{{\mathrm{1}}}}  :  \tau'_{{\mathrm{1}}}  ;  \ottnt{v'_{{\mathrm{2}}}}  :  \tau'_{{\mathrm{2}}}  )} \llbracket  \tau_{{\mathrm{1}}}  ;  \tau_{{\mathrm{2}}} \rrbracket }} \\[2mm]
 (  \ottnt{v_{{\mathrm{1}}}}  ,  \ottnt{v_{{\mathrm{2}}}}  ) \in \rellV ^{(  \ottnt{v'_{{\mathrm{1}}}}  :  \tau'_{{\mathrm{1}}}  ;  \ottnt{v'_{{\mathrm{2}}}}  :  \tau'_{{\mathrm{2}}}  )} \llbracket  \mathsf{Nat}  ;  \mathsf{Nat} \rrbracket  &\defeq &\exists i, \ottnt{v_{{\mathrm{1}}}} = \ottnt{v_{{\mathrm{2}}}} = i\\[2mm]
 (  \langle  \ottnt{v_{{\mathrm{1}}}}  \ottsym{,}  \ottnt{v_{{\mathrm{2}}}}  \rangle  ,  \ottnt{v_{{\mathrm{3}}}}  ) \in \rellV ^{(  \ottnt{v'_{{\mathrm{1}}}}  :  \tau'_{{\mathrm{1}}}  ;  \ottnt{v'_{{\mathrm{2}}}}  :  \tau'_{{\mathrm{2}}}  )} \llbracket  \tau_{{\mathrm{1}}}  \times  \tau_{{\mathrm{2}}}  ;  \tau_{{\mathrm{3}}} \rrbracket  & \defeq &
   (  \ottnt{v_{{\mathrm{1}}}}  ,  \ottnt{v_{{\mathrm{3}}}}  ) \in \rellV ^{(  \ottnt{v'_{{\mathrm{1}}}}  :  \tau'_{{\mathrm{1}}}  ;  \ottnt{v'_{{\mathrm{2}}}}  :  \tau'_{{\mathrm{2}}}  )} \llbracket  \tau_{{\mathrm{1}}}  ;  \tau_{{\mathrm{3}}} \rrbracket \\
  ~~&&\land~~ (  \ottnt{v_{{\mathrm{2}}}}  ,  \ottnt{v_{{\mathrm{3}}}}  ) \in \rellV ^{(  \ottnt{v'_{{\mathrm{1}}}}  :  \tau'_{{\mathrm{1}}}  ;  \ottnt{v'_{{\mathrm{2}}}}  :  \tau'_{{\mathrm{2}}}  )} \llbracket  \tau_{{\mathrm{2}}}  ;  \tau_{{\mathrm{3}}} \rrbracket  \\[2mm]
 (  \ottnt{v_{{\mathrm{3}}}}  ,  \langle  \ottnt{v_{{\mathrm{1}}}}  \ottsym{,}  \ottnt{v_{{\mathrm{2}}}}  \rangle  ) \in \rellV ^{(  \ottnt{v'_{{\mathrm{1}}}}  :  \tau'_{{\mathrm{1}}}  ;  \ottnt{v'_{{\mathrm{2}}}}  :  \tau'_{{\mathrm{2}}}  )} \llbracket  \tau_{{\mathrm{3}}}  ;  \tau_{{\mathrm{1}}}  \times  \tau_{{\mathrm{2}}} \rrbracket  & \defeq &
   (  \ottnt{v_{{\mathrm{3}}}}  ,  \ottnt{v_{{\mathrm{1}}}}  ) \in \rellV ^{(  \ottnt{v'_{{\mathrm{1}}}}  :  \tau'_{{\mathrm{1}}}  ;  \ottnt{v'_{{\mathrm{2}}}}  :  \tau'_{{\mathrm{2}}}  )} \llbracket  \tau_{{\mathrm{3}}}  ;  \tau_{{\mathrm{1}}} \rrbracket \\
  ~~&&\land~~ (  \ottnt{v_{{\mathrm{3}}}}  ,  \ottnt{v_{{\mathrm{2}}}}  ) \in \rellV ^{(  \ottnt{v'_{{\mathrm{1}}}}  :  \tau'_{{\mathrm{1}}}  ;  \ottnt{v'_{{\mathrm{2}}}}  :  \tau'_{{\mathrm{2}}}  )} \llbracket  \tau_{{\mathrm{3}}}  ;  \tau_{{\mathrm{2}}} \rrbracket  \\[2mm]
 (  \ottnt{v_{{\mathrm{1}}}}  ,  \ottnt{v_{{\mathrm{2}}}}  ) \in \rellV ^{(  \ottnt{v'_{{\mathrm{1}}}}  :  \tau'_{{\mathrm{1}}}  ;  \ottnt{v'_{{\mathrm{2}}}}  :  \tau'_{{\mathrm{2}}}  )} \llbracket  \tau_{{\mathrm{1}}}  \rightarrow  \tau_{{\mathrm{2}}}  ;  \tau \rrbracket  & \defeq &~\text{\hspace{18mm}(\(\tau\) not function type)} \\
\multicolumn{3}{r}{\forall \ottnt{c}, \ottnt{c}  \vdash  \tau'_{{\mathrm{1}}} \, \triangleright \, \tau_{{\mathrm{1}}}\Then (  \ottnt{v_{{\mathrm{1}}}} \, \ottsym{(}  \ottnt{c} \, \ottnt{v'_{{\mathrm{1}}}}  \ottsym{)}  ,  \ottnt{v_{{\mathrm{2}}}}  ) \in \rellE^{(  \ottnt{v'_{{\mathrm{1}}}}  :  \tau'_{{\mathrm{1}}}  ;  \ottnt{v'_{{\mathrm{2}}}}  :  \tau'_{{\mathrm{2}}}  )} \llbracket \tau_{{\mathrm{2}}}  ;  \tau \rrbracket }\\[2mm]
 (  \ottnt{v_{{\mathrm{1}}}}  ,  \ottnt{v_{{\mathrm{2}}}}  ) \in \rellV ^{(  \ottnt{v'_{{\mathrm{1}}}}  :  \tau'_{{\mathrm{1}}}  ;  \ottnt{v'_{{\mathrm{2}}}}  :  \tau'_{{\mathrm{2}}}  )} \llbracket  \tau  ;  \tau_{{\mathrm{1}}}  \rightarrow  \tau_{{\mathrm{2}}} \rrbracket  & \defeq & \text{\hspace{18mm}(\(\tau\) not function type)} \\
\multicolumn{3}{r}{\forall \ottnt{c}, \ottnt{c}  \vdash  \tau'_{{\mathrm{2}}} \, \triangleright \, \tau_{{\mathrm{1}}}\Then (  \ottnt{v_{{\mathrm{1}}}}  ,  \ottnt{v_{{\mathrm{2}}}} \, \ottsym{(}  \ottnt{c} \, \ottnt{v'_{{\mathrm{2}}}}  \ottsym{)}  ) \in \rellE^{(  \ottnt{v'_{{\mathrm{1}}}}  :  \tau'_{{\mathrm{1}}}  ;  \ottnt{v'_{{\mathrm{2}}}}  :  \tau'_{{\mathrm{2}}}  )} \llbracket \tau  ;  \tau_{{\mathrm{2}}} \rrbracket }\\[2mm]
 (  \ottnt{v_{{\mathrm{1}}}}  ,  \ottnt{v_{{\mathrm{2}}}}  ) \in \rellV ^{(  \ottnt{v'_{{\mathrm{1}}}}  :  \tau'_{{\mathrm{1}}}  ;  \ottnt{v'_{{\mathrm{2}}}}  :  \tau'_{{\mathrm{2}}}  )} \llbracket  \tau_{{\mathrm{1}}}  \rightarrow  \tau_{{\mathrm{2}}}  ;  \tau_{{\mathrm{3}}}  \rightarrow  \tau_{{\mathrm{4}}} \rrbracket  & \defeq & \\
\multicolumn{3}{r}{\begin{array}{c@{\quad}l}
       & \left(
  \begin{array}{l}
    \begin{aligned}
      &\forall \ottnt{v}~\ottnt{v'}~\tau_{{\mathrm{0}}}~\tau'_{{\mathrm{0}}}~\ottnt{c}~\ottnt{c'}~\ottnt{d}~\ottnt{d'}~\tau_{{\mathrm{01}}}~\tau_{{\mathrm{02}}},\\
      &\ottnt{d}  \vdash  \tau_{{\mathrm{01}}} \, \triangleright \, \tau_{{\mathrm{1}}}\Then \ottnt{d'}  \vdash  \tau_{{\mathrm{02}}} \, \triangleright \, \tau_{{\mathrm{3}}}\Then (  \ottnt{d} \, \ottnt{v}  ,  \ottnt{d'} \, \ottnt{v'}  ) \in \rellE^{(  \ottnt{v}  :  \tau_{{\mathrm{01}}}  ;  \ottnt{v'}  :  \tau_{{\mathrm{02}}}  )} \llbracket \tau_{{\mathrm{1}}}  ;  \tau_{{\mathrm{3}}} \rrbracket \Then\\
      &\ottnt{c}  \vdash  \tau'_{{\mathrm{1}}} \, \triangleright \, \tau_{{\mathrm{01}}}  \rightarrow  \tau_{{\mathrm{0}}}\Then \ottnt{c'}  \vdash  \tau'_{{\mathrm{2}}} \, \triangleright \, \tau_{{\mathrm{02}}}  \rightarrow  \tau'_{{\mathrm{0}}}\Then\\
      & (  \ottnt{v_{{\mathrm{1}}}} \, \ottsym{(}  \ottnt{d} \, \ottnt{v}  \ottsym{)}  ,  \ottnt{v_{{\mathrm{2}}}} \, \ottsym{(}  \ottnt{d'} \, \ottnt{v'}  \ottsym{)}  ) \in \rellE^{(  \langle  \ottnt{v_{{\mathrm{1}}}} \, \ottsym{(}  \ottnt{d} \, \ottnt{v}  \ottsym{)}  \ottsym{,}  \ottnt{c} \, \ottnt{v'_{{\mathrm{1}}}} \, \ottnt{v}  \rangle  :  \tau_{{\mathrm{2}}}  \times  \tau_{{\mathrm{0}}}  ;  \langle  \ottnt{v_{{\mathrm{2}}}} \, \ottsym{(}  \ottnt{d'} \, \ottnt{v'}  \ottsym{)}  \ottsym{,}  \ottnt{c'} \, \ottnt{v'_{{\mathrm{2}}}} \, \ottnt{v'}  \rangle  :  \tau_{{\mathrm{4}}}  \times  \tau'_{{\mathrm{0}}}  )} \llbracket \tau_{{\mathrm{2}}}  ;  \tau_{{\mathrm{4}}} \rrbracket 
    \end{aligned}
  \end{array}\right) \\
\wedge & \forall \ottnt{c}, \ottnt{c}  \vdash  \tau'_{{\mathrm{1}}} \, \triangleright \, \tau_{{\mathrm{1}}}\Then  (  \ottnt{v_{{\mathrm{1}}}} \, \ottsym{(}  \ottnt{c} \, \ottnt{v'_{{\mathrm{1}}}}  \ottsym{)}  ,  \ottnt{v_{{\mathrm{2}}}}  ) \in \rellE^{(  \ottnt{v'_{{\mathrm{1}}}}  :  \tau'_{{\mathrm{1}}}  ;  \ottnt{v'_{{\mathrm{2}}}}  :  \tau'_{{\mathrm{2}}}  )} \llbracket \tau_{{\mathrm{2}}}  ;  \tau_{{\mathrm{3}}}  \rightarrow  \tau_{{\mathrm{4}}} \rrbracket  \\
\wedge &  \forall \ottnt{c}, \ottnt{c}  \vdash  \tau'_{{\mathrm{2}}} \, \triangleright \, \tau_{{\mathrm{3}}}\Then  (  \ottnt{v_{{\mathrm{1}}}}  ,  \ottnt{v_{{\mathrm{2}}}} \, \ottsym{(}  \ottnt{c} \, \ottnt{v'_{{\mathrm{2}}}}  \ottsym{)}  ) \in \rellE^{(  \ottnt{v'_{{\mathrm{1}}}}  :  \tau'_{{\mathrm{1}}}  ;  \ottnt{v'_{{\mathrm{2}}}}  :  \tau'_{{\mathrm{2}}}  )} \llbracket \tau_{{\mathrm{1}}}  \rightarrow  \tau_{{\mathrm{2}}}  ;  \tau_{{\mathrm{4}}} \rrbracket  \\[2mm]
\end{array}}
\\[2mm]
 (  \ottnt{v_{{\mathrm{1}}}}  ,  \ottnt{v_{{\mathrm{2}}}}  ) \in \rellV ^{(  \ottnt{v'_{{\mathrm{1}}}}  :  \tau'_{{\mathrm{1}}}  ;  \ottnt{v'_{{\mathrm{2}}}}  :  \tau'_{{\mathrm{2}}}  )} \llbracket  \tau_{{\mathrm{1}}}  ;  \tau_{{\mathrm{2}}} \rrbracket  &\defeq&\mathmakebox[0pt]{~~~\mathsf{true}}\text{\hspace{18mm}(otherwise)} \\ \\
\multicolumn{3}{l}{\boxed{ (  \ottnt{e_{{\mathrm{1}}}}  ,  \ottnt{e_{{\mathrm{2}}}}  ) \in \rellE^{(  \ottnt{e'_{{\mathrm{1}}}}  :  \tau'_{{\mathrm{1}}}  ;  \ottnt{e'_{{\mathrm{2}}}}  :  \tau'_{{\mathrm{2}}}  )} \llbracket \tau_{{\mathrm{1}}}  ;  \tau_{{\mathrm{2}}} \rrbracket }} \\[2mm]
 (  \ottnt{e_{{\mathrm{1}}}}  ,  \ottnt{e_{{\mathrm{2}}}}  ) \in \rellE^{(  \ottnt{e'_{{\mathrm{1}}}}  :  \tau'_{{\mathrm{1}}}  ;  \ottnt{e'_{{\mathrm{2}}}}  :  \tau'_{{\mathrm{2}}}  )} \llbracket \tau_{{\mathrm{1}}}  ;  \tau_{{\mathrm{2}}} \rrbracket  & \defeq & \\
\multicolumn{3}{r}{\exists \ottnt{v_{{\mathrm{1}}}}\,\ottnt{v_{{\mathrm{2}}}}\,\ottnt{v'_{{\mathrm{1}}}}\,\ottnt{v'_{{\mathrm{2}}}},
  \begin{array}[t]{r}\ottnt{e_{{\mathrm{1}}}}  \longrightarrow^{*}  \ottnt{v_{{\mathrm{1}}}}~\land~\ottnt{e_{{\mathrm{2}}}}  \longrightarrow^{*}  \ottnt{v_{{\mathrm{2}}}} \land~\ottnt{e'_{{\mathrm{1}}}}  \longrightarrow^{*}  \ottnt{v'_{{\mathrm{1}}}}~\land~\ottnt{e'_{{\mathrm{2}}}}  \longrightarrow^{*}  \ottnt{v'_{{\mathrm{2}}}}  \\
   \wedge \:  (  \ottnt{v_{{\mathrm{1}}}}  ,  \ottnt{v_{{\mathrm{2}}}}  ) \in \rellV ^{(  \ottnt{v'_{{\mathrm{1}}}}  :  \tau'_{{\mathrm{1}}}  ;  \ottnt{v'_{{\mathrm{2}}}}  :  \tau'_{{\mathrm{2}}}  )} \llbracket  \tau_{{\mathrm{1}}}  ;  \tau_{{\mathrm{2}}} \rrbracket 
   \end{array}}
\end{array}
\end{equation*}
\end{minipage}
}
\caption{Logical Relations}\label{fig:logrel}
\end{figure}

Figure~\ref{fig:logrel} shows the \calculus definitions of the logical relations
$\mathcal{V}$ for values and $\mathcal{E}$ for terms. These definitions
reconcile the context-dependent nature of modus ponens with the structural
recursion needed for the well-foundedness of the definitions.

The intent of the relations is of course the same as in \necolus, to only allow
semantically indistinguishable overlap. Also the three general principles at
the basis of their definition are preserved: 1) two values of type $ \mathsf{Nat} $
have indistinguishable overlap only when they are identical, 2) related
functions should take related inputs to related outputs, and 3) the coercion
compatibility property should
hold.

The first novelty to deal with modus ponens is that the family of value
relations is equipped with additional indices $\ottnt{v'_{{\mathrm{1}}}}:\tau'_{{\mathrm{1}}}$ and
$\ottnt{v'_{{\mathrm{2}}}}:\tau'_{{\mathrm{2}}}$ to keep track of the values that $\ottnt{v_{{\mathrm{1}}}}$ and $\ottnt{v_{{\mathrm{2}}}}$ are
components of; we call these indices the respective \emph{contexts} of
$\ottnt{v_{{\mathrm{1}}}}$ and $\ottnt{v_{{\mathrm{2}}}}$. To be precise, the intention is that
we use $ (  \ottnt{v_{{\mathrm{1}}}}  ,  \ottnt{v_{{\mathrm{2}}}}  ) \in \rellV ^{(  \ottnt{v'_{{\mathrm{1}}}}  :  \tau'_{{\mathrm{1}}}  ;  \ottnt{v'_{{\mathrm{2}}}}  :  \tau'_{{\mathrm{2}}}  )} \llbracket  \tau_{{\mathrm{1}}}  ;  \tau_{{\mathrm{2}}} \rrbracket $ only when there exist
coercions $\ottnt{c_{{\mathrm{1}}}}$ and $\ottnt{c_{{\mathrm{2}}}}$ such that firstly $\ottnt{c_{{\mathrm{1}}}}  \vdash  \tau'_{{\mathrm{1}}} \, \triangleright \, \tau_{{\mathrm{1}}}$ and
$\ottnt{c_{{\mathrm{2}}}}  \vdash  \tau'_{{\mathrm{2}}} \, \triangleright \, \tau_{{\mathrm{2}}}$, and secondly $\ottnt{c_{{\mathrm{1}}}} \, \ottnt{v'_{{\mathrm{1}}}}  \longrightarrow^{*}  \ottnt{v_{{\mathrm{1}}}}$ and $\ottnt{c_{{\mathrm{2}}}} \, \ottnt{v'_{{\mathrm{2}}}}  \longrightarrow^{*}  \ottnt{v_{{\mathrm{2}}}}$. When the contexts and the values are the same, we sometimes omit
those additional indices, i.e., we abbreviate $ (  \ottnt{v_{{\mathrm{1}}}}  ,  \ottnt{v_{{\mathrm{2}}}}  ) \in \rellV ^{(  \ottnt{v_{{\mathrm{1}}}}  :  \tau_{{\mathrm{1}}}  ;  \ottnt{v_{{\mathrm{2}}}}  :  \tau_{{\mathrm{2}}}  )} \llbracket  \tau_{{\mathrm{1}}}  ;  \tau_{{\mathrm{2}}} \rrbracket $ to $ (  \ottnt{v_{{\mathrm{1}}}}  ,  \ottnt{v_{{\mathrm{2}}}}  ) \in \rellV \llbracket  \tau_{{\mathrm{1}}}  ;  \tau_{{\mathrm{2}}} \rrbracket $. The term relation $\mathcal{E}$ is
equipped with similar context indices; the one difference is that, just like
the elements in the term relation, these indices are terms rather than values.


\paragraph{Unmodified Cases}
The first three cases of the $\mathcal{V}$ definition are essentially the same as
those of \necolus: Two natural numbers are related if they are identical, which
is a direct implementation of the first general principle. Also, a pair is
related to another value if both components of the pair are related to that
value, which is a similar specialisation of the coercion compatibility property
to coercions $\pi_1$ and $\pi_2$.

\paragraph{New Modus Ponens Cases}

The next two cases that relate functions to non-functions are different from
\necolus because of the addition of the modus ponens coercion.  
In \necolus all functions are related to all non-functions because the
only common types they can be coerced to are top-like types whose values are
indistinguishable.

This is not so
in \target, where the \rlabel{S-mp} rule can be used to derive a
non-function value from a function. Hence, our logical relation states that a
function and non-function are related only if all \rlabel{S-mp}-enabled
applications of the function are related to the non-function value. Using modus ponens
implies that not only the function, but also its argument is derived from the original
value. This is why the two symmetric cases of the relation universally quantify over
all coercions that derive an argument of the appropriate input type.

\paragraph{New Function Case}

The one-but-last case, which relates two functions, is the most complicated
one. It states that two functions are related iff they satisfy three conditions.
These three conditions concern the two different ways in which the functions
can be applied to an argument.

The second and third condition are a new way compared to \necolus and analogous
to function/non-function cases; these are the cases where one of the two
functions is applied to an argument through \rlabel{S-mp}. They expose that
$\langle   \lambda \ottmv{x} : \mathsf{Nat} .\,  \lambda \ottmv{y} :  \mathsf{Bool}  .\, \ottmv{y}    \ottsym{,}   1   \rangle$ can be coerced to $ \lambda \ottmv{y} :  \mathsf{Bool}  .\, \ottmv{y} $
and thus has distinguishable overlap with, for instance, $ \lambda \ottmv{y} :  \mathsf{Bool}  .\,  \mathsf{true}  $.

The first condition corresponds to the \necolus condition for function types
and captures the main design principle that related functions take related
inputs to related outputs. However, as a
consequence of modus ponens subtyping, the formulation is much more elaborate.
First, we need to consider the context indices.
An explicit function application demarcates the point where the implicit subtyping
derivation ends and, therefore, the previous context is abandoned. One might
thus expect the following much simpler formulation of the condition, where
the result of the function application becomes its own new context:
\begin{equation*}
  \forall 
     (  \ottnt{v}  ,  \ottnt{v'}  ) \in \rellV ^{(  \ottnt{v}  :  \tau_{{\mathrm{1}}}  ;  \ottnt{v'}  :  \tau_{{\mathrm{3}}}  )} \llbracket  \tau_{{\mathrm{1}}}  ;  \tau_{{\mathrm{3}}} \rrbracket ,~
     (  \ottnt{v_{{\mathrm{1}}}} \, \ottnt{v}  ,  \ottnt{v_{{\mathrm{2}}}} \, \ottnt{v'}  ) \in \rellE^{(  \ottnt{v_{{\mathrm{1}}}} \, \ottnt{v}  :  \tau_{{\mathrm{2}}}  ;  \ottnt{v_{{\mathrm{2}}}} \, \ottnt{v'}  :  \tau_{{\mathrm{4}}}  )} \llbracket \tau_{{\mathrm{2}}}  ;  \tau_{{\mathrm{4}}} \rrbracket 
\end{equation*}
Sadly, this formulation does not properly account
for functions that -- unlike $\ottnt{v_{{\mathrm{1}}}}$ and $\ottnt{v_{{\mathrm{2}}}}$ -- can be derived
from the context without them being structural components. Consider the
composition of multiple subcomponents by means of the \rlabel{S-distArr}
rule. For instance, that rule can derive from the context\(\langle   \lambda \ottmv{x} :  \mathsf{Bool}  .\,  1    \ottsym{,}   \lambda \ottmv{x} :  \mathsf{Bool}  .\,  \mathsf{true}    \rangle\) the function \( \lambda \ottmv{x} :  \mathsf{Bool}  .\, \langle   1   \ottsym{,}   \mathsf{true}   \rangle \) which is not a syntactic component of the initial type. In
\necolus, we do not have
to separately account for these cases, because the components of the function
body never
interact and can thus be considered separately. For that reason, the logical
relation of \necolus does not explicitly have to consider \rlabel{S-distArr}.
However, in \calculus matters are different because of \rlabel{S-mp}.
When the context is $\langle   \lambda \ottmv{x} :  \mathsf{Bool}  .\,  \lambda \ottmv{y} :  \mathsf{Bool}  .\,  1     \ottsym{,}   \lambda \ottmv{x} :  \mathsf{Bool}  .\,  \mathsf{true}    \rangle$, then using \rlabel{S-distArr} we can derive the function $ \lambda \ottmv{x} :  \mathsf{Bool}  .\, \langle   \lambda \ottmv{y} :  \mathsf{Bool}  .\,  1    \ottsym{,}   \mathsf{true}   \rangle $. When this function is explicitly applied
to a value, we get a new original value $\langle   \lambda \ottmv{y} :  \mathsf{Bool}  .\,  1    \ottsym{,}   \mathsf{true}   \rangle$ whose
two components can interact through \rlabel{S-mp}. Thus, just checking \( \lambda \ottmv{x} :  \mathsf{Bool}  .\,  \lambda \ottmv{y} :  \mathsf{Bool}  .\,  1   \) on its own is insufficient. 
To reckon with all ways in which the functions $\ottnt{v_{{\mathrm{1}}}}$ and $\ottnt{v_{{\mathrm{2}}}}$ under
consideration can be combined by means of \rlabel{S-distarr} with other functions
derived from the context, we augment the condition as shown in the following
extended condition:
\begin{align*}
  &\forall (  \ottnt{v}  ,  \ottnt{v'}  ) \in \rellE^{(  \ottnt{v}  :  \tau_{{\mathrm{1}}}  ;  \ottnt{v'}  :  \tau_{{\mathrm{3}}}  )} \llbracket \tau_{{\mathrm{1}}}  ;  \tau_{{\mathrm{3}}} \rrbracket ,\,\forall\ottnt{c}  \vdash  \tau'_{{\mathrm{1}}} \, \triangleright \, \tau_{{\mathrm{1}}}  \rightarrow  \tau_{{\mathrm{0}}},\,\forall\ottnt{c'}  \vdash  \tau'_{{\mathrm{2}}} \, \triangleright \, \tau_{{\mathrm{3}}}  \rightarrow  \tau'_{{\mathrm{0}}},\\
  &~~~ (  \langle  \ottnt{v_{{\mathrm{1}}}} \, \ottnt{v}  \ottsym{,}  \ottnt{c} \, \ottnt{v'_{{\mathrm{1}}}} \, \ottnt{v}  \rangle  ,  \langle  \ottnt{v_{{\mathrm{2}}}} \, \ottnt{v'}  \ottsym{,}  \ottnt{c'} \, \ottnt{v'_{{\mathrm{2}}}} \, \ottnt{v'}  \rangle  ) \in \rellE^{(  \langle  \ottnt{v_{{\mathrm{1}}}} \, \ottnt{v}  \ottsym{,}  \ottnt{c} \, \ottnt{v'_{{\mathrm{1}}}} \, \ottnt{v}  \rangle  :  \tau_{{\mathrm{2}}}  \times  \tau_{{\mathrm{0}}}  ;  \langle  \ottnt{v_{{\mathrm{2}}}} \, \ottnt{v'}  \ottsym{,}  \ottnt{c'} \, \ottnt{v'_{{\mathrm{2}}}} \, \ottnt{v'}  \rangle  :  \tau_{{\mathrm{4}}}  \times  \tau'_{{\mathrm{0}}}  )} \llbracket \tau_{{\mathrm{2}}}  \times  \tau_{{\mathrm{0}}}  ;  \tau_{{\mathrm{4}}}  \times  \tau'_{{\mathrm{0}}} \rrbracket 
\end{align*}
Yet, this formulation breaks the well-foundedness of the recursive definition
because $\tau_{{\mathrm{2}}}  \times  \tau_{{\mathrm{0}}}$ and $\tau_{{\mathrm{4}}}  \times  \tau'_{{\mathrm{0}}}$ are not necessarily smaller than $\tau_{{\mathrm{1}}}  \rightarrow  \tau_{{\mathrm{2}}}$ and
$\tau_{{\mathrm{3}}}  \rightarrow  \tau_{{\mathrm{4}}}$, respectively.
Fortunately, we can mention the additional components only in the
context index to make them available for interaction through \rlabel{S-mp}.
This restores the well-foundedness of the definition.
\begin{align*}
  &\forall (  \ottnt{v}  ,  \ottnt{v'}  ) \in \rellE^{(  \ottnt{v}  :  \tau_{{\mathrm{1}}}  ;  \ottnt{v'}  :  \tau_{{\mathrm{3}}}  )} \llbracket \tau_{{\mathrm{1}}}  ;  \tau_{{\mathrm{3}}} \rrbracket ,\,\forall\ottnt{c}  \vdash  \tau'_{{\mathrm{1}}} \, \triangleright \, \tau_{{\mathrm{1}}}  \rightarrow  \tau_{{\mathrm{0}}},\,\forall\ottnt{c'}  \vdash  \tau'_{{\mathrm{2}}} \, \triangleright \, \tau_{{\mathrm{3}}}  \rightarrow  \tau'_{{\mathrm{0}}},\\
  &~~~ (  \ottnt{v_{{\mathrm{1}}}} \, \ottnt{v}  ,  \ottnt{v_{{\mathrm{2}}}} \, \ottnt{v'}  ) \in \rellE^{(  \langle  \ottnt{v_{{\mathrm{1}}}} \, \ottnt{v}  \ottsym{,}  \ottnt{c} \, \ottnt{v'_{{\mathrm{1}}}} \, \ottnt{v}  \rangle  :  \tau_{{\mathrm{2}}}  \times  \tau_{{\mathrm{0}}}  ;  \langle  \ottnt{v_{{\mathrm{2}}}} \, \ottnt{v'}  \ottsym{,}  \ottnt{c'} \, \ottnt{v'_{{\mathrm{2}}}} \, \ottnt{v'}  \rangle  :  \tau_{{\mathrm{4}}}  \times  \tau'_{{\mathrm{0}}}  )} \llbracket \tau_{{\mathrm{2}}}  ;  \tau_{{\mathrm{4}}} \rrbracket 
\end{align*}
There is one last twist. Rule \rlabel{S-distArr} can still be used on two
function components with different domains. In particular, we can make the
domains agree by
means of the contra-variant part of the \rlabel{S-arr} rule. For instance, we
can derive $ \lambda \ottmv{x} : \mathsf{Nat}  \times   \mathsf{Bool}  .\, \langle   \lambda \ottmv{y} :  \mathsf{Bool}  .\,  1    \ottsym{,}   \mathsf{true}   \rangle $ from the
original value $\langle   \lambda \ottmv{x} :  \mathsf{Bool}  .\,  \lambda \ottmv{y} :  \mathsf{Bool}  .\,  1     \ottsym{,}   \lambda \ottmv{x} : \mathsf{Nat}  \times   \mathsf{Bool}  .\,  \mathsf{true}    \rangle$
by first strengthening the domain of the first component before distributing
the product over the functions. This explains the remaining complication
in the actual condition of Figure~\ref{fig:logrel}.

\paragraph{Top Cases}
Finally, the last case summarizes all combinations of $ \top $ with itself and
with $ \mathsf{Nat} $, where all values are trivially related.

\paragraph{Terms}
Figure~\ref{fig:logrel} also contains the definition of the logical relation
for terms, whose form is fairly standard for strongly normalising languages.
The one extension is that we also normalise the original-term indices and
not just the terms in the relation. 

%% file: sources/termination-proof.mng
\section{Termination of Algorithmic Subtyping}
\label{appendix:termination-proof}

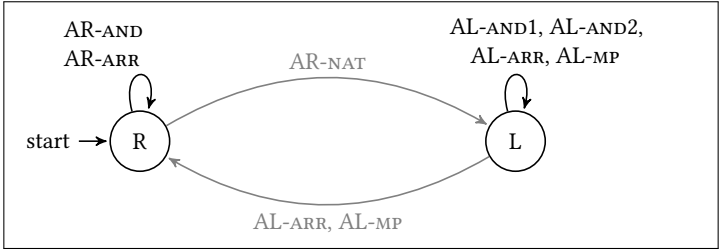
\begin{figure}
  \centering
  \small
  \fbox{
    \begin{tikzpicture}[->,>=stealth',shorten >=1pt,auto,semithick]
      
      \node[initial,state] (R)         {R};
      \node[state]         (L) [right of=R, node distance=5cm] {L};
      
      \path (R) edge [loop above] node {\begin{minipage}{2cm}
          \facerule{AR-and}\\\facerule{AR-arr}
      \end{minipage}} (R)
      edge [bend left, color=gray]  node {\facerule{AR-nat}} (L)
      (L) edge [loop above] node [xshift=0.4cm] {\begin{minipage}{4cm}
          \centering
          \facerule{AL-and1}, \facerule{AL-and2},\\
          \facerule{AL-arr}, \facerule{AL-mp}
      \end{minipage}} (L)
      (L) edge [bend left, color=gray] node {\facerule{AL-arr}, \facerule{AL-mp}} (R);
    \end{tikzpicture}
  }
  \caption{Dependency graph of algorithmic subtyping}
  \label{fig:dep-graphA}
\end{figure}
In this section, we prove that with the use of a loop detection mechanism, the subtyping algorithm presented in
Figure~\ref{fig:algorithmic-subtyping} of Section~\ref{sec:algorithm} terminates for any input types, $\ottnt{A}$ and $\ottnt{B}$.

The proof is partly mechanized in Coq
and partly written in \LaTeX.
The mechanized part is presented here in the form of lemmas, while the written part is contained in the text and the remarks
of this section.
A table at the end of this section links the lemmas presented here with their mechanized counterparts.

\subsection{The proof}

The subtyping algorithm uses a left and a right focusing mode.
Figure~\ref{fig:dep-graphA} shows the dependency graph of the two modes.
(L) and (R) stand for left- and right-focusing, respectively, and the gray edges of the graph are those that switch between
modes.
The algorithm always starts with right-focusing, which decomposes $\ottnt{B}$ in $\noEvAlgSubR{\mathcal{L}}{\ottnt{A}}{\ottnt{B}}$, and
eventually enters the left-focusing phase which works by decomposing $\ottnt{A}$.
In left focusing mode, rules \facerule{AL-arr} and \facerule{AL-mp} recursively continue in this mode, while at the same
time triggering back the right-focusing mode.

In an effort to detect potential diverging behavior, we can start by examining a variation of the algorithm where the
interdependencies of the focusing modes are removed, as shown in the black-colored part of Figure~\ref{fig:dep-graphA},
thus isolating the two modes of the algorithm%
\footnote{%
  In this variation, the right-focusing premises in the left-focusing rules \facerule{AL-mp} and \facerule{AL-arr},
  and the left-focusing premise of the right-focusing rule \facerule{AR-nat} are stripped away.
  This makes the algorithm incomplete w.r.t. the declarative version, but this is not a
  concern here.}%
.
Since both modes follow the structure of their corresponding focal type, they are both terminating when considered in
isolation.
Adding back the left-focusing dependency in rule \facerule{AR-nat} still retains termination, because left-focusing (which
still ignores the right-focusing calls) terminates.
To conclude, the only diverging loops that could possibly occur involve mutual recursion between both focusing modes,
through the rules \facerule{AL-arr} and \facerule{AL-mp}.

\begin{figure}
  \pgfdeclarelayer{fg}    
  \pgfsetlayers{main,fg}  
  \centering
  \begin{tikzpicture}
    \tikzset{
      position/.style args={#1:#2 from #3}{
        at=(#3.#1), anchor=#1+180, shift=(#1:#2)
      }
    }
    \tikzset{
      rfoc/.style={draw=RubineRed, dotted, dash pattern=on 0pt off 1.8\pgflinewidth, line cap=round, line width=2pt},
      lfoc/.style={draw=RoyalPurple, dashed, line cap=round, line width=1.8pt, dash pattern=on 3pt off 4pt},
      judg/.style={rectangle,inner sep=3pt},
      arrloop/.style={rectangle, draw=ForestGreen, outer sep=2pt},
      mploop/.style={rectangle, draw=BurntOrange, outer sep=2pt},
      choice/.style={circle, inner sep=1pt, draw, line width=1pt, outer sep=2pt},
      phantom/.style={draw=none, inner sep=0pt, outer sep=0pt}
    }
    \scriptsize
    \node[judg]    (R)                                       {\(\noEvAlgSubR{[]}{\ottnt{A}}{\ottnt{B}}\)};
    \node[judg]    (AR-and)   [node distance=7mm,above=of R] {\sc{AR-and}};
    \node[judg]    (AR-nat1)  [position=150:8mm from AR-and] {\sc{AR-nat}};
    \node[mploop]  (AL-mp0)   [position=90:6mm from AR-nat1] {\sc{AL-mp}};
    \node[judg]    (AR-nat2)  [position=20:8mm from AR-and] {\sc{AR-nat}};
    \node[choice]  (p2)       [node distance=1.1cm, above of=AR-nat2] {};
    \node[arrloop] (AL-arr)   [position=135:9mm from p2] {\sc{AL-arr}};
    \node[mploop]  (AL-mp)    [position=32:9mm from p2] {\sc{AL-mp}};
    \node[arrloop] (AL-arr2)  [position=80:8mm from AL-mp] {\sc{AL-arr}};
    \node[phantom] (PR-mp0)   [position=90:8mm from AL-mp0] {};
    \node[phantom] (AR-nat0)   [position=150:8mm from AL-mp0] {};
    \node[phantom] (PR-arr)   [position=80:8mm from AL-arr] {};
    \node[phantom] (PL-arr)   [position=150:8mm from AL-arr] {};
    \node[phantom] (PL-mp)    [position=150:8mm from AL-mp] {};
    \node[phantom] (PR-arr2)   [position=80:8mm from AL-arr2] {};
    \node[phantom] (PL-arr2)   [position=150:8mm from AL-arr2] {};
    \node[judg]    (AL-arrNo) [position=90:5mm from PL-arr] {\sc{AL-arr}};
    
    \draw[-, rfoc] (R) -- (AR-and) -- (AR-nat1);
    \draw[-, lfoc] (AR-nat1) -- (AL-mp0);
    \draw[-, rfoc,path fading=west] ([xshift=-3mm]AL-mp0) -- (AR-nat0);
    \draw[-, lfoc] (PL-arr) -- (AL-arrNo);
    \draw[-, lfoc, path fading=north] (AL-mp0) -- (PR-mp0);
    \draw[-, rfoc] (AR-and) -- (AR-nat2);
    \draw[-, lfoc] (AR-nat2) -- (p2);
    \draw[-, lfoc] (p2) -- (AL-arr.south);
    \draw[-, rfoc] ([xshift=-3mm]AL-arr) -- (PL-arr);
    \draw[-, lfoc, path fading=east] (AL-arr) -- (PR-arr);
    \draw[-, lfoc] (p2) -- ([xshift=-1mm]AL-mp.south);
    \draw[-, rfoc, path fading=west] ([xshift=-3mm]AL-mp) -- (PL-mp);
    \draw[-, lfoc] (AL-mp) -- (AL-arr2);
    \draw[-, rfoc, path fading=west] ([xshift=-3mm]AL-arr2) -- (PL-arr2);
    \draw[-, lfoc, path fading=east] (AL-arr2) -- (PR-arr2);

    \begin{pgfonlayer}{fg}[every label/.style={align=left}]
      \matrix [draw,below left, anchor=base, xshift=20mm] at (current bounding box.south east) {
        \node[text width=5mm,label=right:{{A sequence of right-focusing rules}}] {\tikz{\draw[-,rfoc] (0,0) -- (4.5mm,0);}};\\
        \node[text width=5mm,label=right:{{A sequence of left-focusing rules}}] {\tikz{\draw[-,lfoc] (0,0) -- (4.25mm,0);}};\\
        \node [arrloop, text width=3mm, label=right:{Arrow-loop endpoint}] {}; \\
        \node [mploop, text width=3mm, label=right:{{Mp-loop endpoint}}] {}; \\
        \node [text width=4mm,label=right:{Backtracking point}, align=center] {\tikz{\node[choice] {};}};\\
      };
    \end{pgfonlayer}
  \end{tikzpicture}
  \caption{Visualization of the $\arrloopsymbol$ relation}
  \label{fig:ArrLoopViz}
\end{figure}
Figure~\ref{fig:ArrLoopViz} offers a visual aid to closely examine the shape of these loops.
It shows part of an arbitrary computation tree of the algorithmic subtyping process, including backtracking branches.
The algorithm starts by applying a number of right-focusing rules (represented by a
\textcolor{RubineRed}{dotted pink path})
until it reaches an \facerule{AR-nat} rule application.
Now it continues in left-focusing mode (represented by a \textcolor{RoyalPurple}{dashed purple path}),
sometimes igniting a new right-focusing branch through rules \facerule{AL-arr} and \facerule{AL-mp},
as shown in the rightmost left-focusing path of Figure~\ref{fig:ArrLoopViz}.
In the computation tree, a path that has an initial part in right-focusing mode, its remainder in left-focusing mode
and ends on a \facerule{AL-arr} or \facerule{AL-mp} application, represents what we call a
\emph{single loop} of mutual recursion between the two focusing modes.
Informally, it is a computation path of the form
\(\textcolor{RubineRed}{\turnstyleR \, \to \cdots \to \, \turnstyleR \, \to} \,%
\textcolor{RoyalPurple}{\turnstyleL \, \to \cdots \to \, \turnstyleL \, \to} \, \textcolor{RubineRed}{\turnstyleR}\),
where the last node is the right-focusing premise of an \facerule{AL-arr} or an \facerule{AL-mp} rule.
If it ends with an \facerule{AL-arr} rule application, we call it an \emph{arrow-loop}, and otherwise an \emph{mp-loop}.
For instance, in Figure~\ref{fig:ArrLoopViz} all the paths that start at the root and end on an
outlined \facerule{AL-arr} or \facerule{AL-mp} rule are single loops.

It is now clearer that every right-focusing judgment to be resolved, signifies the start of a number (possibly zero)
of sequences (possibly empty) of single loops.
Therefore, a single loop, itself ending in a right-focusing judgment to be resolved, can be considered as the first
loop in a number of sequences of single loops of the same kind (arrow- or mp-loops).
Our termination proof examines the finiteness of the number of single loop sequences that the algorithm might go
though, as well as the finiteness of the size of each of those sequences.

We will show that a) arrow-loops are not a source of non-termination, and b) a loop-detection mechanism that tracks
previously tried mp-loops makes the algorithm terminating.
For that, we define two relations, one for arrow-loops and one for mp-loops,
each capturing the ways in which left-focusing
triggers the right-focusing mode through an application of the \facerule{AL-arr} or the \facerule{AL-mp} rule,
respectively.

\begin{wrapfigure}{l}{47mm}
  \centering
  \vspace{-10pt}
  \begin{setofrules}[45mm]{$A\inAnd B$}{Intersection membership}
    \drule{}{ }{A\inAnd A}\and
    \drule{}{A\inAnd \ottnt{B_{{\mathrm{1}}}}}{A\inAnd \ottnt{B_{{\mathrm{1}}}}  \, \& \,  \ottnt{B_{{\mathrm{2}}}}}\and
    \drule{}{A\inAnd \ottnt{B_{{\mathrm{2}}}}}{A\inAnd \ottnt{B_{{\mathrm{1}}}}  \, \& \,  \ottnt{B_{{\mathrm{2}}}}}
  \end{setofrules}
  \vspace{-5pt}
  \caption{Definition of intersection membership}
  \label{fig:AndMember}
  \vspace{-5pt}
\end{wrapfigure}
Their definition uses an auxiliary intersection-membership relation, written as $\ottnt{A}\inAnd \ottnt{B}$ and defined in
Figure~\ref{fig:AndMember}.
The formula $\ottnt{A}\inAnd \ottnt{B}$ expresses that $\ottnt{B}$ is the type $\ottnt{A}$ possibly intersected with more types.
\[
  \ottnt{B} = (\ldots \& (\ottnt{A}\& \ldots)) \& \ldots
\]
We say that type $\ottnt{A}$ is a component of type $\ottnt{B}$.

The following results on intersection membership are useful for the rest of the proof.
The definition of the left-arrow depth function \(\lad{\cdot}\), here referred to as our \emph{norm}, is given in
Section~\ref{subsec:algorithm:metatheory} of the paper.

\begin{lemma}[Intersection membership decreases the norm]
  If \(\ottnt{A}\inAnd \ottnt{B}\), then \(\lad{\ottnt{A}} \leq \lad{B}\).\\
\end{lemma}
\begin{remark}
  \label{remark:AndMember}
  Every type has a finite number of components. Formally, for a given type $\ottnt{B}$, the set
  \(
    \{ \ottnt{A}\mid \ottnt{A}\inAnd \ottnt{B}\}
  \)
  is finite. We also say that, given type \(\ottnt{B}\), the intersection membership relation produces
  a finite number of elements (that belong to that relation).
\end{remark}

\subsection{Arrow-loops are not a source of non-termination}
\begin{figure}
  \begin{setofrules}{\(\arrloop{\ottnt{A}}{\ottnt{B}}{\ottnt{B'}}{\ottnt{A'}}\)}{Single arrow-loops}
    \drule{LA-base}{\ottnt{B_{{\mathrm{1}}}}  \rightarrow  \ottnt{B_{{\mathrm{2}}}}\inAnd \ottnt{B}}{\arrloop{\ottnt{A_{{\mathrm{1}}}}  \rightarrow  \ottnt{A_{{\mathrm{2}}}}}{\ottnt{B}}{\ottnt{B_{{\mathrm{1}}}}}{\ottnt{A_{{\mathrm{1}}}}}}\and
    \drule{LA-mp}{\arrloop{\ottnt{A_{{\mathrm{2}}}}}{\ottnt{B}}{\ottnt{B'}}{\ottnt{A'}}}{\arrloop{\ottnt{A_{{\mathrm{1}}}}  \rightarrow  \ottnt{A_{{\mathrm{2}}}}}{\ottnt{B}}{\ottnt{B'}}{\ottnt{A'}}}\and
    \drule{LA-arr}{\ottnt{B_{{\mathrm{1}}}}  \rightarrow  \ottnt{B_{{\mathrm{2}}}}\inAnd \ottnt{B} \and \arrloop{\ottnt{A_{{\mathrm{2}}}}}{\ottnt{B_{{\mathrm{2}}}}}{\ottnt{B'}}{\ottnt{A'}}}{\arrloop{\ottnt{A_{{\mathrm{1}}}}  \rightarrow  \ottnt{A_{{\mathrm{2}}}}}{\ottnt{B}}{\ottnt{B'}}{\ottnt{A'}}}\and
    \drule{LA-andL}{\arrloop{\ottnt{A_{{\mathrm{1}}}}}{\ottnt{B}}{\ottnt{B'}}{\ottnt{A'}}}{\arrloop{\ottnt{A_{{\mathrm{1}}}}  \, \& \,  \ottnt{A_{{\mathrm{2}}}}}{\ottnt{B}}{\ottnt{B'}}{\ottnt{A'}}}\and
    \drule{LA-andR}{\arrloop{\ottnt{A_{{\mathrm{2}}}}}{\ottnt{B}}{\ottnt{B'}}{\ottnt{A'}}}{\arrloop{\ottnt{A_{{\mathrm{1}}}}  \, \& \,  \ottnt{A_{{\mathrm{2}}}}}{\ottnt{B}}{\ottnt{B'}}{\ottnt{A'}}}
  \end{setofrules}
  \caption{Single arrow-loops, formalized}
  \label{fig:LoopArr}
\end{figure}
The arrow-loop relation, $\arrloopsymbol$, between right-focusing subtyping judgments, shown in Figure~\ref{fig:LoopArr},
captures single cycles of mutual recursion through an \facerule{AL-arr} rule application.
Informally, $\arrloop{\ottnt{A}}{\ottnt{B}}{\ottnt{B'}}{\ottnt{A'}}$ expresses that resolving $\noEvAlgSubR{[]}{\ottnt{A}}{\ottnt{B}}$,
will trigger resolving $\noEvAlgSubR{[]}{\ottnt{B'}}{\ottnt{A'}}$ through a single arrow loop.

There are many ways in which \facerule{AL-arr} can be reached starting from the right-focusing mode.
The most immediate case is when resolving
$\noEvAlgSubR{[]}{\ottnt{A_{{\mathrm{1}}}}  \rightarrow  \ottnt{A_{{\mathrm{2}}}}}{\ottnt{B_{{\mathrm{1}}}}  \rightarrow  \ottnt{B_{{\mathrm{2}}}}}$, which triggers $\noEvAlgSubR{[]}{\ottnt{B_{{\mathrm{1}}}}}{\ottnt{A_{{\mathrm{1}}}}}$.
We can generalize this to account for intersections of arrow types in the right hand side of the initial subtyping judgment.
Thus, trying to resolve $\noEvAlgSubR{[]}{\ottnt{A_{{\mathrm{1}}}}  \rightarrow  \ottnt{A_{{\mathrm{2}}}}}{\ottnt{B}}$ will ignite a right-focusing call to resolve
$\noEvAlgSubR{[]}{\ottnt{B_{{\mathrm{1}}}}}{\ottnt{A_{{\mathrm{1}}}}}$, for every arrow type component $\ottnt{B_{{\mathrm{1}}}}  \rightarrow  \ottnt{B_{{\mathrm{2}}}}$ of $\ottnt{B}$ (\facerule{LA-base}).
In addition, for every arrow type component $\ottnt{B_{{\mathrm{1}}}}  \rightarrow  \ottnt{B_{{\mathrm{2}}}}\inAnd \ottnt{B}$, if starting from $\noEvAlgSubR{[]}{\ottnt{A_{{\mathrm{2}}}}}{\ottnt{B_{{\mathrm{2}}}}}$
triggers an arrow-loop on $\noEvAlgSubR{[]}{\ottnt{B'}}{\ottnt{A'}}$, then that loop is reached also starting from
$\noEvAlgSubR{[]}{\ottnt{A_{{\mathrm{1}}}}  \rightarrow  \ottnt{A_{{\mathrm{2}}}}}{\ottnt{B}}$.
These are arrow-loops that occur under the left-focusing subderivation of an \facerule{AL-arr} rule (\facerule{LA-arr}).
Similarly, rules \facerule{LA-andL}, \facerule{LA-andR} and \facerule{LA-mp} encode the arrow-loops that occur under an
application of \facerule{AL-and1} or \facerule{AL-and2} or the left-focusing subderivation of \facerule{AL-mp},
respectively.

In Figure~\ref{fig:ArrLoopViz}, 
$\arrloopsymbol$ relates the right-focusing judgment at the root of the tree with the
right-focusing premise of every \facerule{AL-arr} application that is \textcolor{ForestGreen}{outlined with green borders}.
The innermost \facerule{AL-arr} application is not outlined because, starting from the root, it takes more than one
(in particular, it takes two) single mutual recursion cycles to reach it.
Such arrow-loops, nested in the right-focusing subderivation of a previously applied \facerule{AL-arr}, are encoded in the
transitive closure of $\arrloopsymbol$.

\begin{remark}
  \label{remark:LoopArr}
  Given types $\ottnt{A}$ and $\ottnt{B}$, because of Remark~\ref{remark:AndMember}, rule \facerule{LA-base} produces a finite
  number of elements (possibly zero) in the $\arrloopsymbol$-relation.
  Since the rest of the rules in Figure~\ref{fig:LoopArr} are structurally decreasing on argument $\ottnt{A}$, it follows that
  given types $\ottnt{A}$ and $\ottnt{B}$, the set \(\{(\ottnt{B'},\ottnt{A'})\mid \arrloop{\ottnt{A}}{\ottnt{B}}{\ottnt{B'}}{\ottnt{A'}}\})\}\) is
  finite.
  Note that the left premise of rule \facerule{LA-arr} contributes by a finite factor in the number of elements that this
  rule can produce.
\end{remark}

\begin{lemma}[Arrow-loops strictly decrease the norm]
  \label{lemma:ArrLoop_ldepth}
  If \(\arrloop{\ottnt{A}}{\ottnt{B}}{\ottnt{B'}}{\ottnt{A'}}\), then \(\lad{\noEvAlgSubR{[]}{\ottnt{A}}{\ottnt{B}}} < \lad{\noEvAlgSubR{[]}{\ottnt{B'}}{\ottnt{A'}}}\).
\end{lemma}

It follows from the above remark and from Lemma~\ref{lemma:ArrLoop_ldepth} that whenever the algorithm is in right-focusing
mode, it re-enters that mode through \facerule{AL-arr} a finite number of times and with a right-focusing judgment that is
strictly smaller than the previous (w.r.t. our norm).
This ensures that arrow-loops are not a source of non-termination.

\subsection{Loop detection for termination of MP-loops}
\begin{figure}
  \begin{setofrules}{\(\mploop{\ottnt{A}}{\ottnt{B}}{\ottnt{B'}}\)}{Single mp-loops}
    \drule{LM-base}{ }{\mploop{\ottnt{A_{{\mathrm{1}}}}  \rightarrow  \ottnt{A_{{\mathrm{2}}}}}{\ottnt{B}}{\ottnt{A_{{\mathrm{1}}}}}}\and
    \drule{LM-mp}{\mploop{\ottnt{A_{{\mathrm{2}}}}}{\ottnt{B}}{\ottnt{B'}}}{\mploop{\ottnt{A_{{\mathrm{1}}}}  \rightarrow  \ottnt{A_{{\mathrm{2}}}}}{\ottnt{B}}{\ottnt{B'}}}\and
    \drule{LM-arr}{\ottnt{B_{{\mathrm{1}}}}  \rightarrow  \ottnt{B_{{\mathrm{2}}}}\inAnd \ottnt{B} \and \mploop{\ottnt{A_{{\mathrm{2}}}}}{\ottnt{B_{{\mathrm{2}}}}}{\ottnt{B'}}}{\mploop{\ottnt{A_{{\mathrm{1}}}}  \rightarrow  \ottnt{A_{{\mathrm{2}}}}}{\ottnt{B}}{\ottnt{B_{{\mathrm{1}}}}  \rightarrow  \ottnt{B'}}}\and
    \drule{LM-andL}{\mploop{\ottnt{A_{{\mathrm{1}}}}}{\ottnt{B}}{\ottnt{B'}}}{\mploop{\ottnt{A_{{\mathrm{1}}}}  \, \& \,  \ottnt{A_{{\mathrm{2}}}}}{\ottnt{B}}{\ottnt{B'}}}\and
    \drule{LM-andR}{\mploop{\ottnt{A_{{\mathrm{2}}}}}{\ottnt{B}}{\ottnt{B'}}}{\mploop{\ottnt{A_{{\mathrm{1}}}}  \, \& \,  \ottnt{A_{{\mathrm{2}}}}}{\ottnt{B}}{\ottnt{B'}}}
  \end{setofrules}
  \caption{Single mp-loops, formalized}
  \label{fig:LoopMP}
\end{figure}

Figure~\ref{fig:LoopMP} shows the definition of the mp-loop family of relations, which,
similarly to the arrow-loop relation, encodes single cycles of mutual recursion
through \facerule{AL-mp} rule applications.
Informally, $\mploop{\ottnt{A}}{\ottnt{B}}{\ottnt{B'}}$ expresses that resolving the judgment
$\noEvAlgSubR{[]}{\ottnt{A}}{\ottnt{B}}$, generates the call $\noEvAlgSubR{[]}{\ottnt{A}}{\ottnt{B'}}$ through a single mp-loop.

A single right-focusing judgment may generate several mp-loops.
The most immediate case, encoded by rule \facerule{LM-base}, is when resolving \(\noEvAlgSubR{[]}{\ottnt{A_{{\mathrm{1}}}}  \rightarrow  \ottnt{A_{{\mathrm{2}}}}}{\ottnt{B}}\),
which triggers the resolution of \(\noEvAlgSubR{[]}{\ottnt{A_{{\mathrm{1}}}}  \rightarrow  \ottnt{A_{{\mathrm{2}}}}}{\ottnt{A_{{\mathrm{1}}}}}\).
Subsequent applications of the \facerule{AL-mp} rule that occur under the left-focusing branch of a previous
\facerule{AL-mp} rule application are encoded by rule \facerule{LM-mp}.
Similarly, rule \facerule{LM-arr} encodes applications of the \facerule{AL-mp} rule under the left-focusing branch of a
previously applied \facerule{AL-arr} rule.
Lastly, rules \facerule{LM-andL} and \facerule{LM-andR} encode mp-loops that occur under an
application of \facerule{AL-and1} and \facerule{AL-and2}, respectively.

Note that for a type $\ottnt{A}$, \(\mploop{\ottnt{A}}{}{}\) is a set (not a multiset) and, thus, contains unique pairs of related
right-focusing judgments.
Put differently, starting resolution on \(\noEvAlgSubR{[]}{\ottnt{A}}{\ottnt{B}}\), the judgment does not distinguish between
distinct applications of the \facerule{AL-mp} rule that generate the same right-focusing call
\(\noEvAlgSubR{[]}{\ottnt{A}}{\ottnt{B'}}\).
For example, starting with $\noEvAlgSubR{[]}{\ottnt{A_{{\mathrm{1}}}}  \rightarrow  \ottnt{A_{{\mathrm{2}}}}}{\mathsf{Nat}  \, \& \,   \mathsf{Bool}   \rightarrow  \mathsf{Nat}}$, the algorithm branches by rule
\facerule{AR-and} and tries to apply the \facerule{AL-mp} rule within both branches.
Both applications generate the same right-focusing call, $\noEvAlgSubR{[]}{\ottnt{A_{{\mathrm{1}}}}  \rightarrow  \ottnt{A_{{\mathrm{2}}}}}{\ottnt{A_{{\mathrm{1}}}}}$,
but are encoded as one mp-loop through rule \facerule{LM-base}.
This justifies the use of an arbitrary type $\ottnt{B}$ in the left component of the relation \(\mploop{\ottnt{A_{{\mathrm{1}}}}  \rightarrow  \ottnt{A_{{\mathrm{2}}}}}{}{}\),
in the conclusion of the rules \facerule{LM-base} and \facerule{LM-arr} of Figure~\ref{fig:LoopMP}.

\begin{remark}
  \label{remark:LoopMP}
  The rules of the mp-loop relation family are structurally decreasing on the index type $\ottnt{A}$.
  Therefore, taking into consideration Remark~\ref{remark:AndMember} to account for rule \facerule{LM-arr}, it follows that
  for given types $\ottnt{A}$ and $\ottnt{B}$, the set \(\{ B'\mid\mploop{\ottnt{A}}{\ottnt{B}}{\ottnt{B'}}\}\) is finite.
\end{remark}

\begin{lemma}[Mp-loops non-strictly decrease the norm]
  \label{lemma:LoopMP_ldepth}
  If \(\mploop{\ottnt{A}}{\ottnt{B}}{\ottnt{B'}}\), then \(\lad{\noEvAlgSubR{[]}{\ottnt{A}}{\ottnt{B}}} \leq \lad{\noEvAlgSubR{[]}{\ottnt{A}}{\ottnt{B'}}}\).
\end{lemma}

Because a mp-loop signifies the start of a number of sequences of mp-loops to follow, Remark~\ref{remark:LoopMP}
essentially means that when the algorithm is in right-focusing mode it will generate a finite number of mp-loop sequences.
However, Lemma~\ref{lemma:LoopMP_ldepth} does not guarantee that the right-focusing judgment triggered by a mp-loop is
strictly smaller than the initial right-focusing judgment.
As a result, although the \emph{number} of mp-loop sequences is finite, not all sequences are necessarily finite.
Some of them might be infinite, with each mp-loop triggering a right-focusing call on a judgment of equal norm.

In the following subsection, we show that the algorithm's search space, consisting of all possible recursive calls
that the algorithm may trigger given an initial subtyping judgment to resolve, is finite.
As a direct result, the algorithm has a finite number of possible mp-loops to follow.
Then, by the pigeon hole principle and the algorithm's determinism, we can conclude that for an mp-loop sequence to be
infinite, it must contain an infinite repetition of a finite number of finite mp-loop subsequences.
Thus, a mechanism that detects repetitions of mp-loops is sufficient to salvage the algorithm from non-termination.%

\subsection{The search space is finite}
\begin{figure}[h]
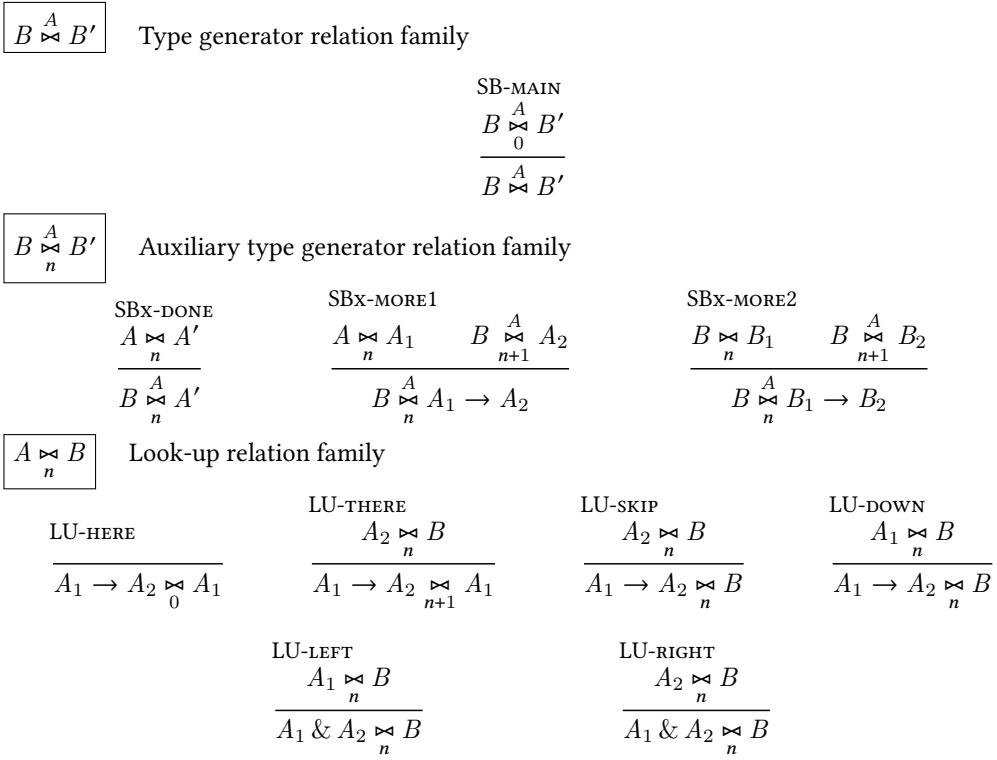

  \begin{setofrules}{\(\SBound{\ottnt{A}}{\ottnt{B}}{\ottnt{B'}}\)}{Type generator relation family}
    \drule{SB-main}{\SBoundAux{0}{\ottnt{A}}{\ottnt{B}}{\ottnt{B'}}}{\SBound{\ottnt{A}}{\ottnt{B}}{\ottnt{B'}}}
  \end{setofrules}
  \begin{setofrules}{\(\SBoundAux{n}{\ottnt{A}}{\ottnt{B}}{\ottnt{B'}}\)}{Auxiliary type generator relation family}
    \drule{SBx-done}{\SLookup{n}{\ottnt{A}}{\ottnt{A'}}}{\SBoundAux{n}{\ottnt{A}}{\ottnt{B}}{\ottnt{A'}}}\and
    \drule{SBx-more1}{\SLookup{n}{\ottnt{A}}{\ottnt{A_{{\mathrm{1}}}}}\and \SBoundAux{n+1}{\ottnt{A}}{\ottnt{B}}{\ottnt{A_{{\mathrm{2}}}}}}{\SBoundAux{n}{\ottnt{A}}{\ottnt{B}}{\ottnt{A_{{\mathrm{1}}}}  \rightarrow  \ottnt{A_{{\mathrm{2}}}}}}\and
    \drule{SBx-more2}{\SLookup{n}{\ottnt{B}}{\ottnt{B_{{\mathrm{1}}}}}\and \SBoundAux{n+1}{\ottnt{A}}{\ottnt{B}}{\ottnt{B_{{\mathrm{2}}}}}}{\SBoundAux{n}{\ottnt{A}}{\ottnt{B}}{\ottnt{B_{{\mathrm{1}}}}  \rightarrow  \ottnt{B_{{\mathrm{2}}}}}}
  \end{setofrules}
  \begin{setofrules}{\(\SLookup{n}{\ottnt{A}}{\ottnt{B}}\)}{Look-up relation family}
    \drule{LU-here}{ }{\SLookup{0}{\ottnt{A_{{\mathrm{1}}}}  \rightarrow  \ottnt{A_{{\mathrm{2}}}}}{\ottnt{A_{{\mathrm{1}}}}}}\and
    \drule{LU-there}{\SLookup{n}{\ottnt{A_{{\mathrm{2}}}}}{\ottnt{B}}}{\SLookup{n+1}{\ottnt{A_{{\mathrm{1}}}}  \rightarrow  \ottnt{A_{{\mathrm{2}}}}}{\ottnt{A_{{\mathrm{1}}}}}}\and
    \drule{LU-skip}{\SLookup{n}{\ottnt{A_{{\mathrm{2}}}}}{\ottnt{B}}}{\SLookup{n}{\ottnt{A_{{\mathrm{1}}}}  \rightarrow  \ottnt{A_{{\mathrm{2}}}}}{\ottnt{B}}}\and
    \drule{LU-down}{\SLookup{n}{\ottnt{A_{{\mathrm{1}}}}}{\ottnt{B}}}{\SLookup{n}{\ottnt{A_{{\mathrm{1}}}}  \rightarrow  \ottnt{A_{{\mathrm{2}}}}}{\ottnt{B}}}\and
    \drule{LU-left}{\SLookup{n}{\ottnt{A_{{\mathrm{1}}}}}{\ottnt{B}}}{\SLookup{n}{\ottnt{A_{{\mathrm{1}}}}  \, \& \,  \ottnt{A_{{\mathrm{2}}}}}{\ottnt{B}}}\and
    \drule{LU-right}{\SLookup{n}{\ottnt{A_{{\mathrm{2}}}}}{\ottnt{B}}}{\SLookup{n}{\ottnt{A_{{\mathrm{1}}}}  \, \& \,  \ottnt{A_{{\mathrm{2}}}}}{\ottnt{B}}}
  \end{setofrules}
  \caption{An over-approximation of mp-loop sequences}
  \label{fig:SBound}
\end{figure}

To establish the finiteness of the algorithm's search space, we only need to consider the subspace generated by rule
\facerule{AL-mp}, since it is the only rule that threatens the algorithm's termination.
For that, we devise the \emph{type generator} relation family, shown at the top of Figure~\ref{fig:SBound},
which uses two auxiliary definitions, \(\SBoundAux{n}{\ottnt{A}}{\ottnt{B}}{\ottnt{B'}}\) and \(\SLookup{n}{\ottnt{A}}{\ottnt{B}}\).

Although an understanding of the two auxiliary definitions is necessary to understand what the type generator family
relates, for the purposes of this proof we only need to realize two properties it satisfies.
First, the type generator relation family is an over-approximation of all possible mp-loop sequences, given an initial
right-focusing subtyping judgment to resolve, and second, we need a finiteness property similar to that in
Remarks~\ref{remark:AndMember},~\ref{remark:LoopArr} and~\ref{remark:LoopMP}.
The first property establishes that the type generator family contains an encoding of the algorithm's
subspace that is generated by the \facerule{Al-mp} rule (or, equivalently, by mp-loops).
Then, the finiteness property allows us to conclude that the algorithm's subspace generated by mp-loops is finite.

\subsubsection{An over-approximation of mp-loop sequences}
\begin{figure}[h]
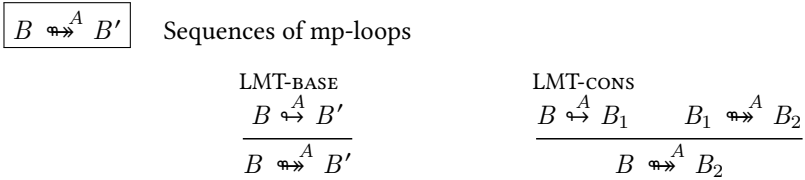

  \begin{setofrules}{\(\mploopseq{\ottnt{A}}{\ottnt{B}}{\ottnt{B'}}\)}{Sequences of mp-loops}
    \drule{LMT-base}{\mploop{\ottnt{A}}{\ottnt{B}}{\ottnt{B'}}}{\mploopseq{\ottnt{A}}{\ottnt{B}}{\ottnt{B'}}}\and
    \drule{LMT-cons}{\mploop{\ottnt{A}}{\ottnt{B}}{\ottnt{B_{{\mathrm{1}}}}} \and \mploopseq{\ottnt{A}}{\ottnt{B_{{\mathrm{1}}}}}{\ottnt{B_{{\mathrm{2}}}}}}{\mploopseq{\ottnt{A}}{\ottnt{B}}{\ottnt{B_{{\mathrm{2}}}}}}
  \end{setofrules}
  \caption{Sequences of mp-loops, formalized}
  \label{fig:MPTrans}
\end{figure}

Before we proceed, we formalize the notion of a sequence of mp-loops.
Naturally, sequences of mp-loops are modeled by the transitive closure of the mp-loops relation family,
as shown in Figure~\ref{fig:MPTrans}.
A member \(\mploopseq{\ottnt{A}}{\ottnt{B}}{\ottnt{B''}}\) of the \(\mploopseq{\ottnt{A}}{}{}\)-relation models a sequence of mp-loops that
starts with a \(\noEvAlgSubR{[]}{\ottnt{A}}{\ottnt{B}}\) and ends with a \(\noEvAlgSubR{[]}{\ottnt{A}}{\ottnt{B''}}\) right-focusing
judgment, after possibly going through intermediate mp-loops of the form \(\noEvAlgSubR{[]}{\ottnt{A}}{\ottnt{B'}}\).

Lemma~\ref{MPFinite} justifies that the type generator family is an over-approximation of all possible mp-loop sequences.
\begin{lemma}
  \label{MPFinite}
  If \(\mploopseq{\ottnt{A}}{\ottnt{B}}{\ottnt{B'}}\), then \(\SBound{\ottnt{A}}{\ottnt{B}}{\ottnt{B'}}\).
\end{lemma}

\subsubsection{A finiteness property for the over-approximation}
First, we examine the look-up relation family, on which the definition of the type generator family depends.
\begin{remark}
  \label{remark:SLookup}
  Consider a member \(\SLookup{n}{\ottnt{A}}{\ottnt{A'}}\) of the look-up relation family.
  The inductive rules of this definition are decreasing on the lexicographic order of the pair $(n, \ottnt{A})$.
  Therefore, given a natural number $n$ and a type $\ottnt{A}$, the set \(\{\ottnt{A'}\mid \SLookup{n}{\ottnt{A}}{\ottnt{A'}}\}\)
  is finite.
\end{remark}

Next, we examine the auxiliary type generator family.
Consider an element \(\SBoundAux{n}{\ottnt{A}}{\ottnt{B}}{\ottnt{B'}}\).
From Remark~\ref{remark:SLookup}, it follows that given $n$, $A$ and $B$,
rule \facerule{SBx-done} produces a finite number of elements in the auxiliary type generator family.
Similarly, given $n$, $A$ and $B$, the left premise of the two remaining rules contributes by a finite factor on the number
of elements that each of those rules can produce.
In particular, the $\ottnt{A_{{\mathrm{1}}}}$ and $\ottnt{B_{{\mathrm{1}}}}$ types in the conclusion of rules \facerule{SBx-more1} and \facerule{SBx-more2},
respectively, are produced by the look-up relation family (on the left premise of the corresponding rule) and thus they can
only be finite in number.
Yet, we are left with the possibility to choose arrow types of unconstrained size for the $\ottnt{A_{{\mathrm{2}}}}$'s and $\ottnt{B_{{\mathrm{2}}}}$'s of those
two rules.
At first sight, we can have countably infinite choices for the right hand side component of this family relation.

\begin{figure}
  \begin{minipage}[b]{0.48\textwidth}
      \begin{align*}
        \tyrange{ \mathsf{Nat} } &= \tyrange{ \mathsf{top} } = 0\\
        \tyrange{\ottnt{A_{{\mathrm{1}}}}  \rightarrow  \ottnt{A_{{\mathrm{2}}}}} &= \max(\tyrange{\ottnt{A_{{\mathrm{1}}}}}, \tyrange{\ottnt{A_{{\mathrm{2}}}}}+1)\\
        \tyrange{\ottnt{A_{{\mathrm{1}}}}  \, \& \,  \ottnt{A_{{\mathrm{2}}}}} &= \max(\tyrange{\ottnt{A_{{\mathrm{1}}}}}, \tyrange{\ottnt{A_{{\mathrm{2}}}}})
      \end{align*}
    \caption{The range of a type}
    \label{fig:tyrange}
  \end{minipage}
  \begin{minipage}[b]{0.48\textwidth}
    \begin{mathpar}
      \inferrule*{\SLookup{n}{\ottnt{A}}{\ottnt{A_{{\mathrm{1}}}}}\and
        \inferrule*{{\SLookup{n+1}{\ottnt{A}}{\ottnt{A_{{\mathrm{2}}}}}}\and
          \inferrule*[vdots=1cm]{}{\SBoundAux{n+k}{\ottnt{A}}{\ottnt{B}}{B_{k+1}'}}
        }{\SBoundAux{n+1}{\ottnt{A}}{\ottnt{B}}{\ottnt{B'_{{\mathrm{2}}}}\to\cdots\to B_{k+1}'}}
      }{\SBoundAux{n}{\ottnt{A}}{\ottnt{B}}{\ottnt{B'_{{\mathrm{1}}}}\to\cdots\to B_{k+1}'}}
    \end{mathpar}
    \caption{Example derivation}
    \label{fig:many_more1}
  \end{minipage}
\end{figure}

Fortunately, the following lemma limits the size of the arrow types that can be chosen.
It expresses that given a natural number index $n$ and types \(\ottnt{A}\) and \(\ottnt{B}\), if $n$ is greater than the range of
both types, then there is no type \(\ottnt{B'}\) such that \(\SBoundAux{n}{\ottnt{A}}{\ottnt{B}}{\ottnt{B'}}\).
The range $\tyrange{\ottnt{A}}$ of a type \(\ottnt{A}\), defined in Figure~\ref{fig:tyrange} is a function in the role of a norm
for types.

\begin{lemma}
  If $max(\tyrange{\ottnt{A}}, \tyrange{\ottnt{B}}) < n$, then
  \(\SBoundAux{n}{\ottnt{A}}{\ottnt{B}}{\ottnt{B'}}\) does not hold 
  for any type $\ottnt{B'}$.
\end{lemma}

This lemma does not directly limit the choices for $\ottnt{B'}$ to a finite number, because it does not impose any constraint
on the $\ottnt{B'}$ component of the auxiliary type generator family. Rather, it constrains $n$ to be necessarily less than
or equal to the range of at least one of the two given types.

However, specializing $\ottnt{B'}$ to be an arrow type, since we are interested in applications of rules \facerule{SBx-more1}
and \facerule{SBx-more2},
note that these two rules increase the natural number index of the auxiliary type generator relation in
their right premise.
As exemplified by the derivation in Figure~\ref{fig:many_more1}, the size of the arrow type \(\ottnt{B'}\) is limited so that
if its derivation contains a branch with $k$ applications of those two rules, then
$n + k \leq \max(\tyrange{\ottnt{A}}, \tyrange{\ottnt{B}})$.

With this, we can conclude that given $n$, $A$ and $B$, rules \facerule{SBx-more1} and \facerule{SBx-more2} generate a
finite number of elements in the auxiliary type generator relation. Altogether, given $n$, $\ottnt{A}$ and $\ottnt{B}$, the set
$\{B'\mid\SBoundAux{n}{\ottnt{A}}{\ottnt{B}}{\ottnt{B'}}\}$ is finite.

Finally, the finiteness result about the type generator family, or to say, about the over-approximation of the algorithm's
possible mp-loop sequences, follows directly: given types $\ottnt{A}$ and $\ottnt{B}$, the set
$\{B'\mid\SBound{\ottnt{A}}{\ottnt{B}}{\ottnt{B'}}\}$ is finite.
\qed